%% file: dissertation_compact.tex
\pdfoutput=1

\documentclass{Dissertate_compact}

\graphicspath{ {figures/} }

\begin{document}

\tikzstyle{startstop} = [rectangle, rounded corners, minimum width=2cm, minimum height=0.5cm,text centered, draw=black]
\tikzstyle{io} = [trapezium, trapezium left angle=70, trapezium right angle=110, minimum width=3cm, minimum height=1cm, text centered, draw=black]
\tikzstyle{process} = [rectangle, minimum width=2cm, minimum height=0.5cm, text centered, draw=black, align=center]
\tikzstyle{decision} = [ellipse, minimum width=2cm, minimum height=1cm, text centered, draw=black]
\tikzstyle{arrow} = [thick,<->,>=stealth]
\tikzstyle{darrow} = [thick,<->,>=stealth,dashed]
\tikzstyle{sarrow} = [thick,->,>=stealth]

\input{frontmatter/personalize}

\frontmatter


\setstretch{1.2}


\setcounter{chapter}{0}  
\include{chapters/introduction}

\include{chapters/chapter1}

\include{chapters/chapter2}

\include{chapters/chapter3}
\include{chapters/chapter4}

\include{chapters/chapter5}
\include{chapters/conclusion}


\addcontentsline{toc}{chapter}{References}
\bibliography{references}
\phantomsection

\end{document}

%% file: frontmatter/personalize.tex
\title{Performance Comparison of Dual Connectivity and Hard Handover for LTE-5G Tight Integration in mmWave Cellular Networks  }
\author{Michele Polese}

\advisor{Michele Zorzi}

\committeeInternalOne{Person Inside One}
\committeeInternalTwo{Person Inside Two}

\coadvisorOne{Marco Mezzavilla}
\coadvisorTwo{Equally D. Researcher}
\committeeInternal{Person Inside}

\degree{Master of Science in Telecommunication Engineering}
\field{Engineering}
\degreeyear{2016}
\degreemonth{July}
\department{Engineering}

%% file: chapters/introduction.tex
\chapter{Introduction}\label{chap:intro}

\begin{figure}[t]
	\centering
 	\includegraphics[width=0.9\textwidth]{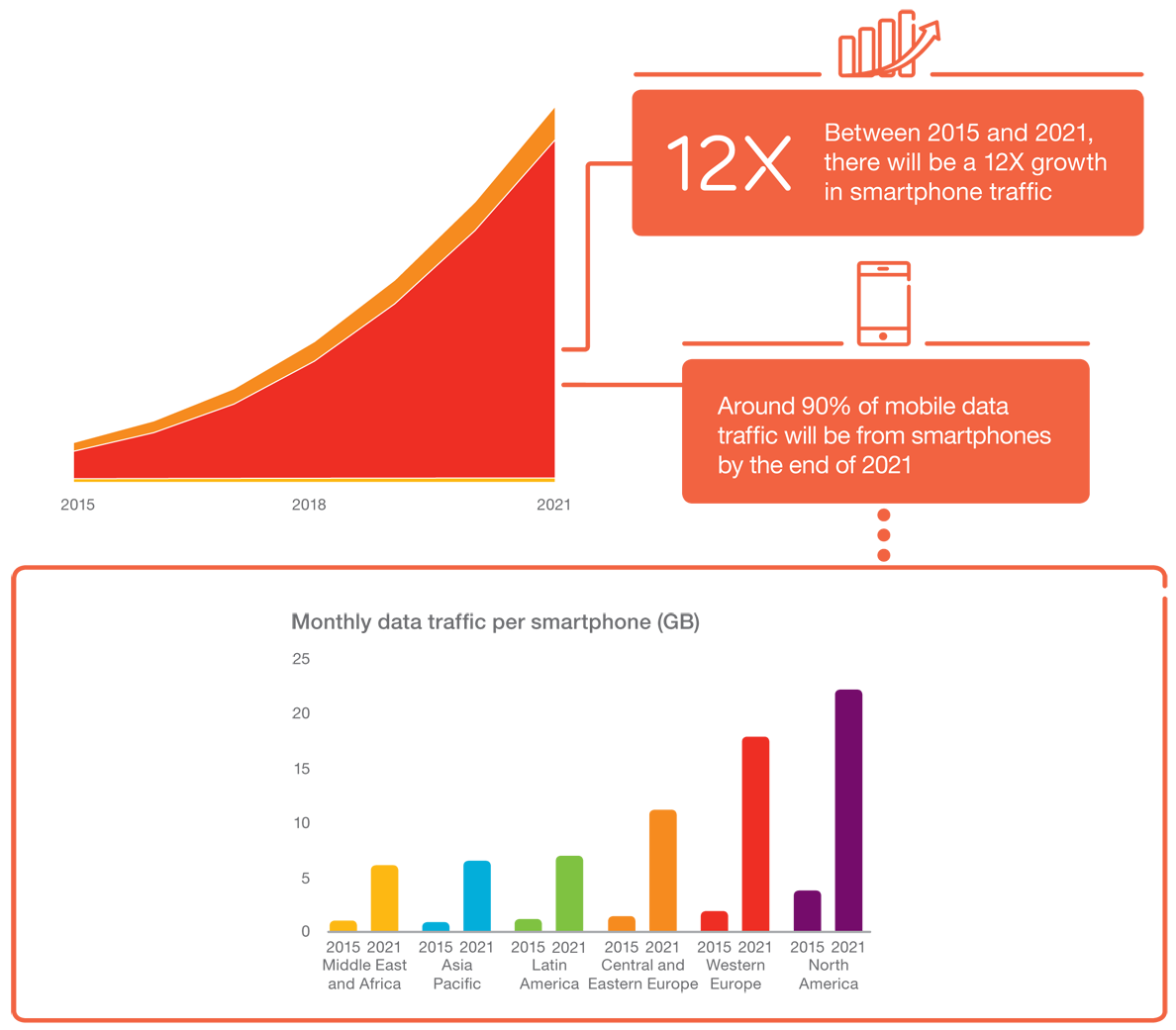}
	\caption{Ericsson Mobility Report mobile traffic outlook, from~\cite{ericmore}}
	\label{fig:ericsson} 
\end{figure}

The next generation mobile network (5G) will become a reality before 2020, driven by an increase in mobile traffic demand and by a variety of use cases that cannot be satisfied by the current LTE networks. 
According to the latest Ericsson Mobility Report~\cite{ericmore}, the smartphone traffic on mobile networks is expected to increase by 12 times before 2021. As shown in Fig.~\ref{fig:ericsson}, the monthly traffic per smartphone in Europe and the United States will be greater than 15 GB. 

The 5G cellular network is required to address these traffic demands, a growth of connected devices, and to define new business models for network operators. It will be designed with a holistic approach, considering different use cases in order to provide natively an optimized experience for each of them. According to the guidelines in~\cite{ngmn5g}, 5G networks should support:
\begin{itemize}
	\item a cell--edge rate of 50 Mbit/s or more, and in general a cell throughput higher than 1 Gbit/s, in order to support 4K video streaming and a large number of connected users; 
	\item an ultra-low end-to-end latency, preferably below 10 ms, with a stricter requirement of 1 ms latency for specific applications (tactile internet, remote industrial controls);
	\item ultra-high service availability, with high reliability and a consistent user experience in the network;
	\item a massive deployment of Machine Type Communications (MTC) devices, which have to be energy efficient and use a very low power.
\end{itemize}

In the last few years, the research on 5G became a hot topic in the telecommunication area. Indeed there are several challenges to address in order to satisfy these requirements. The low latency objective, for example, may require a re-design of the core network. The massive MTC deployment will need cheap electronics and simple networking procedures. 

The main challenge is however to reach the ultra-high throughput objective. A possible enabler is the use of mmWave frequencies. Indeed, the spectrum at lower microWave frequencies is very fragmented, and the allocation of large chunks of spectrum (in order to obtain large available bandwidths) is not possible. On the contrary, in the mmWave band there is a chance to allocate gigahertz bandwidths to network operators~\cite{rangan14}. 

However, several issues must be faced when using carrier frequencies greater than 10 GHz: (i) high isotropic pathloss; (ii) blockage from buildings and also from the human body; (iii) attenuation given by foliage and heavy rain~\cite{samsungmmb}. 

Therefore, mmWave links may provide a very high throughput, but their quality is variable. In particular, a User Equipment (UE) may experience an outage, or an SINR too low to communicate with the mmWave evolved Node Base (eNB). A possible solution is to use the LTE network, which operates on microWave frequencies, as a fallback. In current mobile networks the usual procedure used to fallback is a handover. However, the conventional LTE procedure may be too slow, and there may be an interval in which the cellular service is unavailable. 

In this Thesis, an alternative to the standard handover is investigated. Firstly, a more general topic is discussed and analyzed, i.e., the integration between LTE and 5G networks. In an integrated system, a UE is in connected state to both LTE and mmWave eNBs. Therefore, this is called a \textit{dual connected} setup. Secondly, this system will be analyzed for the usage of fast switching, i.e., only one of the two eNBs serves data to the UE, but it is possible to switch from one Radio Access Technology (RAT) to the other with a single control message, without the involvement of the core network. There are already Dual Connectivity (DC) solutions standardized by 3GPP~\cite{36300}, and in some papers as~\cite{dasilva},~\cite{cloud5G} there are proposals on how LTE and 5G should integrate. The main contributions of this Thesis are (i) the evaluation of a possible architecture for integration at the Packet Data Convergence Protocol (PDCP) layer; (ii) the proposal of new network procedures to enable this solution; and (iii) an implementation of this system for the ns--3 simulator, in order to assess its performance with a thorough simulation campaign.

\vspace{2em}

The thesis is organized as follows:
\begin{itemize}
	\item Chapter~\ref{chap:5G} describes the enabling technologies for 5G networks, with a particular focus on mmWave communications;
	\item Chapter~\ref{chap:integration} reviews which is the state of the art on LTE-5G tight integration. Moreover, the 3GPP proposals on DC for LTE are illustrated. A brief introduction on the LTE protocol stack is also given, and LTE standard handover procedures are shown;
	\item Chapter~\ref{chap:ns3} introduces the New York University (NYU) mmWave module for ns--3, by describing in detail the channel model employed and the functionalities provided. Moreover, the LTE module for ns--3 is briefly described;
	\item Chapter~\ref{chap:impl} describes the proposed architecture and our new procedures for LTE-5G tight integration with dual connectivity. Then, our new implementation of this architecture in ns--3 is detailed, along with the implementation of the baseline handover setup;
	\item Chapter~\ref{chap:results} outlines the simulation scenario, presents figures and comments the results obtained;
	\item Finally, Chapter~\ref{chap:conclusion} draws some conclusions and suggests possible future research topics that will continue the work of this Thesis.
\end{itemize}

%% file: chapters/chapter1.tex

\chapter{5G Cellular Systems}\label{chap:5G}

\newthought{The next generation mobile networks} will be standardized before 2020, according to the 3GPP road map~\cite{3gppp5g}. As described in Chapter~\ref{chap:intro}, research on 5G is driven by forecasts that predict an increase of mobile internet traffic, both human generated and machine generated. There are many technologies that have been identified as enablers by several papers that propose guidelines and research directions for 5G networks. In the following sections, we will briefly provide an introduction to the technologies that will make the 5G vision become reality.

\section{5G Technology Enablers}
\begin{figure}
 	\centering
 	\includegraphics[width=0.95\textwidth]{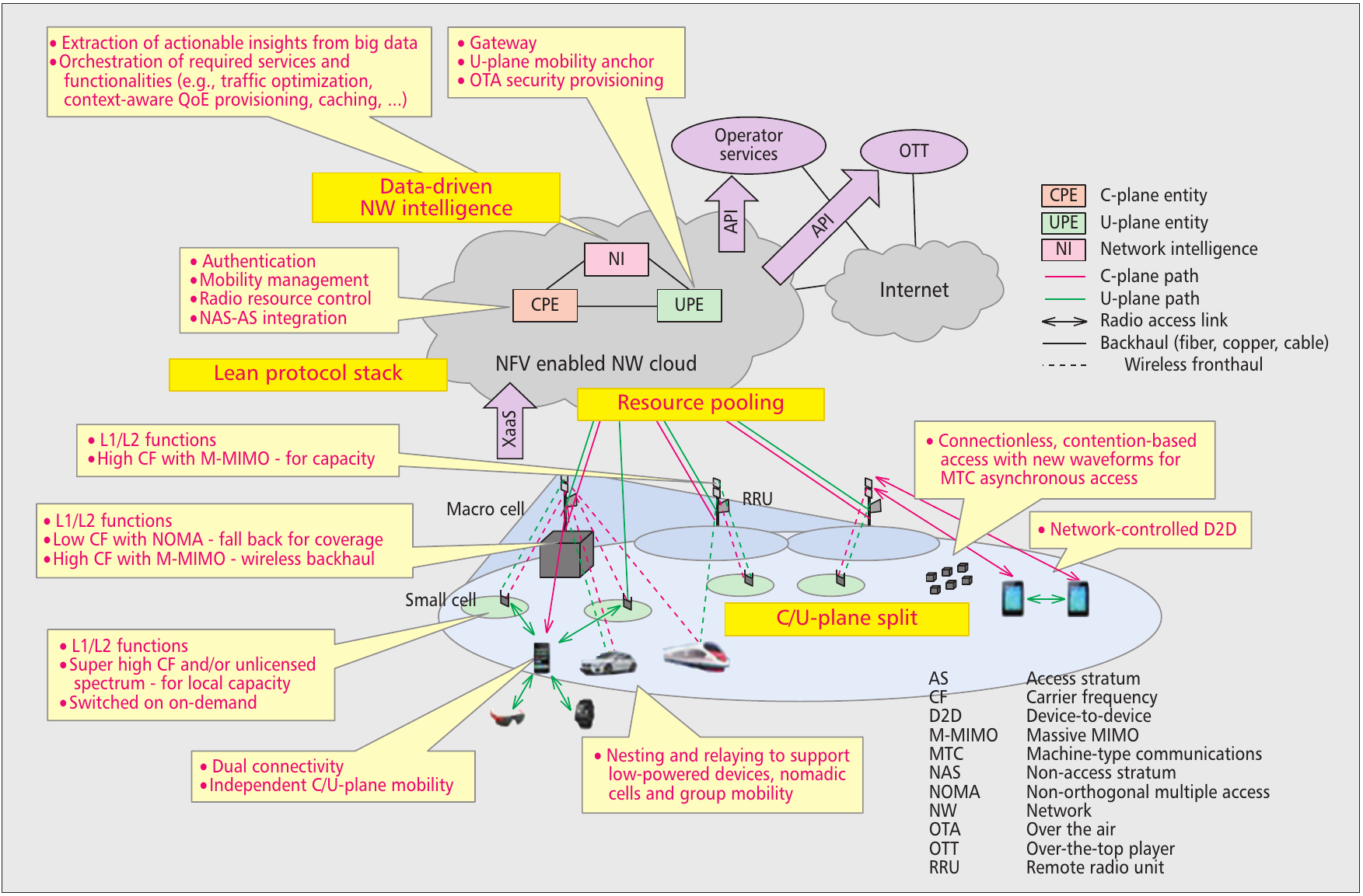}
 	\caption{5G mobile network vision and potential technology enablers, from~\cite{agyapong}}
 	\label{fig:enablers}
 \end{figure}

The ambitious goals upon which 5G network design is based require both an evolution of the current LTE 4G radio access and core network, and new disruptive technologies. Challenges such as a 1000x increase in capacity, a 100x increase in data rate, latency below 10 ms~\cite{metisd64}, along with sustainable costs and a consistent Quality of Experience (QoE), can be addressed only by using a combination of different solutions, based on ground breaking technologies and on refinements of robust and known systems. In particular, the authors of the survey in~\cite{boccardi} list as potential enablers the usage of mmWave frequencies, massive multiple-input multiple-output (MIMO), smart infrastructures, and native support for the different use cases (mobile broadband, massive M2M, ultra-low latency). Other papers agree with this point of view and also add control and user plane split, software defined networking (SDN)~\cite{agyapong}, full duplex radio~\cite{green5g} and heterogeneous networks. Fig.~\ref{fig:enablers} shows a complete set of potential enablers and details their role with respect to the whole system.

The following paragraphs therefore describe how some of these technologies can contribute to the development of 5G networks:
\begin{itemize}
	\item \textbf{mmWave frequencies} can offer large chunks of free unused spectrum that can be allocated to telecom operators. Propagation is harder at these frequencies but, with the exception of the sensitivity to blockage, the conditions are very similar to the ones of microwaves. However, this particular enabler will be discussed in detail in Sec.~\ref{sec:mmw};
	\item \textbf{Heterogeneous networks} allow to increase the capacity of the radio access network with small cells (known as \textit{picocells} and \textit{femtocells}), deployed more densely, but with smaller coverage area and transmission power. These cells will require a coverage layer provided by legacy 4G macro cells or by 5G cells operating on microWave frequencies, in order to avoid service interruptions. As part of the HetNet proposal, the usage of \textbf{U/C plane split} means that user plane functionalities can be provided by mmWave 5G small cells, while control plane messages are sent by using the coverage layer, allowing to increase the reliability of the connection;
	\item \textbf{Massive MIMO} refers to the use of a system in which the number of antennas at the base station (BS) is much larger than the number of devices per signalling resource~\cite{marzetta}. By operating in the mmWave frequency band, it is possible to pack more smaller antennas inside a UE or in a BS. With massive MIMO it is possible to have very narrow beams, which allow to exploit spatial multiplexing and increase the throughput. A main limitation is the need for a timely channel estimation in order to track the user mobility, however as mentioned in~\cite{agyapong} a dual connectivity solution could be used to provide an immediate fallback to another link, whose aim is to provide constant coverage;
	\item \textbf{Support for different use cases} is expected to be empowered by the use of (i) a configurable frame scheme at the Physical (PHY) and Medium Access Control (MAC) layer, based on Orthogonal Frequency Division Multiplexing (OFDM) or on one of its variants; (ii) an adaptive core network that can meet the QoS required for each data flow. This proposal is part of an approach that wants to harmonize the Radio Access Technology of 5G networks with the current LTE and Wi-Fi OFDM-based RATs~\cite{metis-harm};
	\item \textbf{Full duplex radio} technology has been thoroughly studied in recent years, and can be enabled by self interference cancellation techniques, thanks to the increased computational power available at both mobile terminals and base stations. It can be used either in the radio access network or for backhaul links between base stations~\cite{fd};
	\item \textbf{Smart infrastructures} are key to fully exploit the new opportunities and the increase in performance given by the other enablers. Smart infrastructure means the usage of caching at the edge of the network, a core network which can be reconfigured and is able to serve users with different requirements, with SDN and a lean design. Another proposal is network slicing, i.e., different functionalities of the network are offered by different service providers that interface with one another~\cite{networkSlicing}. A smart infrastructure can also offer different business opportunities to telecom operators.
\end{itemize}

\section{MmWave Technology And Its Adoption In 5G Networks}\label{sec:mmw}
\begin{figure}[b]
	\centering
	\includegraphics[width = 0.8\textwidth]{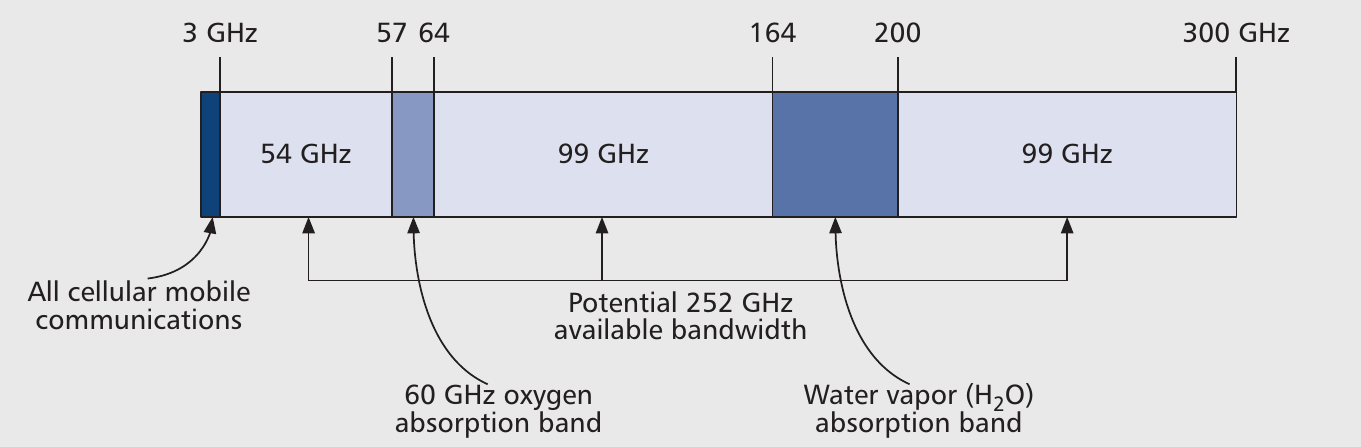}
	\caption{Spectrum in the range [0, 300] GHz, from~\cite{samsungmmb}}
	\label{fig:mmWaveBand}
\end{figure}

As mentioned in the previous section the adoption of millimeter wave (mmWave) frequencies communications in 5G networks is seen as a way to reach the throughput and capacity increase goals. Millimeter wave frequencies are the ones in the 3-300 GHz band, where the wavelength is indeed in the 1-100 millimeter range. They are mostly unlicensed, or lightly licensed~\cite{rappaport14}, and the International Telecommunication Union (ITU) will define which are the most suitable bands for 5G radio access networks in the next few years. Fig.~\ref{fig:mmWaveBand} immediately shows why these systems appeal to telecommunication researchers: the potential spectrum that can be allocated to 5G systems is very large. The potential carrier frequencies studied by a team at NYU are 28 GHz and 73 GHz~\cite{rappaport1}. 

There are several benefits given by the adoption of such high frequencies, as well as some drawbacks. The main pros are (i) the very large available bandwidth; (ii) the possibility of packing more antennas in a mobile terminal, with respect to the ones that a microWave system allows; (iii) an improved relative power consumption, with respect to lower frequencies~\cite{rappaportPower}, i.e., the power spent to transmit each bit is lower for mmWave than for typical LTE bands; (iv) the possibility of using very narrow beams in order to limit the interference toward other base stations and terminal devices, and to improve coverage. 

Among the main cons, there are (i) the limitations in coverage, in particular in urban environments, where mmWave signals suffer from blockage; (ii) the absolute power consumption. These issues, however, have been recently studied and addressed by several papers, that will be summed up in the following paragraphs. 

\subsection{MmWave Radio Propagation} 
Measurements of mmWave-band outdoor propagation have been conducted only in recent years, while the indoor case was extensively covered since the 1980s~\cite{andrewsModel} and the usage of mmWave for indoor communications is already part of a standard~\cite{wifi}. 
The authors in~\cite{samsungmmb} propose to use the mmWave frequencies in mobile networks; outdoor measurements followed soon, and the main preliminary results are reported in~\cite{rappaport1, rappaport14}.

Some general considerations can be made on the propagation of mmWave frequencies:
\begin{itemize}
	\item While the omni-directional \textbf{propagation loss} obeys Friis Law, and increases with the square of the frequency, when considering mmWave link budget also the antenna gain must be taken into account. Given the same antenna aperture area, the gain increases with the frequency. Therefore this factor compensates the free space pathloss in the link budget. Moreover, with mmWave more directional antennas can be created in a small space, thus allowing high beamforming gain, provided that the beam can track the mobile terminal~\cite{rangan14};
	\item The main concern for mmWave frequencies is \textbf{shadowing}. Materials as such as brick exhibit an attenuation factor in the range of 40-80 dB, and also the human body can attenuate mmWave signals up to 35 dB~\cite{rangan14}. However, a higher reflection facilitates non-line-of-sight communications. Also foliage and heavy rain can cause severe attenuation in mmWave bands. 
	The attenuation given by foliage increases with the frequency and with the foliage depth: for example, at 80 GHz a depth of 10 m is enough to attenuate the signal by 23.5 dB~\cite{samsungmmb}. 

	In addition, even rain attenuates the mmWave signals, because the wavelength is comparable to the size of a rain drop, thus causing scattering of the radio signal. The attenuation due to rain is measured in dB/km and strongly depends on the intensity of the rain in mm/hour. In the case of light rain (2.5 mm/hour), the attenuation is small (1 dB/km), in particular when considering the expected typical maximum range of mmWave cells (200 m). However there may be particular cases (such as monsoons) in which mmWave communication can be disrupted by very heavy rain~\cite{samsungmmb}. 
\end{itemize}

\begin{figure}
	\centering
	\includegraphics[width=0.65\textwidth]{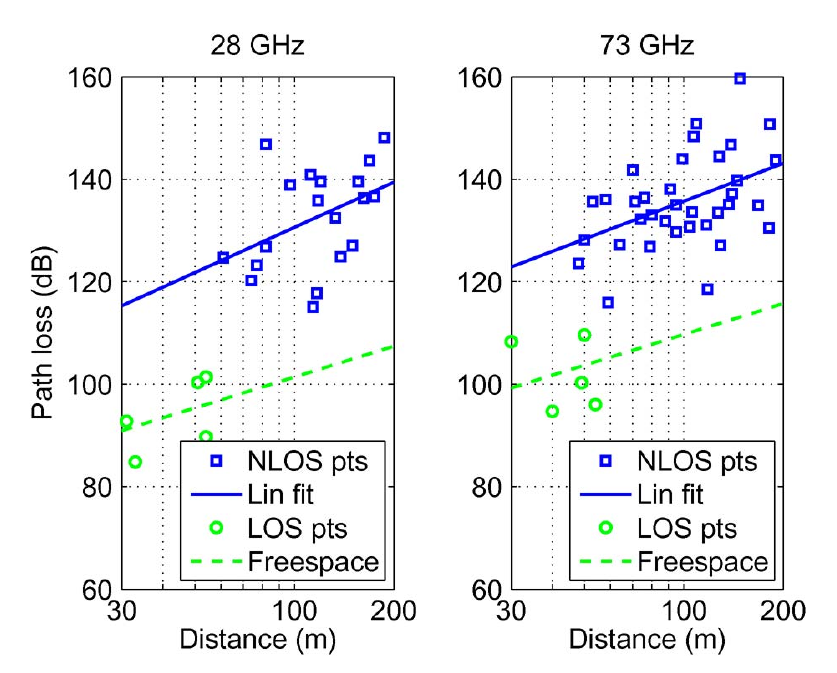}
	\caption{Pathloss for 28 GHz and 73 GHz, from~\cite{rappaport1}}
	\label{fig:mmwprop}
\end{figure}

The measurements of~\cite{rappaport1} corroborate these general considerations. They were performed in New York, using highly directional antennas at 28 GHz and 73 GHz. It can be seen from Fig.~\ref{fig:mmwprop} that Friis Law (freespace line) fits the measurements for the line-of-sight (LOS) case, while the non-line-of-sight (NLOS) scenario exhibits a linear behavior in the distance, with an additional attenuation of 20 dB with respect to the LOS case. The maximum distance considered in Fig.~\ref{fig:mmwprop} is 200 m, since at a higher distance no signal was measured (varying the transmission power from 15 dBm to 30 dBm). This case is considered as outage, i.e., the mobile terminal cannot receive a signal from the base station. This distance is the actual limit of the radius of mmWave small cells, which will have to be densely deployed in order to provide uniform coverage.

\subsection{MmWave Directional Transmission}
As mentioned in the previous section, the high isotropic propagation loss can be compensated by directional antennas with high beamforming gain. This, however, defines another challenge, i.e., directionality for the UE must be tracked and accounted for at the eNB~\cite{rangan14}. 

Moreover, highly directional transmissions create issues for broadcast signals and synchronization for initial cell search. As explained in~\cite{zorziMac}, there is a \textit{directionality} trade-off. With omnidirectional communications, the range that each mmWave eNB can cover is limited, but, at the same time, it is possible for all the devices under coverage to receive broadcast informations. On the other hand, semi or highly directional solutions allow to increase the transmission range, and reduce the interference, but then a spatial search is needed when accessing the network. Besides, if broadcasts are omnidirectional and data transmission is instead directional, there may be a mismatch between the area in which synchronization and broadcast control informations can be received, and the area in which data transmissions are supported, as shown in~\cite{coverage}. A directional procedure for Initial Access (IA), on the other hand, may introduce additional latencies~\cite{rangan14}. The delay and coverage issues for IA are evaluated in~\cite{giordaniIA}, while in~\cite{abbasIA} the performance of different solutions to avoid a greedy spatial search is evaluated. 

\subsection{MmWave Power Consumption}
\begin{figure}[t]
	\centering
	\includegraphics[width=0.95\textwidth]{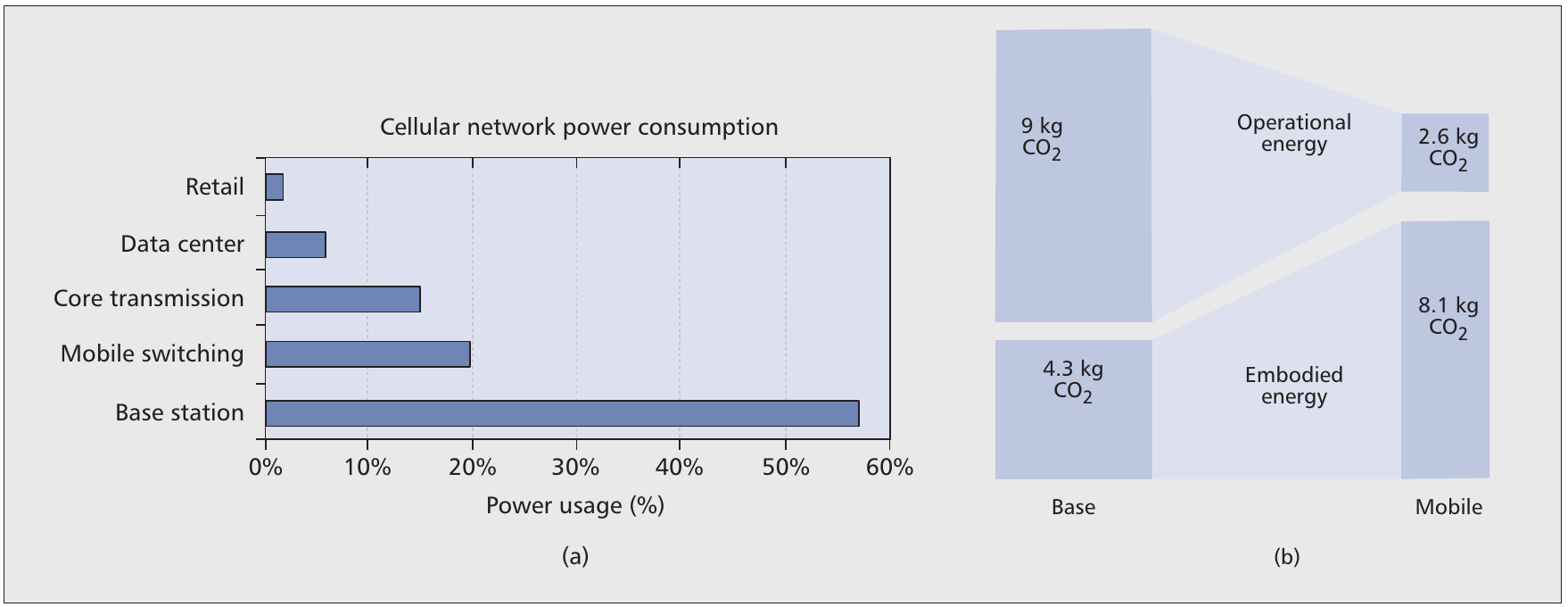}
	\caption{Typical power consumption in a current mobile network, from~\cite{power_bs}}
	\label{fig:power}
\end{figure}

Another issue that must be addressed when considering mmWave communications and the very high bandwidth employed is power consumption. In current cellular networks, as Fig.~\ref{fig:power} shows, the energy consumption of base stations accounts for nearly 60\% of the electric energy bill of a typical telecom operator. Since it is expected that the number of cells deployed will increase to account for the smaller coverage of mmWave frequencies~\cite{rangan14}, it is necessary to adopt an energy efficient approach when designing and planning 5G networks.

\begin{figure}[t]
	\centering
	\includegraphics[width=0.7\textwidth]{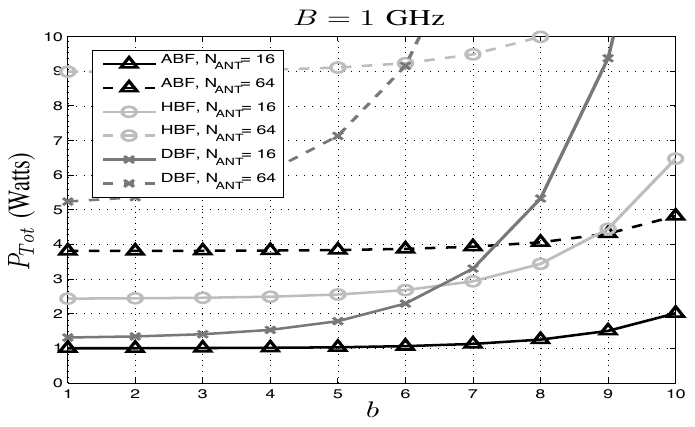}
	\caption{$P_{tot}$ for $B=1$ GHz, different beamforming schemes and number of antennas $N_{ANT}$, from~\cite{abbasPower}}
	\label{fig:dbf}
\end{figure}

Particular attention must be given to the design of analog to digital converters and processing units. Indeed, the power consumption of an A/D converter scales linearly with the rate considered. For example, a state of the art circuit operating at 100 Ms/s~\cite{cmos} can require up to 250 mW when operating, thus causing a too high power consumption in mmWave mobile terminals~\cite{rangan14}.
It is generally expected that digital beamforming (DBF) solutions, which employ two A/D converters for each antenna, have a higher power consumption than hybrid beamforming (HBF) systems, where a lower number of A/D converters is used, at the price of a lower flexibility. However, in~\cite{abbasPower}, the performance of different beamforming schemes in terms of power consumption $P_{tot}$ is assessed. In particular, the authors consider all the elements in a mmWave receiver, i.e., not only the A/D converters, but also combiners, mixers, low noise amplifiers, different bandwidths $B$ and number of bits $b$ for the analog to digital conversion.
As shown in Fig.~\ref{fig:dbf}, there are some values of $b$ for which the power consumption of a receiver with DBF is smaller than that of a receiver with HBF. Analog beamforming (ABF), instead, always has the lowest power consumption, given the same number of antennas used.

\begin{figure}[t]
	\centering
	\includegraphics[width=0.7\textwidth]{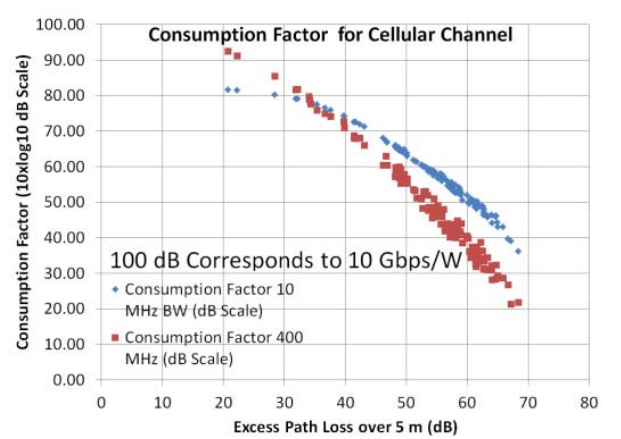}
	\caption{$CF$ for a 38 GHz system, with a bandwidth $B=10$ MHz or $B=400$ MHz, from~\cite{rappaportPower}}
	\label{fig:cf}
\end{figure}

The power consumption of mmWave systems has been studied in relation to the achievable rate in~\cite{rappaportPower}, in order to understand whether LOS or NLOS conditions have a role in the power consumption and how much bandwidth should be allocated in the two different scenarios. In particular the \textit{consumption factor} ($CF$) is defined as 
\begin{equation}
	CF = \frac{R_{max}}{P_{consumed, min}}
\end{equation}
where $R_{max}$ is the maximum rate achievable given a certain communication system and can be computed using Shannon's theory, and $P_{consumed, min}$ is the power consumption. In Fig.~\ref{fig:cf} there is a comparison between the $CF$ that can be obtained by a system with 10 MHz and 400 MHz bandwidth, for different pathlosses. It can be seen that in a LOS setting it is preferable from the point of view of the $CF$ to use larger bandwidths, while in NLOS (higher pathloss) a smaller bandwidth is more efficient.

%% file: chapters/chapter2.tex

\chapter{LTE-5G Tight Integration}\label{chap:integration}
As seen in Chapter~\ref{chap:5G}, the next generation of mobile networks will be a combination of an evolution of legacy 4G networks and new disruptive technologies. However, since telecom operators have recently put a lot of effort in deploying LTE networks, it will make sense to exploit them as part of the new 5G generation. In particular, 4G can provide a coverage layer and make 5G networks more robust to link outages and service unavailability. 

There is a case for a tight integration between these two networks. Indeed, the 5G physical layer is expected to be OFDM based, with different numerologies to account for different use cases~\cite{nokiaNum}. Moreover, while the medium access control operations will have to be adapted to the new physical requirements~\cite{zorziMac}, the higher layers of the mobile network protocol stack are expected to be in common between LTE (and its evolutions) and 5G. 

In the following sections the state on the art on these topics will be described. Firstly, the current LTE protocol stack and the LTE network architecture will be introduced, and from this starting point the main proposals of integration with the 5G stack will be discussed. Secondly, details on DC and Handover in LTE will be given.

\section{The LTE Protocol Stack}\label{sec:ltestack}
\begin{figure}
	\centering
	\includegraphics[width=0.9\textwidth]{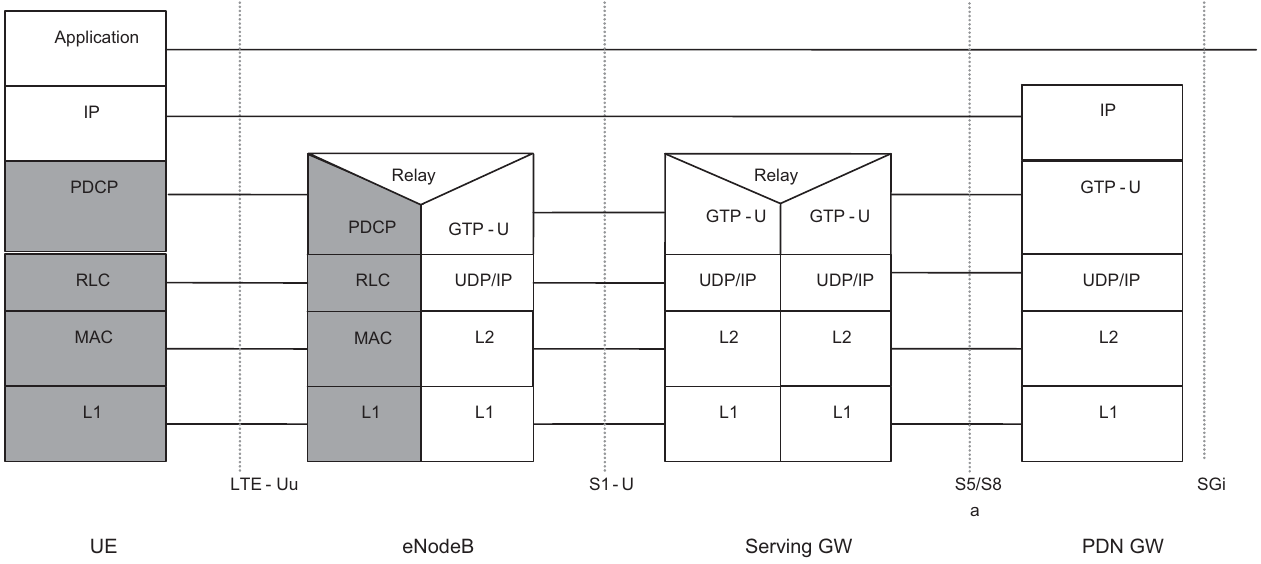}
	\caption{LTE protocol stack, from~\cite{sesia}}
	\label{fig:lte_stack}
\end{figure}

A first comprehensive view of the LTE protocol stack and of the main network nodes is in Fig.~\ref{fig:lte_stack}. The mobile LTE stack is used to provide effective communications between the mobile terminals and the eNBs, and it interfaces with the IP layer. In the following paragraphs, the functionalities offered by the PHY and the MAC layers will be briefly introduced, while the Radio Link Control (RLC) and the Packet Data Convergence Protocol (PDCP) layer will be described in details.

\subsection{LTE Physical and Medium Access Control layers}
The LTE PHY layer provides the low level functionalities (modulation, framing) which are needed for the transmission of data and control packets over the wireless medium. An LTE system can be configured as either Time Division Duplexing (TDD) or Frequency Division Duplexing (FDD), and there are different specifications for the framing in the PHY layer accordingly to the chosen configuration. It is also responsible for Adaptive Modulation and Coding (AMC), power control, and it provides measurements to the Radio Resource Control (RRC) layer for procedures like initial cell search and synchronization. 

The MAC layer is in charge of mapping the data received from higher layers to physical transport channels, thus performing multiplexing and demultiplexing of higher layer Packet Data Units (PDUs) into a single MAC Service Data Unit (SDU). It also performs scheduling at the eNB side and reporting of buffer status from the UE to the eNB. Additionally, the Hybrid Automatic Repeat reQuest (HARQ) mechanism offers error correction via retransmission~\cite{mac}.

The PHY and MAC layers are also responsible for the Random Access (RA) procedure, upon triggering from the RRC layer. There is a single PHY and MAC layer instance for each device (either eNB or UE).

\subsection{Radio Link Control Layer}
The RLC layer~\cite{rlc} is the one above the MAC layer, and it forwards and receives data from the MAC layer through \textit{logical channels}. In both the UE and the eNB there is an RLC entity for each Evolved Packet System (EPS) bearer, i.e., for each data or signalling flow. The RLC layer acts as an interface between the PDCP layer and the MAC layer, since it buffers the data coming from the PDCP layer and receives transmission opportunities (in terms of bytes that can be transmitted) from the lower layer. Therefore it segments and/or concatenates PDCP PDUs into an RLC PDU that can fit into the transmission opportunity, and at the receiver side it performs the inverse process in order to retrieve the original packets. Moreover, the RLC protocol is designed to reorder RLC PDUs in case they are received out of order, for example because of HARQ retransmissions at the MAC layer. 

\begin{figure}[t]
	\centering
	\includegraphics[width=0.65\textwidth]{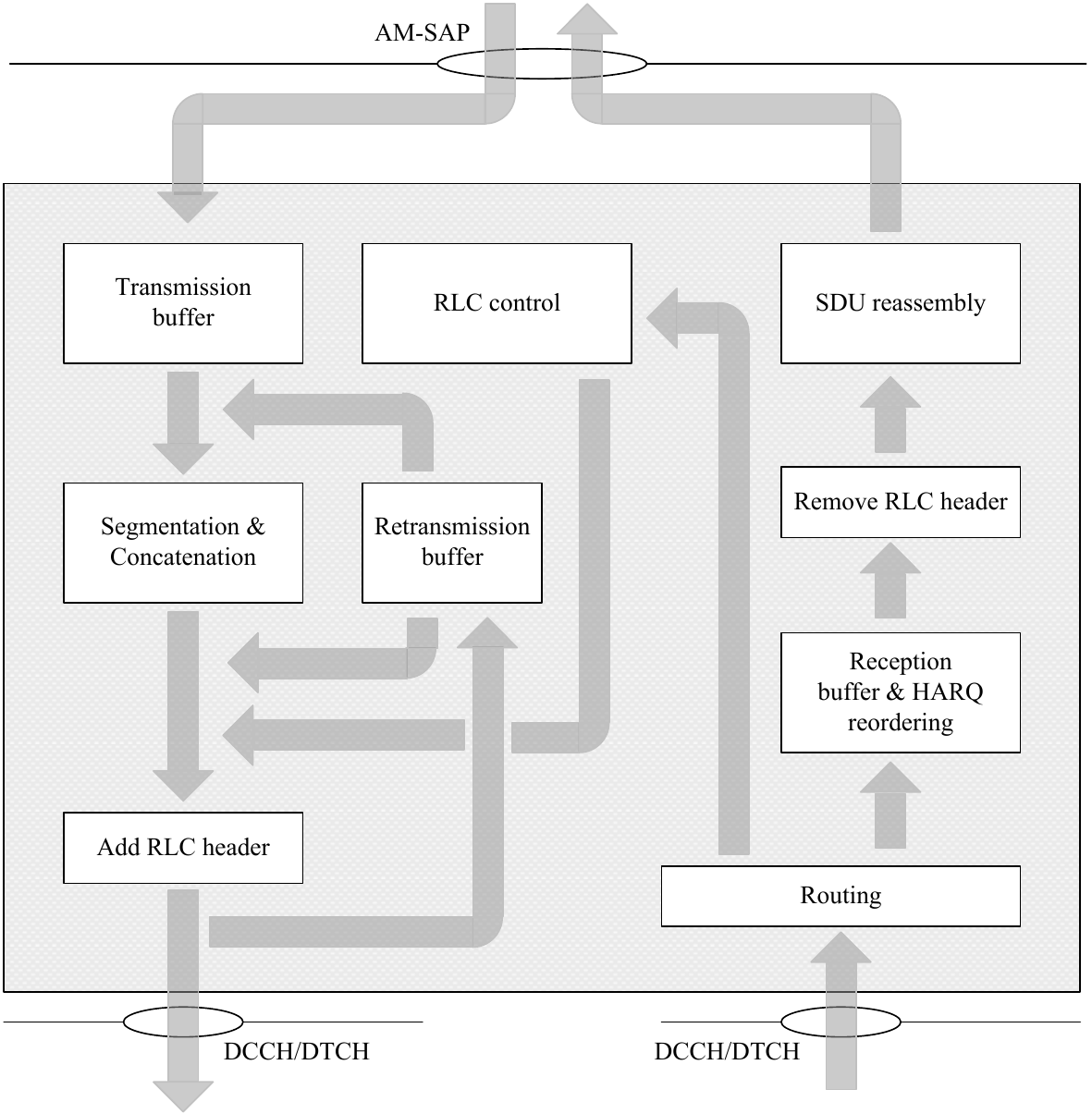}
	\caption{RLC AM block diagram, from~\cite{rlc}}
	\label{fig:rlcam}
\end{figure}

There are three different possible configurations for the RLC layer:
\begin{itemize}
	\item RLC Transparent Mode (TM), which simply maps RLC SDUs (i.e., PDCP PDUs) into RLC PDUs. It cannot be used for data transmission in LTE, but only for operations such as the transmission of System Information Broadcast messages, the first messages in the RRC configuration (RRC Connection Request and RRC Connection Setup) and paging;
	\item RLC Unacknowledged Mode (UM), which performs segmentation and concatenation of RLC SDUs at the transmitter side, reassembly and reordering at the receiver side, and packet loss detection. No retransmission is performed, and packets are simply declared lost (even if a single segment of the entire packet is missing). This configuration is used for delay sensitive applications, that need very low latency (and this does not allow to use retransmissions), at the price of packet losses. Notice that the MAC layer offers a retransmission mechanism (HARQ), which is however limited by a maximum number of retransmissions, typically 3;
	\item RLC Acknowledged Mode (AM), that has the same functionalities of UM, and adds a retransmission mechanism. The receiver entity periodically sends to the transmitting one a status report that contains information on which packets were lost, and they are retransmitted as soon as the MAC layer signals a suitable transmission opportunity. Packets and fragments can be fragmented once again, and reconstructed at the receiver side. The transmitter can also poll for a status report, in case it has completed the transmission of buffered packets. The block diagram of a transmitter and receiver RLC AM entities is shown in Fig.~\ref{fig:rlcam}.
\end{itemize}

RLC PDUs carry one or more (possibly fragmented) RLC SDUs and an RLC header, which contains the sequence number and control information on the payload.

\subsection{Packet Data Convergence Protocol Layer}
The PDCP layer~\cite{pdcp} collects data and signalling packets from the upper layers and forwards them to the associated RLC entity. It provides the first entry point for packet streams to the LTE mobile protocol stack, and there is a PDCP instance for each EPS bearer. It provides in order delivery to upper layers, and discards user plane data if a timeout expires. Its main functionalities are however header compression (upper-layer static header parts are not transmitted for each packet, thus reducing the overhead) and security (ciphering and integrity protection).

\subsection{Radio Resource Control Protocol}
The RRC protocol provides the control functionalities for eNBs and UEs, and it supports the communication of control-related information either in broadcast from the eNB or in an exchange with a single UE. In particular, the services that it offers are related to~\cite{sesia, rrc}:
\begin{itemize}
	\item Broadcast and reception of System Information (SI), which includes initial configurations of the eNB that UEs need to start a connection;
	\item Establishment, maintenance, modification and release of an RRC connection between an eNB and a UE. The RRC protocol has primitives for the setup of Data and Signalling Radio Bearers (DRB and SRB), for connection reconfiguration during handovers and configuration of lower layers;
	\item Inter-RAT mobility, with context transfer, security functionalities, cell handover commands;
	\item Collection of measurements from PHY layer (at the UE) and reporting to the eNB.
\end{itemize}
The RRC messages are sent over SRBs. Signalling Radio Bearer 0 configuration is fixed and known to all the LTE devices, it uses RLC TM, and it is responsible for the exchange of the first RRC messages at the beginning of a connection setup. SRB1 and SRB2 are respectively for the normal-priority and the low-priority RRC messages. Both these SRBs use RLC AM in order to reliably deliver the message to the other endpoint.

\section{LTE Network Architecture}
A brief introduction to LTE network architecture will help understand the description of dual connectivity and handover that will be given in the next sections. The LTE standard provides specifics on the Evolved Universal Terrestrial Radio Access Network (E-UTRAN), which is the radio access part and is used in conjunction with the Evolved Packet Core (EPC) network. Together they form the EPS~\cite{36300}.

The entry point to this network is the Packet Data Network Gateway (P-GW), which has a link to the Service Gateway (S-GW). This node, which is sometimes co-located with the P-GW, has knowledge of which eNB a certain UE (mapped to an IP address) is connected to, thanks to the interaction with the Mobility Management Entity (MME). The MME node is in charge of tracking the UE mobility and updating the path for each UE in the S-GW. 

\begin{figure}
	\centering
	\includegraphics[width=0.75\textwidth]{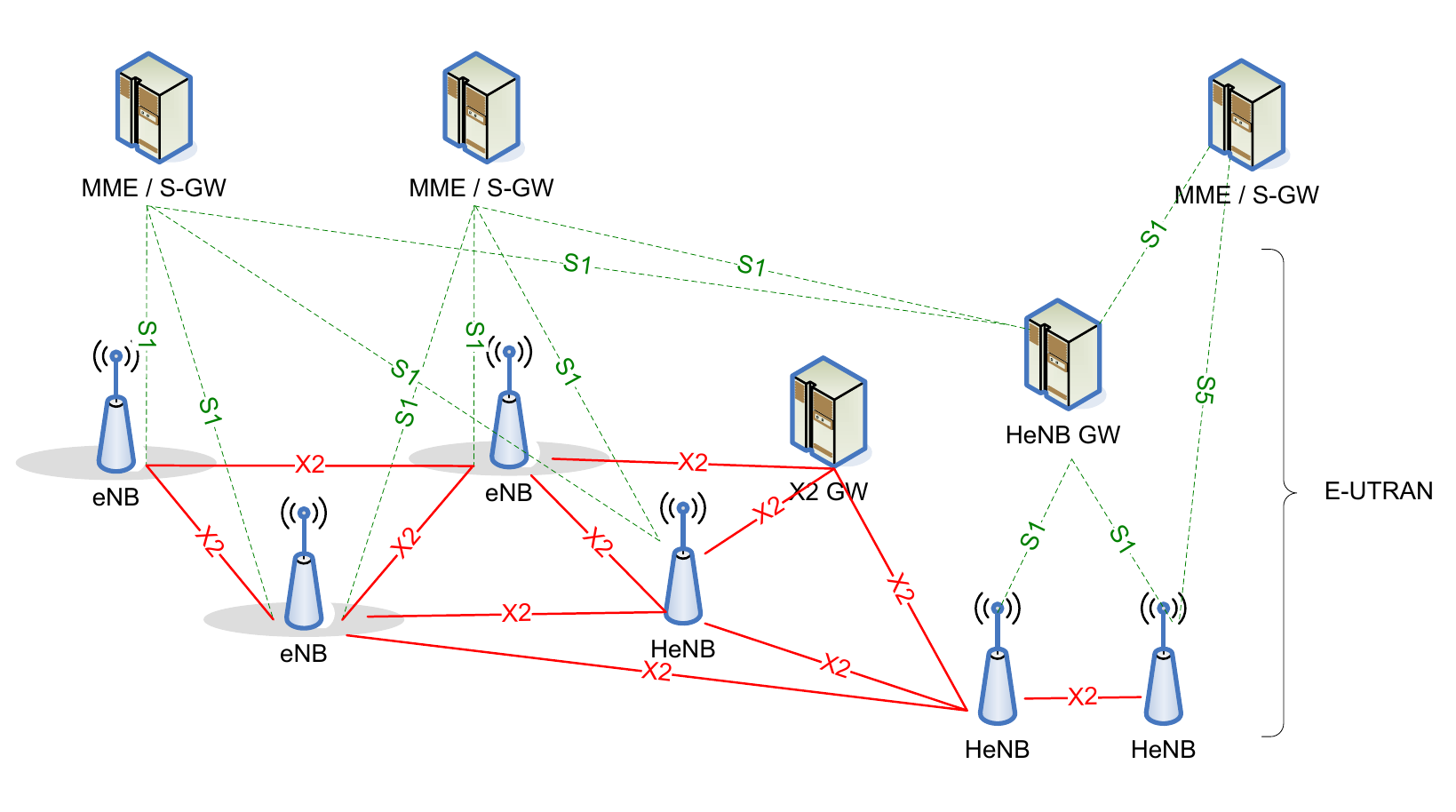}
	\caption{The LTE access network, composed of EPC and E-UTRAN, from~\cite{36300}}
	\label{fig:epc}
\end{figure}

As shown in Fig.~\ref{fig:epc}, the base stations, namely eNBs, are connected to S-GW and MME via the S1 interface, which is split into S1-MME (for the control channel to the MME) and S1-U (for the data channel, to the S-GW). In the eNBs the data packets are forwarded in the PDCP layer of the radio stack. eNBs are connected to their neighbors with the X2 interface, which is used to trasfer handover commands, data during handovers and load information~\cite{sesia}. There are also additional components (X2 gateways, Home eNB gateways) that enable the EPC to provide heterogeneous networking functionalities.

\section{LTE-5G Tight Integration} 
As shown in Chapter~\ref{chap:5G}, mmWave communications can enable very high throughput, but they also suffer from the high variability of quality of the received signal, and from outages due to buildings and obstacles. Additionally, a very dense deployment of base stations is expected. This introduces some key challenges: (i) frequent handovers between mmWave cells, or to legacy RATs, due to user mobility, and (ii) exposure to Radio Link Failure (RLF), which triggers time and energy consuming random access procedures. This is why the integration between a legacy RAT such as LTE and the new 5G air interface has been recently proposed by the major players.

\subsection{The METIS Vision}

In~\cite{metisInt}, the European project Mobile and wireless communications Enablers for the Twenty-twenty Information Society (METIS) considers 5G as a set of evolved versions of existing RATs (such as, for example, LTE) and new wireless functionalities suited for different use cases. Therefore, there will be a need for new architectures to manage this multi-RAT system, in terms of coordination, inter-networking, radio resource management. The METIS final report on architecture~\cite{metisd64} suggests that the LTE Advanced radio access can be used as a coverage layer to improve reliability and ease the deployment of 5G networks. In particular, an integration of LTE and 5G can bring benefits to different applications, e.g.:
\begin{itemize}
	\item Unified system access, with broadcast messages with common information for different RATs sent only with LTE, common paging, high resiliency to mobility thanks to the better propagation at LTE frequencies;
	\item User plane aggregation, either with the possibility of transmitting on multiple links in order to maximize throughput, or on a link at a time but with the potential to quickly switch from one RAT to another;
	\item Common control plane, possibly on lower frequencies, in order to provide a more robust system.  
\end{itemize}
The report does not specify the final architecture that should be adopted, but offers some considerations on the requirements of different integration solutions. In the mobile stack, some functions need synchronization, i.e., different layers must cooperate with a tight time schedule, and others can be asynchronous. Therefore, synchronous functionalities (such as, for example, the ones provided by the MAC layer) must be RAT dependent and deployed in each eNB, while asynchronous ones (i.e., higher layer services) can be centralized, or common to the different RATs. Another consideration is about the possibility of co-locating the access points (i.e., the base stations) of the different RATs. This would be a more expensive solution to deploy, but would offer the possibility of integrating the synchronous services of different RATs.

\subsection{Different Architectures to Enable Tight Integration}\label{sec:dasilva}

The layer at which the LTE and the 5G protocol stacks will converge is defined as the \textit{integration layer}. This layer has an interface to lower layers which belong to different radio access technologies, but offer the same services to the integration layer. The latter will deliver packets from upper layers to the different RATs, and collect the traffic coming from the different lower layers. 

In~\cite{dasilva} there is an analysis of the main pros and cons of using the PHY, MAC, RLC or PDCP layers as integration points:
\begin{itemize}
	\item Common PHY layer: this solution should be viable in principle, since OFDM or one of its variants are expected to be the basis for the 5G physical layer. However, very different frame structures and numerologies are expected to be used in 5G, with multiple numerologies to account for different use cases. Therefore, integrating LTE and 5G at the PHY layer is a very challenging task, and the benefits would be limited. Moreover, the usage of a common PHY layer limits the possibility to change the upper layers stack in order to adapt it to 5G requirements. Finally, operations at the PHY layer must be tightly synchronized in this case, and this prevents a non co-located deployment of eNBs for different RATs;
	\item Common MAC layer: integration at the MAC layer could enable high coordination gains. A possible option for MAC aggregation is carrier aggregation, which is already standardized for LTE~\cite{carrier}. At this level, it is possible to coordinate the scheduling of resources to the different RATs, to perform HARQ on different carriers, and to avoid the complexity of context transfers between RLC and PDCP entities, since there would be a single instance of both, for each bearer. However, as for the PHY layer, the operations at the MAC layer are synchronized, allowing only the deployment of co-located RATs. Moreover, LTE and 5G may be designed with different duplexing solutions, and different time and frequency resource allocation schemes. Therefore, while the possible gains are very appealing, the integration at the MAC layer would limit the possibilities of designing 5G medium access control differently from LTE, not allowing a brand new design that addresses the peculiarities of mmWave communications;
	\item Common RLC layer: also this choice presents some limitations that would prevent a non co-located deployment. Indeed, the RLC layer receives from the MAC layer scheduler indications on the transmission opportunities, i.e., how many bytes are available for transmission during the next slot. This communication cannot be subjected to the additional latencies of a MAC-RLC communication between remote locations. Moreover, segmentation and reassembly would work only in the presence of a common scheduler. Finally, the main benefit of integration at the RLC layer is the presence of a single transmission and, for RLC AM, retransmission buffer, and this allows to increase the coordination between the two RATs;
	\item Common PDCP layer: as shown in Sec.~\ref{sec:ltestack}, the PDCP layer has no strict synchronization requirements and therefore can be a suitable candidate as the integration layer when a non co-located approach is desired. Integration at the PDCP level allows a clean slate design of the PHY, MAC and RLC layers, so that they can be adapted to the new requirements of 5G networks. 
\end{itemize}

The authors of~\cite{dasilva} also propose a common RRC protocol. Its functionalities do not require synchronization, and having a single RRC protocol allows to optimize the control functionalities of the overall system.

\subsection{LTE As 5G Backup: The SDN Point Of View}

In~\cite{cloud5G} and~\cite{sdr5g} there is a case for integration of 5G and LTE from a software defined networking point of view.

One of the main reasons behind LTE and 5G integration is economic: 5G will probably be developed on top of existing and already deployed LTE infrastructure~\cite{cloud5G}. The 3GPP too is currently studying this topic in the 5G standardization process~\cite{3gppCoext}. This is why 5G protocol layers should be able to integrate and coexist with the LTE stack, in particular at upper layers. A multi-connectivity solution must be designed in order to (i) support a flexible and possibly dynamic centralization of certain Radio Access Network (RAN) services; (ii) take into account the capacity of backhaul, and the computational power available in the distributed nodes (eNBs, coordinators); (iii) provide an interface for a network controller, if SDN is employed. 

A software defined network can be used also to enable the possibility of orchestrating the access to one or another RAT, in case the mobile terminal is under the coverage of different technologies (LTE, 5G, Wi-Fi)~\cite{sdr5g}. Multi-connectivity would allow also to perform load balancing and assign resources to the different RATs according to traffic needs and signal quality over the various links.

\subsection{Expected Benefits of LTE-5G Tight Integration}
The integration of the new radio interface of 5G, which will probably work at mmWave frequencies, with the already deployed LTE, at microWave frequencies, can improve the performance of 5G networks. The benefits can be summarized in two categories:
\begin{itemize}
	\item Robustness-oriented;
	\item Throughput-oriented.
\end{itemize}
For example, the reliability of both user plane and control plane communications can be enhanced by a \textit{fast switching} (FS) mechanism, in which both the LTE and the 5G radios are in connected mode, but only one of the two at a time is actually used. If the quality of the signal on the link that the UE is currently using degrades below a certain threshold, the mobile equipment can simply switch to the other link by receiving a command from the eNB. In the current mobile networks this is done with a handover, which however requires a long procedure that may introduce significant latency. Another approach is based on transmit \textit{diversity}, with the same packet sent on both links, but this would limit the system to the LTE data rates.

Instead, throughput-oriented solutions make use of both links at the same time in order to increase the bandwidth and thus the throughput available to the UE.

Finally, by using a multi-connected device it is possible to transmit system information on all the RATs on a single radio interface (for example, LTE), and turn off all the others when not used. This reduces both energy consumption and broadcast overhead in the air interfaces not used for SI transmission~\cite{dasilva}. Moreover, the transmission of broadcast information on LTE bands is seen in~\cite{zorziMac} as a possible way to tackle the issue of directional transmissions when performing IA.

\section{LTE Dual Connectivity}\label{sec:ltedc}
The 3GPP has proposed a Dual Connectivity solution for LTE systems in Release 12. 

In~\cite{36300}, there is a basic description of the functionalities needed to support DC.
In particular it is specified that ``\textit{E-UTRAN supports Dual Connectivity operation whereby a multiple Rx/Tx UE in RRC\_CONNECTED is configured to utilise radio resources provided by two distinct schedulers, located in two eNBs connected via a non-ideal backhaul over the X2 interface}''. 
An eNB involved in a DC connection may be a Master (MeNB) or a Secondary (SeNB), and the UE in DC is connected to one MeNB and one SeNB at a time.

\begin{figure}
	\centering
	\includegraphics[width = 0.7\textwidth]{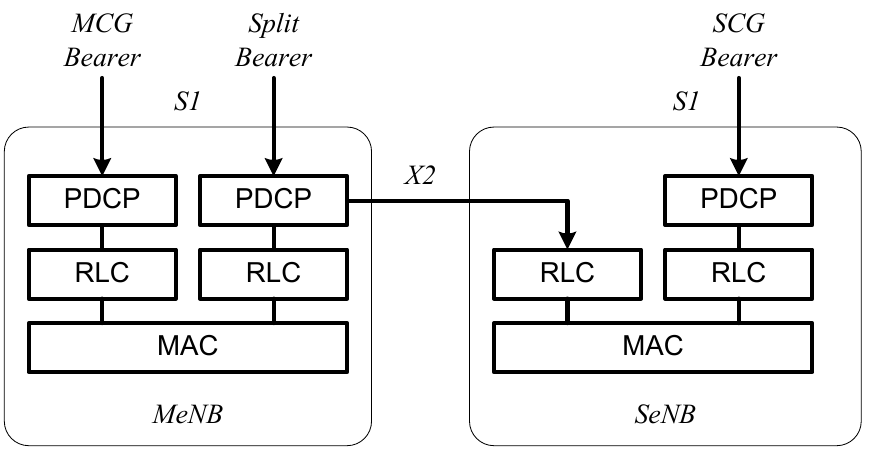}
	\caption{Radio Protocol Architecture for DC, from~\cite{36300}}
	\label{fig:rpcdc}
\end{figure}

There are 3 different kinds of EPS bearers that can be set up:
\begin{itemize}
	\item Master Cell Group (MCG) bearer;
	\item Secondary Cell Group (SCG) bearer;
	\item Split bearer;
\end{itemize}
The three different configurations are shown in Fig.~\ref{fig:rpcdc}. A MCG bearer is an end-to-end bearer that uses the Master eNB, while a SCG uses the Secondary eNB. In order to support these bearers, both the eNBs have a termination to an S1 interface to the S-GW and P-GW. A split bearer, instead, is a single flow that is forwarded from the core network to the MeNB PDCP, which splits the traffic into the MeNB RLC and the SeNB RLC. The connection between the PDCP and the remote RLC is an X2 link.

In this proposal, there is only one RRC entity, which is located in the MeNB. SRBs are thus always configured as MCG bearers and only use the radio resources of the MeNB. Therefore there is only one connection from the RAN to the MME per DC UE. Each base station should be able to handle UEs independently, i.e., to serve as Master to some UEs and as Secondary to others. Each eNB involved in DC for a certain UE controls its radio resources and is primarily responsible for allocating radio resources in its cell. Coordination between MeNB and SeNB is performed with X2 signalling.

In the 3GPP report~\cite{36842} there is another study on different possible configurations for a Dual Connectivity setup in a heterogeneous network scenario. In particular, it focuses on the Dual Connectivity for the User Plane, and lists some options, namely 1A, 2A, 2B, 2C, 2D, 3A, 3B, 3C and 3D. The numbers represent different choices in the configuration of the S1-U interface termination at Master and Secondary eNBs:
\begin{enumerate}[label=\textbf{\arabic*}]
	\item S1-U interface terminates both at MeNB and SeNB;
	\item S1-U interface terminates at MeNB, but no bearer split is performed in the Radio Access Network (i.e., two independent bearers are carried over S1-U to the MeNB, and one of the two is forwarded to SeNB via X2);
	\item S1-U terminates in MeNB and bearer split is performed in RAN, i.e., there is a single bearer for each dual-connected UE and its flow is split in the MeNB.
\end{enumerate}
Each option is completed by a letter (A, B, C, D), where 
\begin{enumerate}[label=\textbf{\Alph*}]
	\item stands for Independent PDCP layers, i.e., there are independent user plane endpoints in MeNB and SeNB;
	\item stands for Master-Slave PDCP layers, i.e., a part of the PDCP layer is in MeNB, and another, which acts as a slave, is in SeNB. However the report does not specify the details of the functional split between the two PDCP layers;
	\item stands for Independent RLC layers, i.e., there is a single PDCP layer which is located in the MeNB, and two independent RLC layers in the Master and Secondary cell;
	\item stands for Master-Slave RLCs, i.e., as in option C there is a single PDCP layer and a master RLC layer in the MeNB. The latter can forward some RLC PDUs (i.e., packets ready for transmission to the MAC layer, already segmented and with sequence numbers assigned by the Master) to a slave RLC layer in the SeNB.
\end{enumerate}

These 9 options are further analyzed with pros and cons. In particular, the report considers implementation issues and impact to the standard. For example, all the options require an extension of the X2 interface between eNBs in order to support signalling and coordination, and the transmission of packets (either as PDCP PDUs or SDUs, or as RLC SDUs). Options C and D require also a remote coordination between the PDCP and the RLC layers. Another aspect that is considered is security. Alternative A, for example, requires two different encryptions at MeNB and SeNB (since this functionality is located at the PDCP layer), while B, C and D do not. Finally, the modifications to transmission and reception mechanisms are taken into account. In particular, the number of PDCP and RLC entities needed for each bearer and the level of coordination needed are considered: alternative 3A requires two PDCPs even for split bearers, thus it needs a new layer above PDCP in which the flows can be split, and this has a large impact in terms of both standardization and efficiency. Option 1A, instead, can serve each of the two independent bearer flows with one of the two independent PDCPs and thus it does not require changes to the protocol stack. 

Other aspects that are considered are the SeNB mobility and its transparency to the core network, and the service interruptions required. For example, alternative 1A would need a complete path switch in the core network with the involvement of the MME. The same holds for dynamic offloading, i.e., for the possibility of dynamically routing traffic in the two eNBs according to radio conditions, congestion, etc. Alternative 1 requires the intervention of the MME to change the flow allocation between the two bearers, while alternatives 2 and 3 do not need to involve the MME and the dynamic offloading happens at the eNB level. 

Finally, the processing power of the MeNB is taken into account. With alternative 1, the MeNB does not process any of the SeNB traffic. Option 2 also allows only a lightweight operation (routing of packets to the SeNB), while alternative 3 is the most computationally expensive for the MeNB, since it has to process all the SeNB traffic, and in some cases (3B, 3C, 3D) it has also to cypher and decypher it. This has an impact also on the size of the buffers of PDCPs and RLCs entities.

The report concludes that only 1A and 3C can be considered for further studies and performance evaluations, since alternative 1A does not require a backhaul and a coordination between eNBs, but it is expected to provide a lower performance gain, while option 3C needs a fast backhaul but promises a higher performance gain thanks to a greater coordination between MeNB and SeNB.

It then defines 2 alternative schemes for RRC, the first with the RRC layer only in MeNB and the second with RRC in both eNBs. A single RRC layer simplifies the UE protocol stack, but needs forwarding of RRC-related messages of the SeNB to the MeNB via X2.

\begin{figure}[t]
	\centering
	\includegraphics[width=0.75\textwidth]{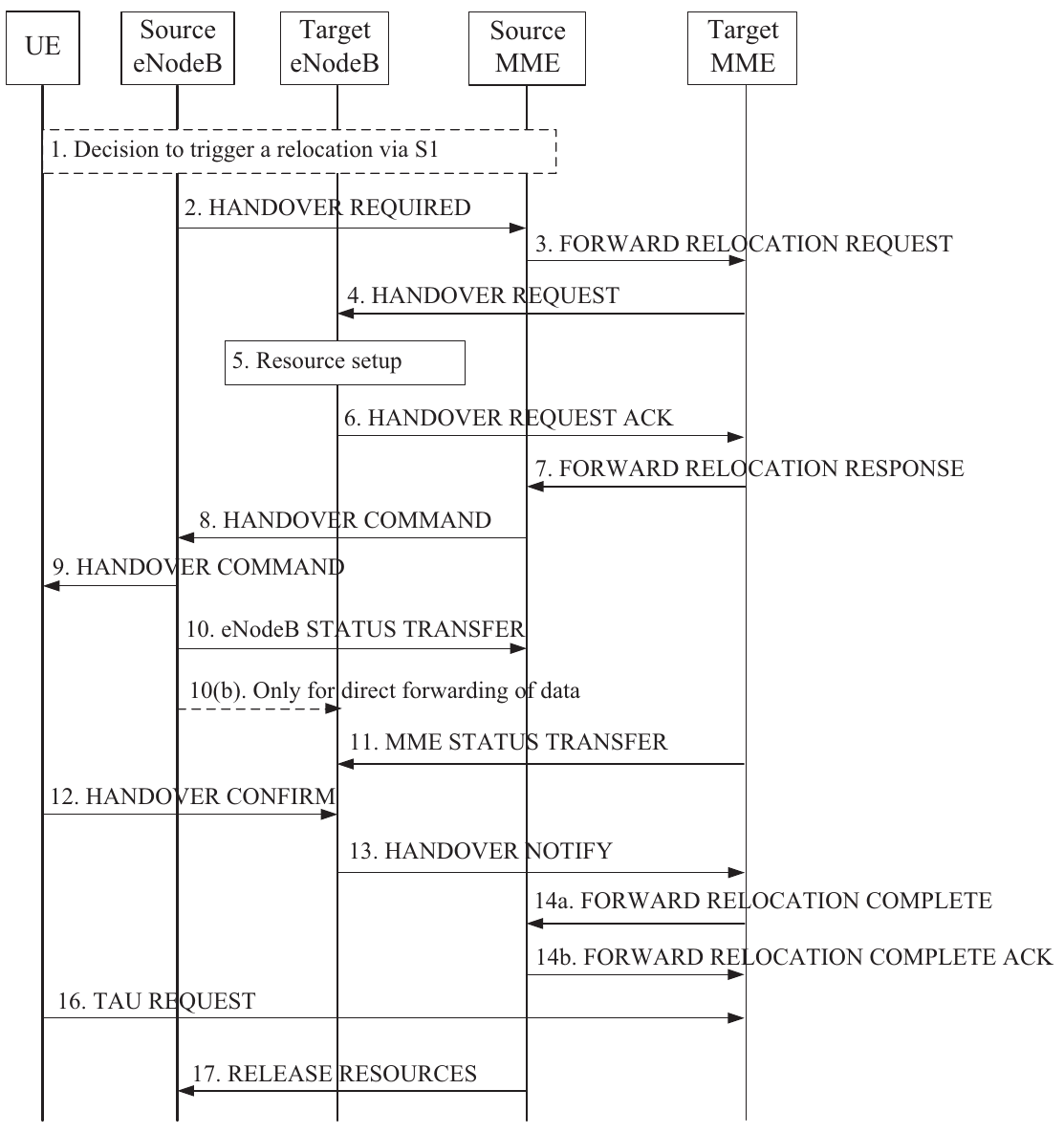}
	\caption{S1-based handover procedure, from~\cite{sesia}}
	\label{fig:handovers1}
\end{figure}

\section{Handover In LTE}\label{sec:llho}
LTE supports handover inside the E-UTRAN and also to other legacy RATs (UMTS, CDMA2000, GSM). There are an X2-based handover procedure and an S1-based handover procedure. The first is used for intra-RAT handovers only, and is based on the interaction between the source and the target eNB. The second, instead, is used when there is no X2 link between eNBs or when the handover is toward another RAT. The inter-RAT handover, indeed, requires the relocation of the UE to a different MME that handles the mobility of the other Radio Access Network, and this service is provided by the S1-based handover~\cite{sesia}.

The S1-based handover procedure is shown in Fig.~\ref{fig:handovers1}.
It involves the exchange of several messages with core network nodes, and this could increase the latency of the operation with a service interruption of up to 300 ms~\cite{36133}. 

\begin{figure}[t]
	\centering
	\includegraphics[width=0.75\textwidth]{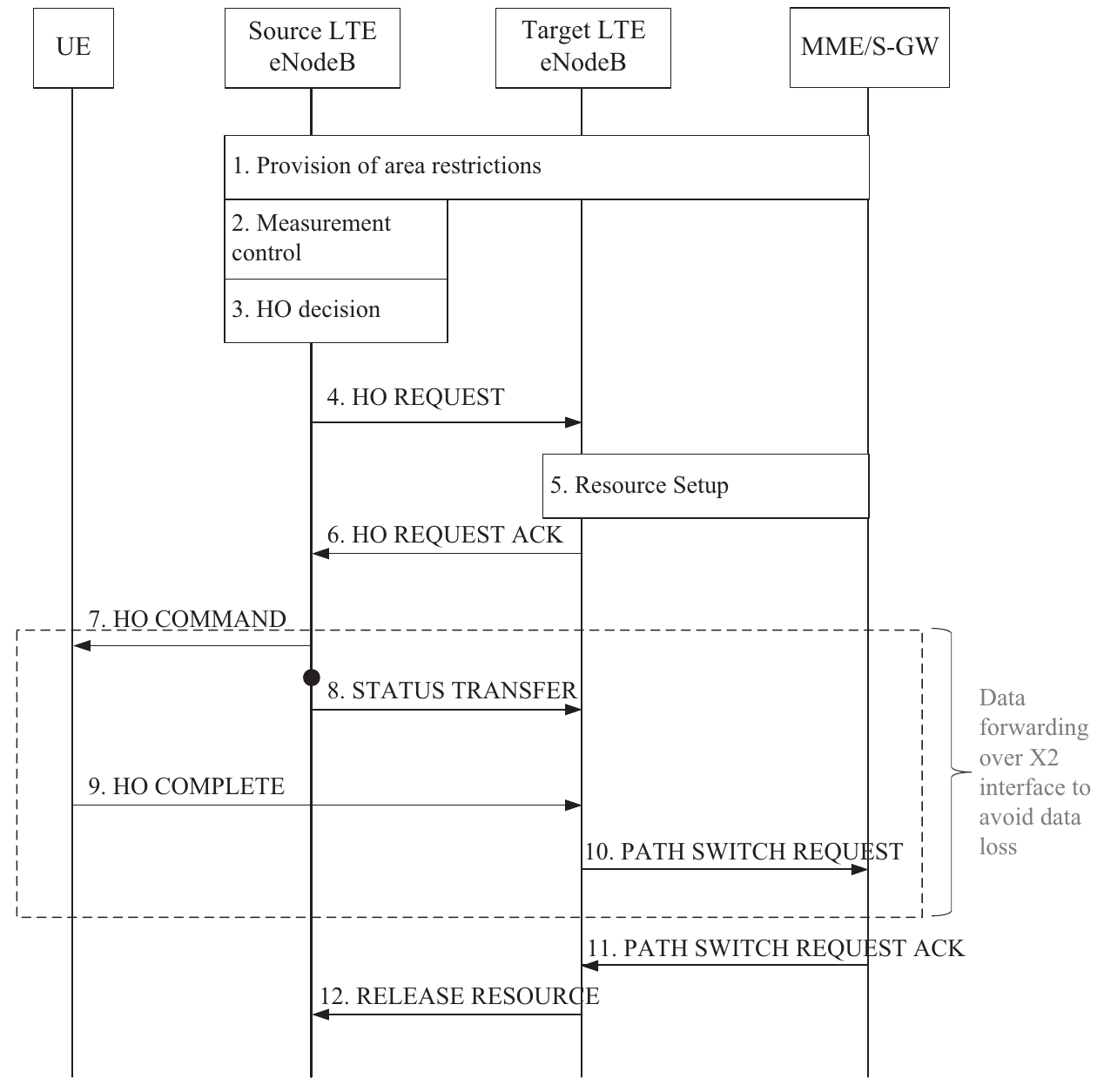}
	\caption{X2-based handover procedure, from~\cite{sesia}}
	\label{fig:handoverx2}
\end{figure}

The X2-based handover request, instead, as shown in Fig.~\ref{fig:handoverx2}, involves the core network only at the end, in order to switch the path from the S-GW to the target eNB. It is designed in order to limit the data loss during handovers. Notice that, after the reception of the handover command, the UE has to go through a complete Random Access procedure. However, it is a Non Contention Based RA, i.e., the target eNB reserves a preamble ID for the incoming UE, which is notified to the UE with the handover command from the source eNB. 

There are also two different PDCP handover modes: \textbf{seamless} and \textbf{lossless} handover. 
From the time at which the source eNB receives the Handover Request ACK from the target eNB and sends the handover command to the UE, the packets that arrive to the source eNB from the core network are forwarded via X2 to the target eNB. This forwarding ends when the Path Switch Request message is received by the S-GW, which in turn starts forwarding packets for the UE to the target eNB. 
Moreover, also RLC buffers are forwarded from source to target eNB, in two different ways, according to the handover mode chosen. Seamless HO is used for bearers which use RLC UM, and packets already processed by the PDCP and in the RLC or lower layer buffers are not forwarded, and thus are lost if not already transmitted to the UE. This is used for example in delay sensitive but loss tolerant applications. Instead, when RLC AM is used, and lossless HO is chosen, packets in the RLC AM buffers are actually forwarded to the target eNB and must be sent to the UE before any other packet is actually sent. The buffers that are forwarded are those with PDCP PDUs not yet transmitted, those transmitted but not yet acknowledged ones, and those with packets waiting for retransmission.

%% file: chapters/chapter3.tex

\chapter{Network Simulator 3}\label{chap:ns3}

Network Simulator 3 (ns--3) is an open source discrete-event simulator that aims to provide an advanced tool for research, development and educational use~\cite{whatis}. 

In particular, it focuses on networking research, and thanks to the contribution of an active community it provides modules for the simulation of several network standards and protocols. It is developed in C++ and Python and, thanks to the high level of detail that can be obtained when implementing a particular protocol, it offers the possibility of studying the performance of complex systems, where a mathematical analysis is impractical. 

In this Thesis, the simulations use:
\begin{itemize}
	\item the LTE module (LENA,~\cite{lena}), which models a realistic LTE radio access network and offers some elements of the EPC network;
	\item the mmWave module which is being developed by NYU~\cite{mmwaveSim} and that was released in its first version in May 2015. It offers the channel model, the PHY and the MAC layer for 5G mmWave protocol stack, and relies on the LTE module for the upper layers;
	\item the Building module, which allows to add buildings to the simulation, with different kind of walls, sizes, number of floors~\cite{building_model};
	\item the core simulator modules that offer TCP/IP connectivity, discrete-event simulation functionalities and tracing features.
\end{itemize}

These modules combined allow to simulate complex scenarios with base stations and mobile terminals in an environment with buildings, streets and obstacles, and with realistic applications on top of the transport and network layers.

The version of ns--3 on which the research of this Thesis is based is 3.25, with the addition of the NYU mmWave module. In the following sections, the features and the modeling choices of the mmWave module will be described, then some details on which functionalities are available in the LTE module will be given. 

\section{NYU mmWave Module for ns--3}

The module is the first open source framework that allows to simulate end to end mmWave systems, and its main strength are (i) a fully customizable physical layer, where carrier frequency, bandwidth, frame structure and OFDM numerology can be changed in order to test different PHY and MAC configurations; (ii) a channel model for the 28 GHz and 73 GHz carrier frequencies based on real measurements made in New York~\cite{rappaport14}. The adaptability of the physical layer to different frame structures is a wise implementation choice, since 5G is not yet a standard, and therefore having the possibility of changing the parameters without altering the source code makes the simulator flexible and ideal for research. 

The structure of the classes is based on the ns--3 LTE module, which is implemented with an interface paradigm, i.e., layer \texttt{A} communicates with layer \texttt{B} not directly by calling its methods but using an interface \texttt{I}, which acts as a wrapper on the actual implementation of the functions in \texttt{B}. Therefore, layer \texttt{B} can be swapped with layer \texttt{B\_1}, provided that the minimal requirements of interface \texttt{I} are met by the new layer. 

Thanks to this implementation paradigm, the NYU mmWave module can use the upper layers of the LTE module, as described in~\cite{forde2e}, and this allows to perform end to end simulations with 5G devices. 

\subsection{MmWave Channel Modeling}
The mmWave model offers two different channel models. The first, described in~\cite{mmwaveSim}, is derived from extensive MATLAB traces obtained from the measurements of~\cite{rappaport1}, while the second is based on a third-party ray tracing software. 

\subsubsection*{\textbf{Simulation-Based Statistical Model}} 
This model takes into account several features in order to describe the mmWave channel in a realistic way: firstly, the SINR is computed from pathloss, gain provided by MIMO and interference, and then an error model is applied.

The link budget in dB can be expressed as 
\begin{equation}\label{eq:lb}
	P_{rx} = P_{tx} + G_{MIMO} - L - S
\end{equation}
where $P_{rx}$ and $P_{tx}$ are respectively the received and transmitted power, $G_{MIMO}$ the MIMO gain (MIMO is used for beamforming), $L$ the pathloss and $S$ the shadowing. 

\vspace{10pt}

\textbf{Pathloss and Shadowing}: the mmWave module relies on the ns--3 Building module to create obstacles in the simulation scenario. Then, for each transmitter - receiver pair an imaginary line is drawn in order to decide whether the communication happens in a LOS or NLOS environment. If an obstacle is crossed, then the channel state is set to NLOS, otherwise LOS is assumed. Then the pathloss as a function of the distance $d$ is computed as
\begin{equation}\label{eq:pl}
	L[dB](d) = \alpha + \beta 10 \log_{10} (d) + \xi, \quad \xi \sim N(0, \sigma^2) 
\end{equation}
where $\xi$ is the shadowing, and $\alpha$, $\beta$ and $\sigma$ are parameters that change accordingly to the LOS or NLOS state. These are obtained in~\cite{rappaport1}, by fitting the measurement shown in Fig.~\ref{fig:mmwprop}, and are reported in Table~\ref{table:prop}.

\begin{table}[h!]
\begin{equation}
\centering
	\ra{1.3}
	\begin{array}{@{}llllllll@{}}
	\toprule 
	 & \multicolumn{3}{c}{28 \mbox{ GHz}} & & \multicolumn{3}{c}{73 \mbox{ GHz}}  
	\\ \midrule 
	 & \alpha & \beta & \sigma\mbox{ [dB]} & & \alpha & \beta & \sigma\mbox{ [dB]} \\ \cline{2-4} \cline{6-8}
	 \mbox{LOS} & 61.4 & 2 & 5.8 & & 69.8 & 2 & 5.8 \\ \cline{2-4} \cline{6-8}
	 \mbox{NLOS} & 72 & 2.92 & 8.7 & & 86.6 & 2.45 & 8.0 \\ \bottomrule

	\end{array}
\end{equation}	
\caption{Propagation parameters for Eq.~\eqref{eq:pl}, from~\cite{rappaport1}}
\label{table:prop}
\end{table}

\vspace{10pt}

\begin{figure}[t]
	\centering	
	\includegraphics[width = 0.7\textwidth]{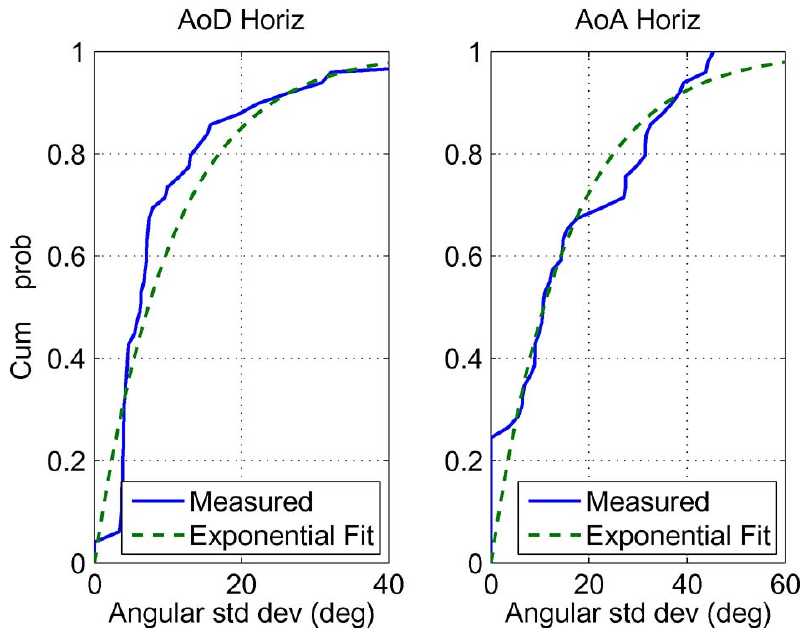}
	\caption{Distribution of the measured beamspread for 28 GHz, and exponential fit, from~\cite{rappaport1}}
	\label{fig:ang}
\end{figure}

\textbf{Channel Matrix}: the channel model is based on the 3GPP/ITU MIMO model. It is modeled as a random number of $K$ path clusters, with each of them representing a macro-scattering path. Each cluster is described by (i) a fraction of the transmission power; (ii) angles of departure (AoD) and arrival (AoA) of the path; (iii) how much the beam is spread around those angles; (iv) the group propagation delay and the delay profile of the cluster. In the NYU mmWave channel model the temporal element (iv) is not taken into account. In~\cite{rappaport1} the clusters of a mmWave transmission on 28 GHz and 73 GHz are defined by fitting real measurements. In particular the parameters of each cluster are:
\begin{itemize}
	\item The number of clusters, estimated with a clustering algorithm, for each measure, and modeled as a Poisson random variable
	\begin{equation}
	 	K = \max\{ 1, Poisson(\lambda)\}
	 \end{equation} 
	with $\lambda = 1.8$ for 28 GHz and $\lambda = 1.9$ for 73 GHz;
	\item The power fraction for each cluster, computed in~\cite{rappaport1} following the 3GPP/ITU MIMO model;
	\item The angular dispersion of each cluster, measured as the Root Mean-Squared (RMS) beamspread around the two angular dimensions (vertical and horizontal), which is modeled as an exponential random variable as shown in Fig.~\ref{fig:ang}.
\end{itemize}

Then, the channel gain matrix is generated as follows~\cite{rappaport1}. At first, realizations of large-scale parameters are drawn (pathloss, number of clusters $K$, power fractions, angular beamspread). Secondly, small-scale fading is taken into account by splitting each cluster into $L$ subpaths, each with a horizontal and a vertical AoA ($\theta_{k,l}^{rx}$, $\phi_{k,l}^{rx})$) and AoD ($\theta_{k,l}^{tx}$, $\phi_{k,l}^{tx})$), with $k = 1, \dots, K$ the cluster index and $l= 1, \dots, L$ the subpath index. These are generated as Gaussian random variables centered around the cluster central angle and with a standard deviation equal to the RMS beamspread of the cluster. Then, given a pair TX-RX with $n_{rx}$ and $n_{tx}$ receiving and transmitting antennas, the channel gain matrix at time $t$ is~\cite{forde2e}
\begin{equation}
	\mathbf{H}(t, f) = \sum_{k=1}^K \sum_{l=1}^L g_{k,l}(t)\mathbf{u}_{rx}(\theta_{k,l}^{rx}, \phi_{k,l}^{rx})\mathbf{u}_{tx}^*(\theta_{k,l}^{tx}, \phi_{k,l}^{tx})
\end{equation}
where $\mathbf{u}_{rx}(\theta, \phi)$ and $\mathbf{u}_{tx}(\theta, \phi)$ are the RX and TX spatial signatures, respectively, and $g_{k,l}(t)$ is the small-scale fading coefficient on the $l$-th subpath of the $k$-th cluster. It describes the sudden fluctuations in the received power due to the self-interference given by the same signal received from the $L$ different subpaths of a cluster. It is computed as~\cite{forde2e}
\begin{equation}\label{eq:small}
 	g_{k,l}(t, f) = \sqrt{L} e^{2\pi i t f_{d, \max} \cos{\omega_{k,l}} - 2 \pi i \tau_{k,l} f}
\end{equation} 
where $f_{d, \max}$ is the maximum Doppler frequency, $\omega_{k,l}$ is the angle of arrival of the subpath relative to the motion direction, $\tau_{k,l}$ the delay spread, $f$ the carrier frequency and $L$ is the pathloss computed in Eq.~\eqref{eq:pl}.

\vspace{10pt}
\textbf{Beamforming}: the NYU mmWave module provides a new antenna model for a Uniform Linear Array (ULA) in the \texttt{AntennaArrayModel} class, that supports analog beamforming. The beamforming vectors are pre-computed using MATLAB and loaded when the simulation is launched. The beamforming assumes perfect Channel Side Information (CSI) and full knowledge of the channel matrix $\mathbf{H}(t,f)$. The result is that the optimal beamforming vector is always chosen for a TX/RX pair $i,j$, and the beamforming gain is computed as~\cite{mmwaveSim}
\begin{equation}\label{eq:bf}
	G(t,f)_{i,j} = | \mathbf{w}_{rx_{i,j}}^* \mathbf{H}(t,f) \mathbf{w}_{tx_{i,j}} |^2
\end{equation}
where $\mathbf{w}_{rx_{i,j}}$ is the beamforming vector of receiver $j$, when the transmitter is $i$, and vice versa $\mathbf{w}_{tx_{i,j}}$ is the beamforming vector of transmitter $i$ when the receiver is $j$.

\begin{figure}
\centering
	\includegraphics[width = 0.7\textwidth]{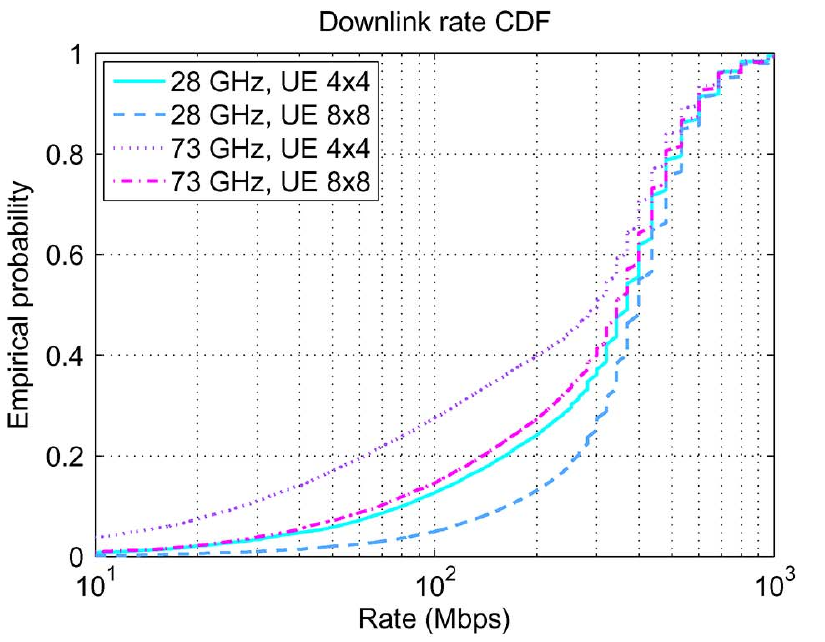}
	\caption{Downlink rate Cumulative Distribution Function (CDF) with 4x4 and 8x8 ULA at the UE side (eNB has a 8x8 ULA), for 28 GHz and 73 GHz, from~\cite{rappaport1}}
	\label{fig:rate_cdf_rapp}
\end{figure}

The available antenna arrays of the \texttt{AntennaArrayModel} class are a 2x2, a 4x4, an 8x8 and a 16x16 array, which offer a good compromise between the size of the array (in order to be placed in a mobile device) and the performance in an urban environment, as shown in Fig.~\ref{fig:rate_cdf_rapp}.

\vspace{10 pt}
\textbf{Channel Configuration for the simulation}: at simulation startup, previously generated channel matrices and beamforming vectors are loaded in the simulator. This helps reduce the computational load of a simulation. As stated in~\cite{forde2e}, due to the lack of understanding of the time dynamics of the mmWave channel, and in order to simulate a time varying channel with large-scale fading effects, the channel matrices are updated periodically for NLOS channels. For the LOS state, instead, the channel is assumed to be much more stable and remains constant. In NLOS, at each update, one of the channel matrix instances is picked at random, thus making each interval independent. As specified in~\cite{forde2e}, this method is not yet validated, but allows to simulate a form of large-scale block fading.

The small-scale fading represented by Eq.~\eqref{eq:small}, instead, is updated at every transmission, using the mobility model of ns--3 to have knowledge of the UE speed and position relative to the mmWave eNB. In particular, the information on the position is used to compute the pathloss $L$. Then, by knowing the UE speed $v$ it is possible to compute the Doppler frequency 
\begin{equation}
	f_{d, \max} = v\frac{f_c}{c}
\end{equation}
with $f_c$ the carrier frequency and $c$ the speed of light. 
The other factors, such as the Doppler shift $\cos{\omega_{k,l}}$ and the delay spread $\tau_{k,l}$, are not based on measurements, and are constant throughout all simulations.

\subsubsection*{\textbf{Ray-Tracing Generated Model}}
The NYU mmWave module offers also the possibility of using traces generated by a third-party ray tracing software, which simulates the radio propagation environment (see~\cite{mmNet} for an example of application to TCP). It is used to generate channel matrices and update them according to the UE mobility, while the beamforming is updated with Eq.~\eqref{eq:bf}. 

\subsection{Error Model}\label{sec:error}
The Error Model is based on standard link-to-system mapping techniques, which allow to map the SINR to an error probability for the whole transport block (TB), taking into account modulation and coding techniques~\cite{mmwaveSim}. 

\begin{figure}[t]
	\centering	
	\includegraphics[width=0.7\textwidth]{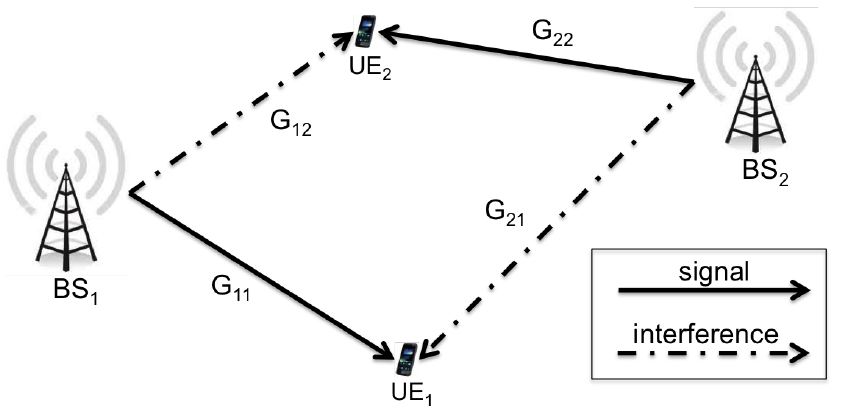}
	\caption{Interference computation example, from~\cite{mmwaveSim}}
	\label{fig:interference}
\end{figure}

The SINR computation is done for each transmission according to the following procedure. Firstly, interference from adjacent eNBs that operate in the same frequency is accounted for: as shown in Fig.~\ref{fig:interference}, in order to compute the SINR for the transmission from base station 1 to UE 1, the beamforming gain $G_{1,1}$ of this pair is computed, as well as the beamforming gain $G_{2,1}$ of the interfering base station 2. Then, the SINR is computed as
\begin{equation}
 	SINR_{1,1} = \frac{\frac{P_{Tx_{1,1}}}{L_{1,1}}G_{1,1}}{\frac{P_{Tx_{2,2}}}{L_{2,1}}G_{2,1} + B N_0}
\end{equation} 
where $P_{Tx_{i,j}}$ is the transmit power the eNB $i$ uses to transmit to UE $j$, $L_{i, j}$ is the pathloss from transmitter $i$ to receiver $j$, and $B N_0$ is the thermal noise. Thanks to the ns--3 Spectrum module, it is possible to compute the SINR of each OFDM subcarrier.

This is used to compute the Mean Mutual Information per coded Bit (MMIB) $\gamma_i$ for each of the $C$ codeblocks (CB) of which the TB is composed. The MMIB is a sample mean of the Mutual Information per Coded Bit (MICB) computed for each subcarrier as described in~\cite{mezzavilla_bler} as a function of the SINR. Then for CB $i$ the block error rate (BLER) is modeled with a Gaussian cumulative model, in order to reduce the simulation complexity:
\begin{equation}
	C_{BLER,i}(\gamma_i) = \frac{1}{2} \left[1 - \mbox{erf} \left(\frac{\gamma_i - b_{C_{SIZE}, MCS}}{\sqrt{2} c_{c_{SIZE}, MCS}}\right) \right]
\end{equation}
where the parameters $b_{C_{SIZE}, MCS}$ and $c_{C_{SIZE}, MCS}$ are the mean and standard deviation of the Guassian distribution, and are computed by numerical fitting of link level error rate curves for each CB size and Modulation and Coding Scheme (MCS). Finally, the TB BLER is
\begin{equation}
	T_{BLER} = 1 - \Pi_{i=1}^C (1 - C_{BLER,i}(\gamma_i)).
\end{equation}

For each transmission, a TB is declared received with error by drawing a uniform random value and comparing it to $T_{BLER}$.

\subsection{mmWave Physical Layer Frame Structure}
Several papers argue that a TDD structure will allow to reduce the latency of the 5G radio interface~\cite{samsungtdd, ghoshTDD, duttaTDD}. Therefore the NYU mmWave module implements a TDD frame structure for the physical layer, which can be configured on several parameters. The default values are shown in Table~\ref{table:sim_param}, and unless specified otherwise, they will be used for the simulations of Chap.~\ref{chap:results}.

\begin{table}[t]
\centering
	\ra{1.3}
	\begin{tabular}{@{}ll@{}}
	\toprule 
	\textbf{Parameter} & \textbf{Value} \\ \midrule
	\textbf{Time-related parameters} & \\ \midrule
	Subframes per frame & 10 \\
	Subframe duration & 100 $\mu$s \\
	OFDM symbols per subframe & 24 \\
	OFDM symbol length & 4.16 $\mu$s \\ 
	UL/DL switching guard period & 4.16 $\mu$s \\ \midrule
	\textbf{Frequency-related parameters} & \\  \midrule
	Number of sub-bands & 72 \\
	Sub-bands bandwidth & 13.89 MHz \\
	Subcarriers in each sub-band & 48 \\ 
	Carrier frequency & 28 GHz (also 73 GHz is supported) \\ \midrule
	\textbf{Processing latencies} & \\ \midrule
	MAC scheduling to transmission delay & 2 subframes \\
	PHY reception to MAC processing delay & 2 subframes \\
	\bottomrule
\end{tabular}
\caption{Default frame structure and PHY-MAC related parameters for ns--3 mmWave module}
\label{table:sim_param}
\end{table}

\begin{figure}[t]
	\centering
	\includegraphics[width = 0.8\textwidth]{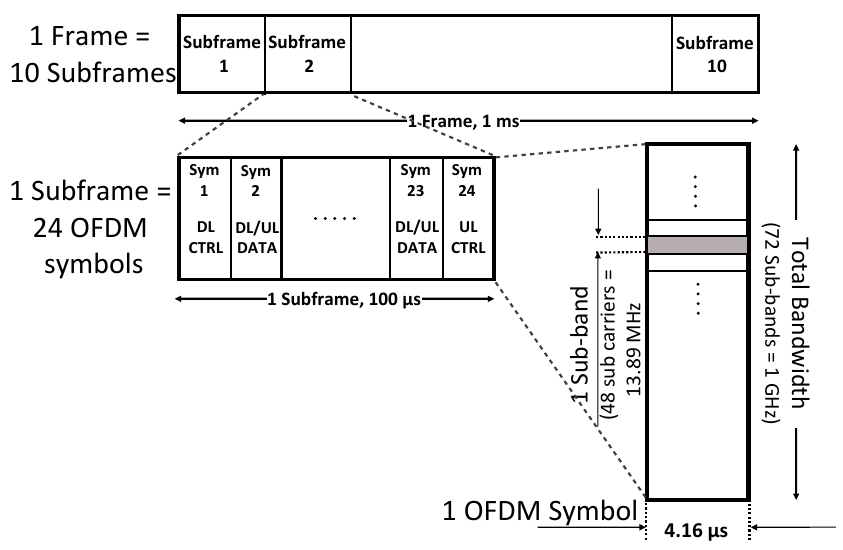}
	\caption{Possible time-frequency structure of a mmWave frame, from~\cite{forde2e}}
	\label{fig:frame}	
\end{figure}

Each slot of a subframe can be assigned either to a downlink (DL) or to an uplink (UL) transmission, with the exception of the first (DL control symbol) and the last (UL control symbol). A guard period between UL and DL symbols is introduced. In the frequency domain, while in principle each sub-band could be assigned to different UEs, the Orthogonal Frequency-Division Multiple Access (OFDMA) of LTE is not implemented, and therefore the whole 1 GHz band is assigned to a certain UE in each slot. 

The frame structure in time and frequency is shown in Fig.~\ref{fig:frame}.

\subsection{mmWave PHY and MAC Layer Operations}
The PHY and MAC layers of the mmWave module are organized as the LTE corresponding classes. 

The mmWave PHY layer of eNBs and UEs provides the following common functionalities:
\begin{itemize}
	\item it receives MAC PDUs and stores them in buffers;
	\item it calls the StartSubFrame end EndSubFrame methods at fixed intervals, as specified in Table~\ref{table:sim_param};
	\item it calls the StartSlot and EndSlot methods, with a slot representing a variable quantity of OFDM symbols of a subframe allocated to DL and UL. The length of a slot is specified by the MAC layer, and in a slot a device may either transmit or receive;
	\item it relies on the \texttt{MmWaveSpectrumPhy} class to perform the transmission, reception, computation of SINR and packet error probabilities as described in Sec.~\ref{sec:error}, and generation of a Channel Quality Indicator (CQI), which is fed back to the eNB MAC layer either in UL data or control slots.
\end{itemize}

The MAC layer main functionalities are instead scheduling in the eNB, AMC and multiprocess stop and wait HARQ. 

\textbf{MAC Scheduling}: the scheduler is described in~\cite{forde2e}. It is a Round-Robin scheduler which assumes a Time Division Multiple Access (TDMA) scheme, which may be a reasonable assumption if analog beamforming is employed\footnote{The transmitter and receiver have to align their antennas in order to experience the maximum gain in a certain direction, and this is possible for a single TX/RX pair at a time with analog beamforming. If digital beamforming is employed, instead, multiple transmissions are possible, however digital beamforming is typically considered to be very costly and power consuming, therefore it may deployed only in the eNBs}. The TDMA scheme is improved with a variable slot duration as described in~\cite{duttaTDD}, in order to maximize resource utilization and account for transport blocks of different sizes. 

The scheduler is triggered by the MAC layer at the beginning of a frame. Firstly, HARQ retransmissions are scheduled, and secondly new data is processed, by dividing the data symbols in the subframe evenly among users. 

\textbf{Adaptive Modulation and Coding}: it mainly reuses the ns--3 LTE module AMC class, with some changes in order to account for a TDMA scheme. It computes the MCS from the CQIs reported by the UE or the SINR measurements at the eNB. Then, it computes the number of symbols needed to serve a TB with a certain MCS. 

At the UE side, there is a method to generate a CQI from a wideband SINR measurement, so that feedback for the downlink channel can be provided to the eNB.

\textbf{Hybrid ARQ retransmission}: also HARQ is based on the ns--3 LTE HARQ. It provides a stop and wait HARQ with the possibility of using up to \\\texttt{NumHarqProcesses} parallel HARQ processes. 

\section{The ns--3 LTE Upper Layers}
The ns--3 LTE module~\cite{lena} implements the LTE stack and some nodes of the EPC network. The main focus of the ns--3 LTE module is the assessment of the system level performance when UEs are in the RRC\_CONNECTED state, therefore the functionalities related to this state are implemented with a high level of detail, while other parts of the LTE standard are not implemented.

The mmWave module uses the RLC, PDCP, RRC layers and core network classes of the LTE module, on top of the custom mmWave PHY and MAC. 

\subsection{The RLC and PDCP Layers}
The PDCP layer implementation mainly offers (i) the creation of PDCP PDUs with an SDU received from upper layers and a PDCP header with a sequence number; (ii) the transmission and reception of data or control packets. It does not offer security-related primitives, nor header compression, in-order delivery and timeout-triggered PDU discard. It offers an interface to the RRC layer (\texttt{PdcpSapProvider}) with the primitive \texttt{TransmitPdcpPdu}, and an interface to any RLC entity associated with it (\texttt{PdcpSapUser}), with the \texttt{ReceivePdcpSdu} method. 

Moreover, the PDCP implementation does not support lossless handover, i.e., during a handover only new incoming packets are forwarded from the source to the target eNB, while the content of RLC buffers is not. 

\begin{figure}
	\centering	
	\includegraphics[width = 0.95\textwidth]{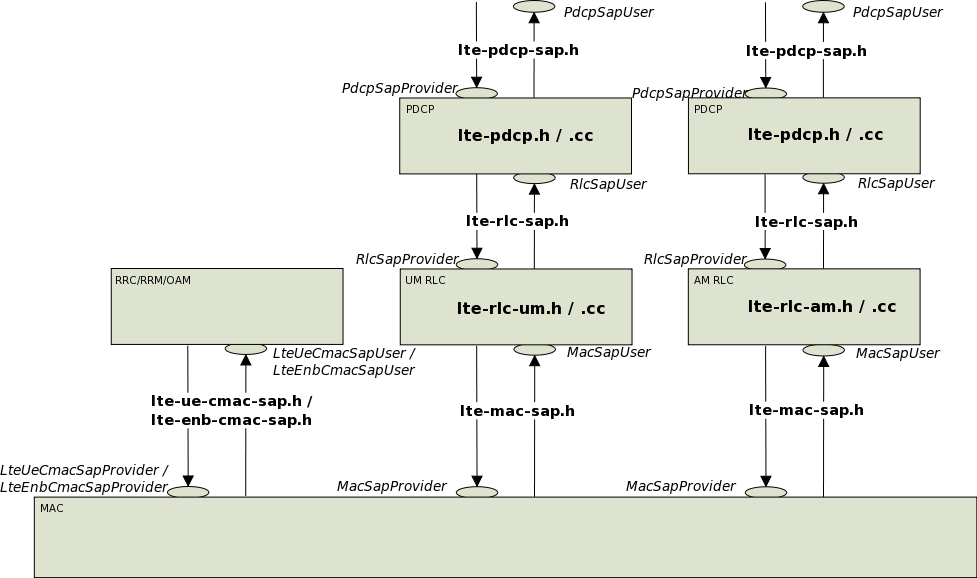}
	\caption{Implementation Model of PDCP and RLC entities and SAPs, from~\cite{ltemodel_rlc}}
	\label{fig:rlc_inter}
\end{figure}

The RLC layer is implemented in the three different versions described in Chapter~\ref{chap:integration}, and offers segmentation/concatenation of PDCP PDUs and retransmission for the AM entities. The interfaces provided by RLC and to RLC are described in Fig.~\ref{fig:rlc_inter}. With respect to the PDCP layer, the RLC interface allows to receive a PDCP PDU and forward to the PDCP layer a PDCP PDU. With respect to the MAC layer, the RLC layer reports the buffer status and forwards RLC PDUs, while the MAC layer notifies transmission opportunities and forwards RLC PDUs to the RLC layer.

The RLC AM supports concatenation and segmentation of PDCP PDUs but not of segments to be retransmitted. Moreover, it does not offer the primitives to signal that the maximum number of retransmissions is reached and to reassemble packets segmented in buffers for lossless handover. The same holds for RLC UM, with the exception of retransmissions. 

The RLC TM, instead, is directly interfaced with the RRC in order to transmit control packets, without modifying them (no segmentation, and the RLC is not added).

For each signalling and data radio bearers, a PDCP and an RLC entity are created both at the eNB and at the UE side.

\subsection{The RRC Layer}
The RRC layer model which is implemented in the ns--3 LTE module is divided among the \texttt{LteEnbRrc} and \texttt{LteUeRrc} classes, with support of the \texttt{LteRrcProtocolIdeal} and \texttt{LteRrcProtocolReal} classes. The implementation has a lot of details, and the main functionalities are 
\begin{itemize}
	\item transmission of SI from the eNB to the UE - only Master Information Block (MIB), System Information Block Type 1 and 2 (SIB1 and SIB2);
	\item LTE initial cell search and synchronization procedures;
	\item three different RRC procedures related to the connection state, i.e., RRC connection establishment, RRC connection reconfiguration (for example for data and signalling radio bearers setup, and for handover), and finally RRC connection re-establishment after handover;
	\item maintaining a list of radio bearers with related PDCP and RLC entities. 
\end{itemize}
 
Each of these functionalities requires methods on both the UE and the eNB RRC layers, and primitives in the protocol classes in order to transmit the messages between the two entities. 

The \texttt{LteRrcProtocolIdeal} class offers an ideal way to forward RRC commands from the eNB to the UE and vice versa, i.e., the methods of the other RRC endpoint are called directly and no packet is sent on physical interfaces. The \texttt{LteRrcProtocolReal} class, instead, models the transmission of RRC messages as it is defined by the LTE standard. In particular, for every RRC message that needs to be sent, an RRC PDU is created by encoding the \textit{Information Elements} (IEs, namely each parameter-value pair of the message) with a ASN.1 encoder, as specified in~\cite{rrc}. Only the IEs that are useful to the simulation are encoded and sent, thus the actual traffic generated by the RRC layer is slightly lower than the one that would be generated in a real system. Then, each encoded RRC PDU, containing the ASN.1 header and the actual payload, is forwarded to the PDCP layer associated with a signalling radio bearer. Only SRB0 and SRB1 are actually modeled, the first uses RLC TM, while the second uses RLC AM. Therefore the PDUs generated by the \texttt{LteRrcProtocolReal} class are subject to the same modeling used for data communications: scheduling and transmission delays, possibility of not receiving the packet, thus retransmissions, and actual radio resource consumption.

The RRC implementation does not model the functionalities associated with the RRC\_IDLE state at the UE side, or needed to reach this state once in the RRC\_CONNECTED state, as for example RLF or RRC connection release. The only occasions in which the RRC exits from state RRC\_CONNECTED are the handovers, for which it switches to RRC\_CONNECTED\_HANDOVER. At the eNB side the release of a UE context is implemented for the handover functionality or if one of the standard-defined timers expires. 

\subsection{Evolved Packet Core Network in ns--3}\label{sec:cn}
The ns--3 LTE module provides a basic modeling of the EPC network. 

The S-GW and P-GW nodes are hosted in the same node, which is connected to the eNBs with point to point links. These are characterized by a limited bandwidth, a latency and a Maximum Transfer Unit (MTU), and are used to transfer data packets from the internet to the LTE UEs. These links are the physical medium upon which the S1-U interface works, which performs tunneling for each data radio bearer. The tunnelling protocol used in the 3GPP LTE standard is the GPRS Tunneling Protocol (GTP). In downlink, the S-GW node adds the GTP-U header to the packets and forwards them to the eNB, in which the class \texttt{EpcEnbApplication} is in charge of delivering the packet to the radio protocol stack, addressing it to the correct UE thanks to the tunneling information. 

The MME, instead, is not modeled as a node but it is simply an object whose methods are invoked when needed. Therefore, also the S1-AP interface is not realistically modeled. The primitives that are supported by the \texttt{EpcMme} class are related to the UE initial setup and to the path switching operations during handovers.

Finally, the X2 interface between eNBs is modeled as a point to point link, with its datarate, latency and MTU, on top of which are exchanged packets with X2 headers and X2-AP PDUs. The X2-C should be implemented using SCTP as transport protocol, however, since this is not available in ns--3, UDP is used. X2-U instead performs tunneling over GTP.

This part of the ns--3 LTE module is not integrated in the mmWave module, which does not support X2-based handover. The implementation of X2-based handover for the mmWave is part of the work of this Thesis and will be discussed in Chapter~\ref{chap:impl}.

%% file: chapters/chapter4.tex

\chapter{LTE-5G Integration Implementation}\label{chap:impl}

In this Chapter, the proposed Dual Connectivity architecture for LTE-5G integration at the PDCP level will be discussed and the implementation in ns--3 will be presented. Firstly, general features and architectural choices will be described. Then, this Chapter will focus on the features needed to support fast switching, i.e., the form of Dual Connectivity in which the UE is connected to both RATs, but uses just one of the two at a time for data transmissions. By being connected to both, the UE will switch between the two with a single RRC message. Moreover, the baseline for comparison, i.e., hard handover (HH) between LTE and 5G, will be described, along with some implementation details.

\section{LTE-5G Multi-Connectivity Architecture: Control Signalling}\label{sec:meas}
A first attempt to describe an LTE-5G architecture from the control point of view was made in~\cite{giordani}. This paper, that is the starting point of the control implementation of this Thesis, considers different aspects of an LTE-5G integrated system, such as: (i) control signalling and coordination between LTE and 5G; (ii) 5G sound reference signals, with analysis of different alternatives.

\subsection{Measurement Collection}
A UE is typically within reach of an LTE eNB, which is designated as Master Cell, in accordance to 3GPP terminology. This cell can act as a coordinator for the mmWave cells which are located under its coverage, but the coordinator entity can be placed also in a different node inside the core network (provided it is close enough to the edge). The mmWave cells act as Secondary Cells, and exchange control information with the coordinator via the X2 interface. 

One of the main functionalities of this architecture is to report the mmWave link signal quality to the coordinator, which selects the best mmWave cell to which the UE should connect. In particular, it is expected that mmWave-capable UEs and eNBs will use directional phase arrays for beamforming. Therefore, each node selects a certain number of directions, or sectors ($N_{UE}$ for the UE and $N_{eNB}$ for the eNB). A measure of the signal quality is needed for each UE-eNB direction pairs, for a total of $N_{UE} \times N_{eNB}$ measures per UE, considering all the mmWave eNBs within reach. These measures are then reported to the coordinator, in a procedure that works as follows:
\begin{enumerate}
  \item The UE broadcasts a reference signal for each of the $N_{UE}$ directions, changing sector at each transmission. The reference signal is known to the eNB and can be used for channel estimation. If analog beamforming is used, each mmWave eNB either scans its $N_{eNB}$ sectors one at a time, or, if digital beamforming is applied, collects measurements from all of them at once. The mmWave eNB fills a \textit{Report Table} (RT) with the SINR and the SINR variance for each UE, in each direction, and sends it to the coordinator;
  \item The coordinator is able to build a Complete Report Table (CRT) for each UE, considering the information coming from all the mmWave eNBs. The optimal eNB and direction for each UE is then selected considering the SINR for each (mmWave eNB, direction) pair;
  \item The LTE eNB (even if not acting as coordinator) reports to the UE which is the (mmWave eNB, direction) pair that yields the best performance. The choice of using the LTE control link is motivated by the fact that the UE may not be able to receive from the optimal mmWave link if not properly configured. The LTE control link, moreover, offers higher stability and reliability.
\end{enumerate}

There is a necessary delay to collect all the measurements for a UE, as described in~\cite{giordani}. The period of transmission of a reference signal is defined as $T_{per}$, and each signal lasts $T_{sig}$. The assumed values are $T_{per} = 200\,\mu$s and $T_{sig} = 10\,\mu$s, in order to maintain an overhead of 5\%. The measurement procedure for each UE requires $N_{eNB}N_{UE}/L$ scans, with $L$ the number of simultaneous directions from which the receiver can receive. For example, with analog beamforming $L=1$, while for an eNB (UE) with digital beamforming $L=N_{eNB}$ ($L=N_{UE}$) respectively. Therefore the delay will be
\begin{equation}
  D = \frac{N_{eNB}N_{UE}T_{per}}{L}.
\end{equation}
Table~\ref{table:srs} reports the delay for different configurations of a system with $N_{UE} = 8$ and $N_{eNB} = 16$, for uplink-based reference signals.

\begin{table}
  \centering
    \ra{1.3}
    \begin{tabular}{@{}lll@{}}
    \toprule
    \multicolumn{2}{l}{\textbf{BF Architecture}} & \multirow{2}{*}{\textbf{Delay} $D$} \\
    \cline{1-2}
    eNB side & UE side & \\ \midrule
    Analog & Analog & 25.6 ms \\ 
    Analog & Digital & 25.6 ms \\ 
    Digital & Analog & 1.6 ms \\ 
    Digital & Digital & 1.6 ms \\ \bottomrule
  \end{tabular}
  \caption{Delay needed to collect measurements for each UE, at each mmWave eNB, for $T_{ref} = 200\,\mu$s, $N_{UE}=8$, $N_{eNB}=16$}
  \label{table:srs}
\end{table}

Once the reporting is done, the UE has to perform initial access to the mmWave eNB, or handover from a mmWave eNB to a new one. These procedures will be described in Sec.~\ref{sec:dc}, by highlighting the details that are added to the scheme in~\cite{giordani} in order to be compatible with the architecture and the implementation of this Thesis. 

Notice that the NYU mmWave module does not implement any kind of sound reference signal transmission, but, when assigning radio resources, it accounts for the overhead that it generates. In order to be able to compute the SINR in the mmWave eNBs, it is possible to exploit the flexibility provided by the fact that ns--3 is a simulator, and not an actual implementation. What was done for this Thesis is to use the same procedure of SINR computation described in Chapter~\ref{chap:ns3}, by adding it in the new \texttt{UpdateSinrEstimate} method of the \texttt{MmWaveEnbPhy} class. Then, every $D$ ms, for each UE in the scenario, the \texttt{MmWaveEnbPhy} class of each mmWave eNB uses this method to compute the SINR and reports it to the mmWave RRC layer.

\section{Implementation of Dual Connectivity}\label{sec:dc}
Sec.~\ref{sec:dasilva} presents an extensive discussion on which layer can be used as the integration layer. In this thesis, the PDCP layer is chosen for evaluation as the candidate integration layer. Indeed, there are several points in favor of this choice. 

\vspace{2pt}

\textit{1) Non co-located deployment} - The first is that synchronization is not required, and therefore a non co-located deployment of the stack is feasible. Since mmWave cells are expected to have a coverage radius of at most 200~m, they will be deployed with a density higher than that of LTE cells (which are already installed)~\cite{rangan14}. It would be costly to install both an LTE and a mmWave eNB in each new site. Moreover, a high density of LTE eNBs implies a smaller coverage area for each of them, in order to avoid inter-cell interference. One of the main features of LTE-5G tight integration is the large coverage layer that LTE macro cell could provide. If the area of each LTE cell is reduced to the same as that of mmWave cells, then the coverage layer would not be effective.

Moreover, the PDCP layer can be moved to the core network, in a new coordinator node, that can act as gateway for a group of LTE eNBs and the mmWave eNBs under their coverage, or can be deployed in a macro LTE eNB.

\vspace{2pt}

\textit{2) No design constraints on 5G PHY to RLC layers} - The second is the possibility of designing the mmWave 5G protocol stack from the PHY layer to the RLC layer without constraints given by the already standardized LTE protocols. This allows to have a clean slate approach that may help addressing 5G performance requirements and tackle the mmWave challenges. For example, a TDD scheme can be employed at the PHY and MAC layers, since it helps reducing the radio access latency~\cite{samsungtdd}. If the integration is performed at the MAC or the PHY layer, for example, the duplexing would have to be the same for LTE and mmWave 5G, and most of the already deployed LTE networks use FDD. 

\vspace{2pt}

\textit{3) Lean and simple solution} - The third is that a dual connectivity solution at the PDCP layer is a simple and lean solution. If the integration happens at the RLC layer, the reassembly process at the receiver would be slowed down by the fact that the fragments sent on the LTE air interface have a higher latency than the one of mmWave fragments, and therefore the latter would have to stay in the buffer and wait for the LTE RLC PDUs with the missing fragments. At the PDCP layer, instead, no fragmentation/reassembly is performed. The PDCP layer may however discard packets due to timeout: in order to account for this problem the timeout has to be set high enough.

\vspace{2pt}

\begin{figure}[t]
  \centering
  \includegraphics[width=0.9\textwidth]{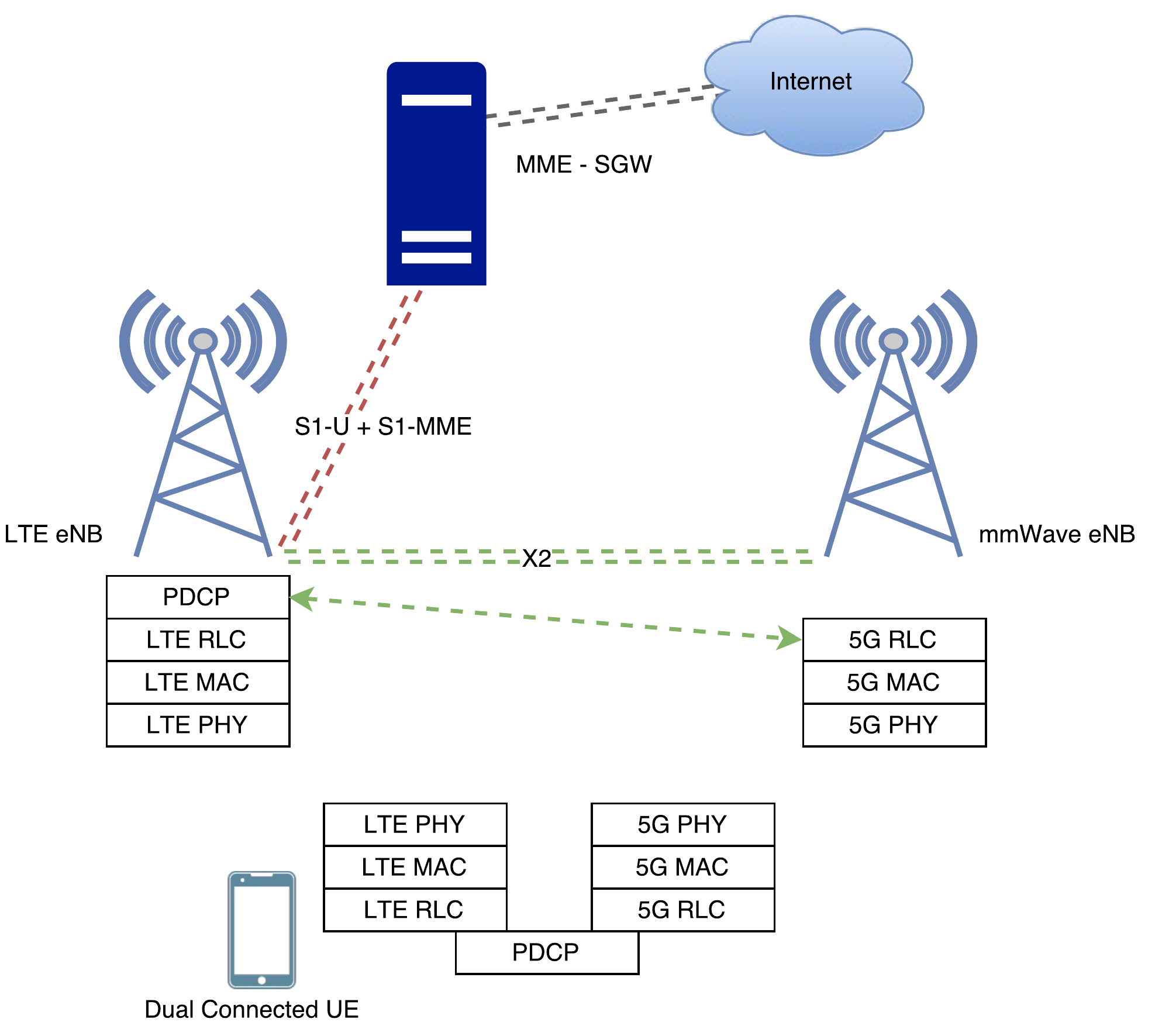}
  \caption{LTE-5G tight integration architecture}
  \label{fig:arch}
\end{figure}

Fig.~\ref{fig:arch} shows a block diagram of the proposed architecture. Notice that it recalls option 3C from the 3GPP report~\cite{36842}, as discussed in Sec.~\ref{sec:ltedc}, but there are some differences that will be described in the following paragraphs. 

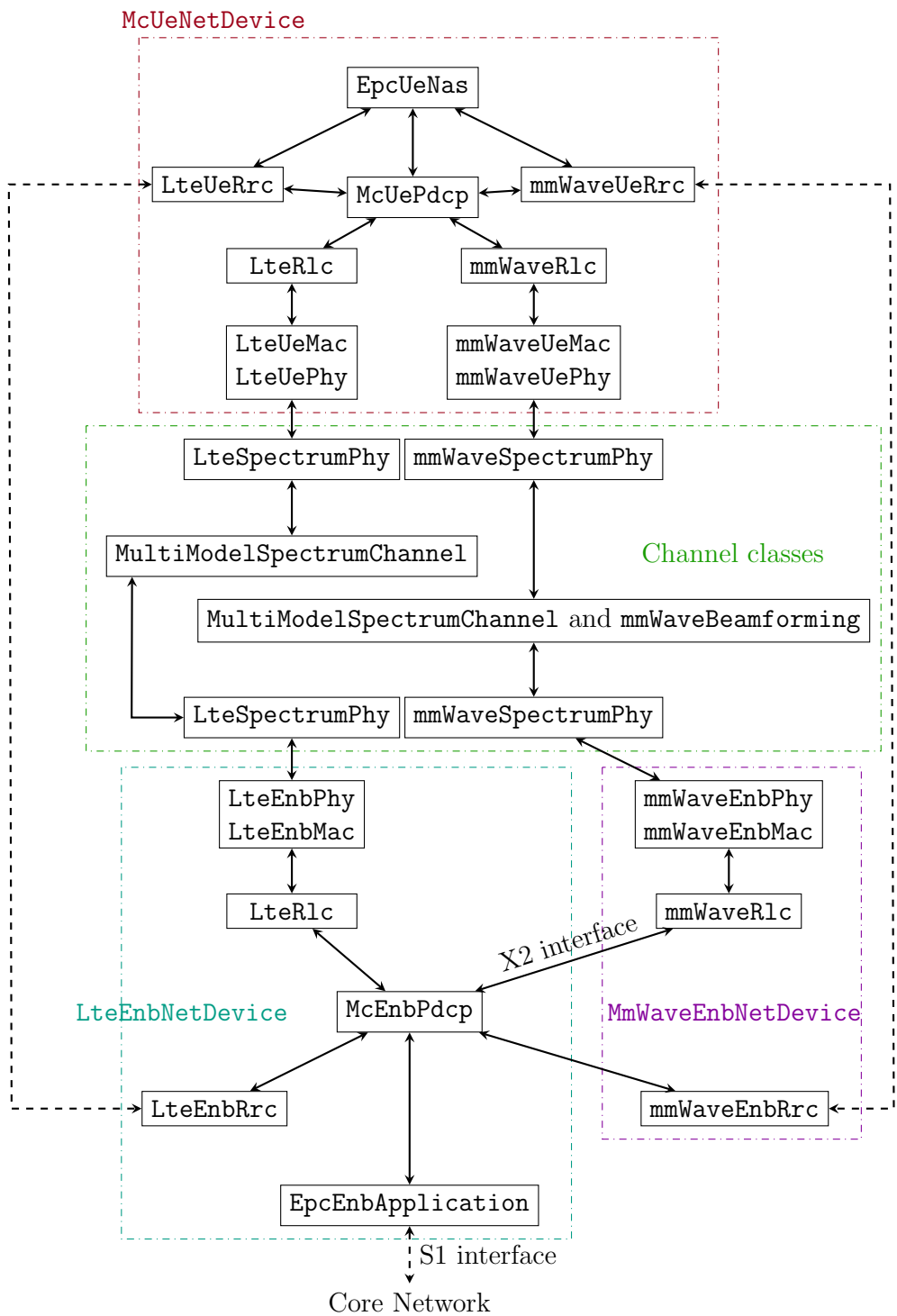
\begin{figure}
\centering
\begin{tikzpicture}[node distance=1.5cm, scale=0.95, every node/.style={scale=0.95}]
  \node (epc) [process] {\texttt{EpcUeNas}};
  \node (rrc) [process, below of=epc, xshift=-3cm] {\texttt{LteUeRrc}};
  \node (mmrrc) [process, below of=epc, xshift=3cm] {\texttt{mmWaveUeRrc}};
  \node (pdcp) [process, below of=epc, yshift=-0.2cm] {\texttt{McUePdcp}};
  \node (rlc) [process, below left of=pdcp, xshift=-0.8cm] {\texttt{LteRlc}};
  \node (mmrlc) [process, below right of=pdcp, xshift=0.8cm] {\texttt{mmWaveRlc}};
  \node (ltephy) [process, below of=rlc] {\texttt{LteUeMac} \\ \texttt{LteUePhy}};
  \node (mmphy) [process, below of=mmrlc] {\texttt{mmWaveUeMac} \\ \texttt{mmWaveUePhy}};

  \draw[chaptergrey,dashdotted] ($(rrc.north west)+(-0.2,2)$) rectangle ($(mmphy.south east)+(1.5,-0.2)$);
  \node (legendMcUe) [above left of=epc, xshift=-2cm]{\textcolor{chaptergrey}{\texttt{McUeNetDevice}}};

  \node (ltephy2) [process, below of=ltephy] {\texttt{LteSpectrumPhy}};
  \node (mmphy2) [process, below of=mmphy] {\texttt{mmWaveSpectrumPhy}};
  \node (channel) [process, below of=mmphy2, yshift=-1cm] {\texttt{MultiModelSpectrumChannel} and \texttt{mmWaveBeamforming}};
  \node (ltechannel) [process, below of=ltephy2] {\texttt{MultiModelSpectrumChannel}};
  \node (enbltephy2) [process, below of=ltechannel, yshift = -1cm] {\texttt{LteSpectrumPhy}};
  \node (enbmmphy2) [process, below of=channel] {\texttt{mmWaveSpectrumPhy}};

  \draw[chaptergreen,dashdotted] ($(ltephy2.north west)+(-1.5,0.2)$) rectangle ($(enbmmphy2.south east)+(3.35,-0.2)$);
  \node (legendChannel) [above right of=channel, xshift=2cm] {\textcolor{chaptergreen}{Channel classes}};

  \node (lteEnbLl) [process, below of=enbltephy2] {\texttt{LteEnbPhy} \\ \texttt{LteEnbMac}};
  \node (mmEnbLl) [process, below of=enbmmphy2, xshift=3cm] {\texttt{mmWaveEnbPhy} \\ \texttt{mmWaveEnbMac}};
  \node (lteEnbRlc) [process, below of=lteEnbLl] {\texttt{LteRlc}};
  \node (mmEnbRlc) [process, below of=mmEnbLl] {\texttt{mmWaveRlc}};
  \node (enbPdcp) [process, below right of=lteEnbRlc, xshift=0.75cm, yshift=-0.5cm] {\texttt{McEnbPdcp}};
  \node (enbrrc) [process, below of=enbPdcp, xshift=-3cm] {\texttt{LteEnbRrc}};
  \node (enbmmrrc) [process, below of=enbPdcp, xshift=5cm] {\texttt{mmWaveEnbRrc}};
  \node (epcEnb) [process, below of=enbPdcp,yshift=-1.5cm] {\texttt{EpcEnbApplication}};
  \node (cn) [below of=epcEnb] {Core Network};

  \draw[chapterpurple,dashdotted] ($(mmEnbLl.north west)+(-0.5,0.2)$) rectangle ($(enbmmrrc.south east)+(0.5,-0.2)$);
  \node (legendChannel) [left of=enbPdcp, xshift=6.5cm] {\textcolor{chapterpurple}{\texttt{MmWaveEnbNetDevice}}};

  \draw[chapterlightgreen,dashdotted] ($(lteEnbLl.north west)+(-1.5,0.2)$) rectangle ($(epcEnb.south east)+(0.5,-0.2)$);
  \node (legendChannel) [left of=enbPdcp, xshift=-2cm] {\textcolor{chapterlightgreen}{\texttt{LteEnbNetDevice}}};


  \draw[arrow] (epc) -- (rrc);
  \draw[arrow] (epc) -- (mmrrc);
  \draw[arrow] (epc) -- (pdcp);
  \draw[arrow] (mmrrc) --  (pdcp);
  \draw[arrow] (rrc) --  (pdcp);
  \draw[arrow] (pdcp) -- (rlc);
  \draw[arrow] (pdcp) -- (mmrlc);
  \draw[arrow] (rlc) -- (ltephy);
  \draw[arrow] (mmrlc) -- (mmphy);
  \draw[arrow] (mmphy) -- (mmphy2);
  \draw[arrow] (ltephy) -- (ltephy2);
  \draw[arrow] (ltephy2) -- (ltechannel);
  \draw[arrow] (mmphy2) -- (channel);
  \draw[arrow] (channel) -- (enbmmphy2);
  \draw[arrow] ([xshift=0.4cm]ltechannel.south west) -- node[anchor=east] {} ([xshift=-0.8cm]enbltephy2.west) -- (enbltephy2.west);
  \draw[arrow] (enbmmphy2) -- (mmEnbLl);
  \draw[arrow] (enbltephy2) -- (lteEnbLl);
  \draw[arrow] (mmEnbLl) -- (mmEnbRlc);
  \draw[arrow] (lteEnbLl) -- (lteEnbRlc);
  \draw[arrow] (lteEnbRlc) -- (enbPdcp);
  \draw[arrow] (mmEnbRlc) -- node[sloped, anchor=center, above] {X2 interface} (enbPdcp);
  \draw[arrow] (enbmmrrc) --  (enbPdcp);
  \draw[arrow] (enbrrc) --  (enbPdcp);
  \draw[arrow] (enbPdcp) -- (epcEnb);
  \draw[darrow] (rrc.west) -- node[anchor=east] {} ([xshift=-2.2cm]rrc.west) -- node[anchor=east] {} ([xshift=-2cm]enbrrc.west) -- (enbrrc.west);
  \draw[darrow] (mmrrc.east) -- node[anchor=east] {} ([xshift=+3cm]mmrrc.east) -- node[anchor=east] {} ([xshift=+1cm]enbmmrrc.east) -- (enbmmrrc.east);
  \draw[darrow] (epcEnb) -- node[anchor=west] {S1 interface} (cn);

\end{tikzpicture}
\caption{Block diagram of a multiconnected device, an LTE eNB and a mmWave eNB}
\label{fig:mcdevice}
\end{figure}

\subsection{The \texttt{McUeNetDevice} Class}\label{sec:mcue}

The core of the Dual Connectivity implementation is the \texttt{McUeNetDevice} class. It is a subclass of the ns--3 \texttt{NetDevice}, which is a basic class that abstracts network devices and provides an interface between the upper layers of the TCP/IP stack and custom lower layers. In particular, the LTE module extends this class with the \texttt{LteUeNetDevice} and the \texttt{LteEnbNetDevice}, the same is done in the mmWave module. The \texttt{NetDevice} holds pointers to the custom lower layer stack classes, and has a \texttt{Send} method that forwards packets to the TCP/IP stack. In both the LTE and mmWave module this method is linked to a callback on the \texttt{DoRecvData} of the \texttt{EpcUeNas} class, which as specified by the LTE standard acts as a connection between the LTE protocol stack and the TCP/IP stack. 

The \texttt{McUeNetDevice} represents a UE with a single \texttt{EpcUeNas}, but with a dual stack from this layer down, and a basic UML diagram can be seen in Fig.~\ref{fig:mcdevice}: there are mmWave PHY, MAC and RRC layers, and LTE PHY, MAC, RRC layers. The \texttt{EpcUeNas} layer has an interface to both RRC entities and is in charge of the exchange of information between them. 

This class can be used to simulate different dual connected modes, i.e., it can support both fast switching and throughput-oriented dual connectivity, according to which RRC and X2 procedures and primitives are implemented.

Each physical layer of the two stacks uses the respective mmWave or LTE channel model. Notice that since the two systems operate on different frequencies, the modeling of interference between the two RATs is not needed. Each of the two channel models can therefore be configured independently. 

In order to use a \texttt{McUeNetDevice} as a mobile terminal in the simulation, several features were added to the \textit{helper} class of the mmWave module: (i) the possibility to create an LTE channel; (ii) a method to install and configure LTE eNBs, so that they can be connected to the LTE stack of the \texttt{McUeNetDevice}; (iii) methods to set up a \texttt{McUeNetDevice} and connect its layers as shown in Fig~\ref{fig:mcdevice}.

\subsection{Dual Connected PDCP Layer}

As mentioned in Chapter~\ref{chap:ns3}, the PDCP layer implementation in ns--3 only performs basic functions. In order to support Dual Connectivity, the \texttt{LtePdcp} class was extended by the \texttt{McEnbPdcp} and \texttt{McUePdcp} classes, respectively at the eNB side and at the UE side.

The \texttt{McUePdcp} simply adds to the PDCP layer implementation a second interface to a lower RLC layer, by storing the \texttt{LteRlcSapProvider} interface offered by the mmWave RLC. Then, when packets have to be sent on the mmWave (LTE) RAT, the PDCP uses the mmWave (LTE) RLC interface. 

The implementation of \texttt{McEnbPdcp} and of the remote RLC on the mmWave eNB, instead, required new interfaces to the \texttt{EpcX2} class methods:
\begin{itemize}
  \item \texttt{EpcX2PdcpSapProvider} offers a \texttt{SendMcPdcpPdu} method that the PDCP can call to send a PDU to the remote RLC layer using the X2 interface;
  \item \texttt{EpcX2PdcpSapUser} offers a \texttt{ReceiveMcPdcpPdu} method that the \texttt{EpcX2} class can call to forward to the PDCP a packet received from the remote RLC;
  \item \texttt{EpcX2RlcSapProvider} has a \texttt{ReceiveMcRlcSdu} method that the specific RLC implementation calls to send a packet received from lower layers to the remote PDCP, via X2;
  \item \texttt{EpcX2RlcSapUser} has a \texttt{SendMcRlcSdu} method that is used by the class implementing the X2 interface to forward packets to the RLC layer in the mmWave eNB, once it has received them from the LTE eNB.
\end{itemize}

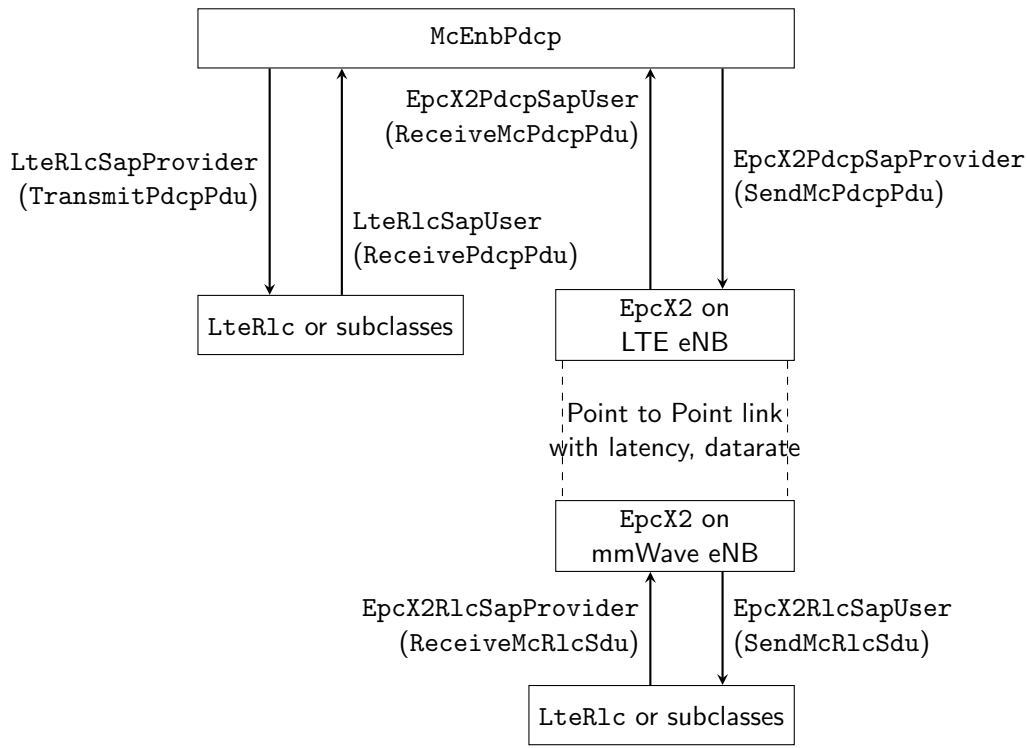
\begin{figure}[ht]
  \centering
  \begin{tikzpicture}[font=\sffamily\small, node distance=1.5cm, scale=0.95, every node/.style={scale=0.95}]
    \node [rectangle, minimum width=20em, minimum height=2em, draw=black, align=center] (pdcp) {\texttt{McEnbPdcp}};
    
    \node [rectangle, minimum width=8em, minimum height=2em, draw=black, anchor=east, align=center] at ($(pdcp.east) - (0, 4)$) (epcx2) {\texttt{EpcX2} on\\LTE eNB};

    \node [rectangle, minimum width=8em, minimum height=2em, draw=black, anchor=east, align=center, below of=epcx2, yshift=-3.5em] (epcx2rem) {\texttt{EpcX2} on\\mmWave eNB};

    \node [align=center, below of=epcx2] (label) {Point to Point link\\with latency, datarate};

    \node [rectangle, minimum width=8em, minimum height=2em, draw=black, anchor=west] at ($(pdcp.west) - (0, 4)$) (localRlc) {\texttt{LteRlc} or subclasses};

    \node [rectangle, minimum width=8em, minimum height=2em, draw=black, anchor=east] at ($(epcx2rem.east) - (0, 2.5)$) (remoteRlc) {\texttt{LteRlc} or subclasses};

    \draw [sarrow] ($(pdcp.south east) - (1, 0)$) -- node[anchor=west, align=left] {\texttt{EpcX2PdcpSapProvider}\\(\texttt{SendMcPdcpPdu})} ($(epcx2.north east) - (1, 0)$);

    \draw [sarrow] ($(epcx2.north east) - (2, 0)$) -- node[anchor=east, align=right,yshift=2em] {\texttt{EpcX2PdcpSapUser}\\(\texttt{ReceiveMcPdcpPdu})} ($(pdcp.south east) - (2, 0)$);

    \draw [sarrow] ($(pdcp.south west) + (1, 0)$) -- node[anchor=east, align=right] {\texttt{LteRlcSapProvider}\\(\texttt{TransmitPdcpPdu})} ($(localRlc.north west) + (1, 0)$);

    \draw [sarrow] ($(localRlc.north west) + (2, 0)$) -- node[anchor=west, align=left,yshift=-2em] {\texttt{LteRlcSapUser}\\(\texttt{ReceivePdcpPdu})} ($(pdcp.south west) + (2, 0)$);

    \draw [sarrow] ($(epcx2rem.south east) - (1, 0)$) -- node[anchor=west, align=left] {\texttt{EpcX2RlcSapUser}\\(\texttt{SendMcRlcSdu})} ($(remoteRlc.north east) - (1, 0)$);

    \draw [sarrow] ($(remoteRlc.north east) - (2, 0)$) -- node[anchor=east, align=right] {\texttt{EpcX2RlcSapProvider}\\(\texttt{ReceiveMcRlcSdu})} ($(epcx2rem.south east) - (2, 0)$);

    \draw [dashed] ($(epcx2.south west) + (0.1, 0)$) -- ($(epcx2rem.north west) + (0.1, 0)$);
    \draw [dashed] ($(epcx2.south east) - (0.1, 0)$) -- ($(epcx2rem.north east) - (0.1, 0)$);

  \end{tikzpicture}
  \caption{Relations between PDCP, X2 and RLC}
  \label{fig:pdcp}
\end{figure}

A basic diagram of interfaces and methods is shown in Fig.~\ref{fig:pdcp}. The \texttt{McEnbPdcp} and the instance of the remote RLC store the LTE eNB and the mmWave eNB cell IDs, and the GTP tunneling identity for the transmission on the X2 interface.

\subsection{RRC Layer}

In this section, the features that were added to the RRC layer will be described. 

Differently from what proposed in~\cite{36842} for the Dual Connectivity option 3C, both eNBs have an RRC layer. At the same time, the UE has an RRC layer for both protocol stacks. This allows to achieve more flexibility, since new features may be added to the RRC protocol for mmWave RATs. 
Moreover, the LTE RRC is used for the management of the LTE connection and to exchange commands related to dual connectivity, while the mmWave RRC is used to manage only mmWave-related communications, and the collection of measurements in the mmWave eNB and the reporting to the coordinator. In order to minimize the communications on the mmWave RRC link, the LTE RRC is in charge also of the handling of the RLC and PDCP entities for both RATs.
By using a dedicated RRC link, the secondary eNB avoids to encode and transmit the control PDU to the master cell, therefore the latency of control commands is reduced. The mmWave signalling radio bearers are used only when a connection to LTE is already established, and this can offer a ready backup in case the mmWave link suffers an outage. 
Finally, only the LTE eNB (or coordinator) has to interact with the core network when dealing with Dual Connected devices, acting as a data plane and mobility anchor for the UE. This allows to reduce the number of connections needed to the MME, but increases the computational and networking load of the LTE eNB (or coordinator).

In the implementation of this Thesis, the coordinator is placed in the LTE eNB and therefore the Master Cell RRC layer is extended in order to support the new coordinator functionalities.

\subsubsection{Collection of measurements from mmWave eNBs}
A new X2 primitive is added for the transmission of reports from secondary eNBs to the master cells. Each report is organized as in Fig.~\ref{fig:report}. Notice that, since the mmWave NYU beamforming always assumes that there is a perfect alignment between the UE and the mmWave eNB, and that directionality measurements are not implemented yet, the SINR that is reported is the one corresponding to the optimal direction, and not an SINR for each eNB direction. Moreover, the variance is not yet accounted for and is left as an extension of the framework, since the evaluation of complex handover algorithms is out of the scope of this research. In the Report Table, each UE is identified by the International Mobile Subscriber Identity (IMSI)\footnote{The IMSI is a unique identifier of a UE in a mobile network. Each UE connected to an eNB has also another identifier, the C-RNTI (Cell Radio Network Temporary Identifier), which is assigned on a cell basis.}.

When the LTE RRC receives a new Report Table it updates the relative entries in the CRT. An example of CRT is shown in Fig.~\ref{fig:crt}. Notice that this CRT is simpler than the one in~\cite{giordani}, since variance and directionality are not accounted for.

\begin{figure}[h]
  \centering
  \sffamily
  \begin{tabular}{|c|c|}
    \hline
    \multicolumn{2}{|c|}{RT for mmWave eNB $i$} \\ \hline
    UE \texttt{imsi} 1 & SINR $\Gamma_{i,1}$ \\
    UE \texttt{imsi} 2 & SINR $\Gamma_{i,2}$ \\
    \multirow{2}{*}{$\vdots$} & \multirow{2}{*}{$\vdots$} \\
    & \\
    UE \texttt{imsi} $N$ & SINR $\Gamma_{i,N}$ \\
    \hline 
  \end{tabular}
  \caption{Report Table for mmWave eNB $i$. There is an entry for each UE, each entry is a pair with the UE IMSI and the SINR $\Gamma$ measured in the best direction between the eNB and the UE}
  \label{fig:report}
\end{figure}

\begin{figure}[h]
  \centering
  \sffamily
  \begin{tabular}{|c|c|c|c|c|c|}
    \hline
    & eNB 1 & eNB 2 & \multicolumn{2}{c|}{$\dots$} & eNB $M$ \\ \hline
    UE \texttt{imsi} 1 & SINR $\Gamma_{1,1}$ & SINR $\Gamma_{2,1}$ & \multicolumn{2}{c|}{$\dots$} & SINR $\Gamma_{N,1}$ \\
    UE \texttt{imsi} 2 & SINR $\Gamma_{1,2}$ & SINR $\Gamma_{2,2}$ & \multicolumn{2}{c|}{$\dots$} & SINR $\Gamma_{N,2}$ \\
    \multirow{2}{*}{$\vdots$} & \multirow{2}{*}{$\vdots$} & \multirow{2}{*}{$\vdots$} & \multicolumn{2}{c|}{$\dots$} & \multirow{2}{*}{$\vdots$} \\
    & & & \multicolumn{2}{c|}{$\dots$} & \\
    UE \texttt{imsi} $N$ & SINR $\Gamma_{1,N}$ & SINR $\Gamma_{2,N}$ & \multicolumn{2}{c|}{$\dots$} & SINR $\Gamma_{N,N}$ \\
    \hline 
  \end{tabular}
  \caption{Complete Report Table available at the LTE eNB (or coordinator). There is an entry for each UE in each mmWave eNB, each entry is a pair with the UE IMSI and the SINR $\Gamma$ measured in the best direction between the eNB and the UE}
  \label{fig:crt}
\end{figure}

\begin{algorithm}[t]
  \sffamily
  \setstretch{1.2}
  \caption{UE Association Algorithm for Dual Connected UEs with fast switching}\label{alg:switch}
  \begin{algorithmic}[1]
    \State The LTE eNB has an updated CRT.
    \For {All the UEs connected to the LTE eNB}
      \State Consider the UE with \texttt{imsi} $i$, currently attached to mmWave \texttt{curreNB} $j$. 

      \State Find the mmWave eNB $\epsilon$ with the best SINR $\Gamma_{opt}$.
      \State Let $\Gamma_{curr}$ be the SINR in the current mmWave eNB $j$.

      \State  \parbox[t]{\dimexpr\linewidth-\algorithmicindent}{Let $\Delta = \Gamma_{opt} - \Gamma_{curr}$ be the gain in SINR that can be obtained by changing the mmWave eNB.\strut}

      \If{$\Gamma_{opt} < \Delta_{LTE}$}
        \State \parbox[t]{\dimexpr\linewidth-\algorithmicindent}{mmWave cell SINR too low, switch to the LTE stack.\strut}
      \ElsIf{the UE is not already performing a handover}
        \If{\parbox[t]{\dimexpr\linewidth-\algorithmicindent}{$ \epsilon \ne j$ and $\Delta > \Delta_{hys}$ and the UE is using the mmWave stack.\strut}}
          \State \parbox[t]{\dimexpr\linewidth-\algorithmicindent}{Switch to the LTE stack.\strut}
          \State \parbox[t]{\dimexpr\linewidth-\algorithmicindent}{Trigger a secondary cell handover with \texttt{SendMcHandoverRequest} primitive.\strut}
        \ElsIf{\parbox[t]{\dimexpr\linewidth-\algorithmicindent}{$ \epsilon \ne j$ and $\Delta > \Delta_{hys}$ and the UE is using the LTE stack\strut}}
          \State \parbox[t]{\dimexpr\linewidth-\algorithmicindent}{Trigger a secondary cell handover with \texttt{SendMcHandoverRequest} primitive.\strut}
        \ElsIf{\parbox[t]{\dimexpr\linewidth-\algorithmicindent}{$ \epsilon == j $ and $\Gamma_{opt} > \Delta_{LTE} + \Delta_{hys}$ and the UE is using the LTE stack\strut}}
          \State \parbox[t]{\dimexpr\linewidth-\algorithmicindent}{Switch to the mmWave stack.\strut}
        \EndIf
      \EndIf
    \EndFor
  \end{algorithmic}
\end{algorithm}

\begin{algorithm}[t]
  \sffamily
  \setstretch{1.2}
  \caption{UE Association Algorithm for Hard Handover with coordinator}\label{alg:hh}
  \begin{algorithmic}[1]
    \State The LTE eNB has an updated CRT.
    \For {All the UEs connected to the LTE eNB}
      \State Consider the UE with \texttt{imsi} $i$. 

      \State Find the mmWave eNB $\epsilon$ with the best SINR $\Gamma_{opt}$.
      \If{The UE is attached to a mmWave eNB}
        \State \parbox[t]{\dimexpr\linewidth-\algorithmicindent}{Let $\Gamma_{curr}$ be the SINR in the current mmWave eNB $j$.}
      \Else
        \State $\Gamma_{curr} = -\inf$
      \EndIf

      \State  \parbox[t]{\dimexpr\linewidth-\algorithmicindent}{Let $\Delta = \Gamma_{opt} - \Gamma_{curr}$ be the gain in SINR that can be obtained by performing handover to mmWave eNB $\epsilon$.\strut}

      \If{$\Gamma_{opt} < \Delta_{LTE}$}
        \State \parbox[t]{\dimexpr\linewidth-\algorithmicindent}{mmWave cell SINR too low, perform a handover to the LTE cell.\strut}
      \ElsIf{\parbox[t]{\dimexpr\linewidth-\algorithmicindent}{The UE is not already performing a handover and is attached to a mmWave eNB}}
        \If{\parbox[t]{\dimexpr\linewidth-\algorithmicindent}{$ \epsilon \ne j$ and $\Delta > \Delta_{hys}$\strut}}
          \State \parbox[t]{\dimexpr\linewidth-\algorithmicindent}{Trigger a mmWave cell handover.\strut}
        \EndIf
      \ElsIf{\parbox[t]{\dimexpr\linewidth-\algorithmicindent}{the UE is not already performing a handover and is attached to the LTE eNB and ${\Gamma_{opt} > \Delta_{LTE} + \Delta_{hys}}$}}    
        \State \parbox[t]{\dimexpr\linewidth-\algorithmicindent}{Perform handover to mmWave eNB $\epsilon$.\strut}
      \EndIf
    \EndFor
  \end{algorithmic}
\end{algorithm}

\subsubsection{Selection of the best mmWave eNB for each UE}\label{sec:selection}

Once all the entries in the CRT are updated, the LTE eNB selects which is the best mmWave cell for each UE. In order to support fast switching, it also detects when a switch to LTE is needed, i.e., when the SINR of the mmWave link is too low to be used for data transmission. The same criterion is used to trigger the handover from mmWave to LTE in the hard handover baseline scenario. The algorithm that selects if a handover or switch is necessary is described by the pseudocode in Alg.~\ref{alg:switch} for the fast switching case and in Alg.~\ref{alg:hh} for the handover scenario. The algorithm to trigger the handover is presented here in order to highlight the fact that the criteria for switching and handover are the same, but the actual practicalities of the handover implementation will be given in Sec~\ref{sec:hhImpl}.

The parameters of the algorithm are the SINR threshold $\Delta_{LTE}$ under which a UE switches or handovers to LTE (e.g., because there may be an outage), which is set to $-5$~dB, and a hysteresis of $\Delta_{hys}$ that avoids to perform handover when the SINR difference is too small, in order to prevent frequent handovers due to small-scale fading. 

Notice that the handover or switch rules are quite simple, but the algorithm is designed in such a way that they can be easily replaced by a more sophisticated handover or cell selection algorithm.

\subsubsection{New Data Structures}
In order to maintain a dual connection, some changes have to be made to the RRC layer of the Master eNB, of the Secondary eNB and of the UE, and some X2 and RRC protocol commands have to be added. These changes embrace new data structures, and new signalling procedures. 

The \texttt{LteEnbRrc} has a private variable that maps \texttt{UeManager} instances to each UE C-RNTI. Each \texttt{UeManager} object represents a UE which is known to the eNB, and stores
\begin{itemize}
  \item the UE RRC state;
  \item a map of \texttt{LteDataRadioBearerInfo} objects, that abstract the information associated with EPS Data Radio Bearers (and contain also pointers to RLC and PDCP entities), as shown in Fig.~\ref{fig:bearer};
  \item a pointer to \texttt{LteSignalingRadioBearerInfo} for SRB0 and SRB1;
  \item \texttt{imsi}, \texttt{rnti} and X2 configurations;
\end{itemize}
The same informations are present in the \texttt{LteUeRrc} class, along with the \textit{CellId} of the current cell. 

In order to support dual connectivity, a \texttt{UeManager} knows whether the UE is a dual connected device or not, and whether the \texttt{LteEnbRrc} class to which it belongs is the LTE or the mmWave one, thus if it is hosted in a coordinator, or in a remote eNB where, for each bearer, only the RLC entity must be managed. In particular, in order to create and manage remote RLCs, a new \texttt{RlcBearerInfo} class is introduced. It is the equivalent of the \texttt{LteDataRadioBearerInfo} class but for remote eNBs, and, as can be seen in Fig.~\ref{fig:bearer}, it stores a pointer to the RLC entity but not to the PDCP one, since it is not needed in the remote eNB. A \texttt{UeManager} of a Dual Connected device that is stored in a mmWave eNB therefore contains also a map of \texttt{RlcBearerInfo} objects. 

\begin{figure}
  \centering
  \sffamily
  \begin{tabular}{|c|c|}
    \hline
    \multicolumn{2}{|c|}{\textbf{Bearer classes}} \\ \hline
    \texttt{LteDataRadioBearerInfo} & \texttt{RlcBearerInfo} \\ \hline
    \multicolumn{2}{|c|}{Bearer ID} \\ \hline
    \multicolumn{2}{|c|}{Type of RLC} \\ \hline
    \multicolumn{2}{|c|}{Logical Channel Identity and Configuration} \\ \hline
    \multicolumn{2}{|c|}{GTP tunneling ID for X2 and S1} \\ \hline
    \texttt{bool} splitBearer & CellId of the LTE cell \\ \hline
    RLC entity & RLC remote entity \\ \hline
    PDCP entity & \\ \hline
    
  \end{tabular}
  \caption{Information of \texttt{LteDataRadioBearerInfo} and \texttt{RlcBearerInfo} classes}
  \label{fig:bearer}
\end{figure}

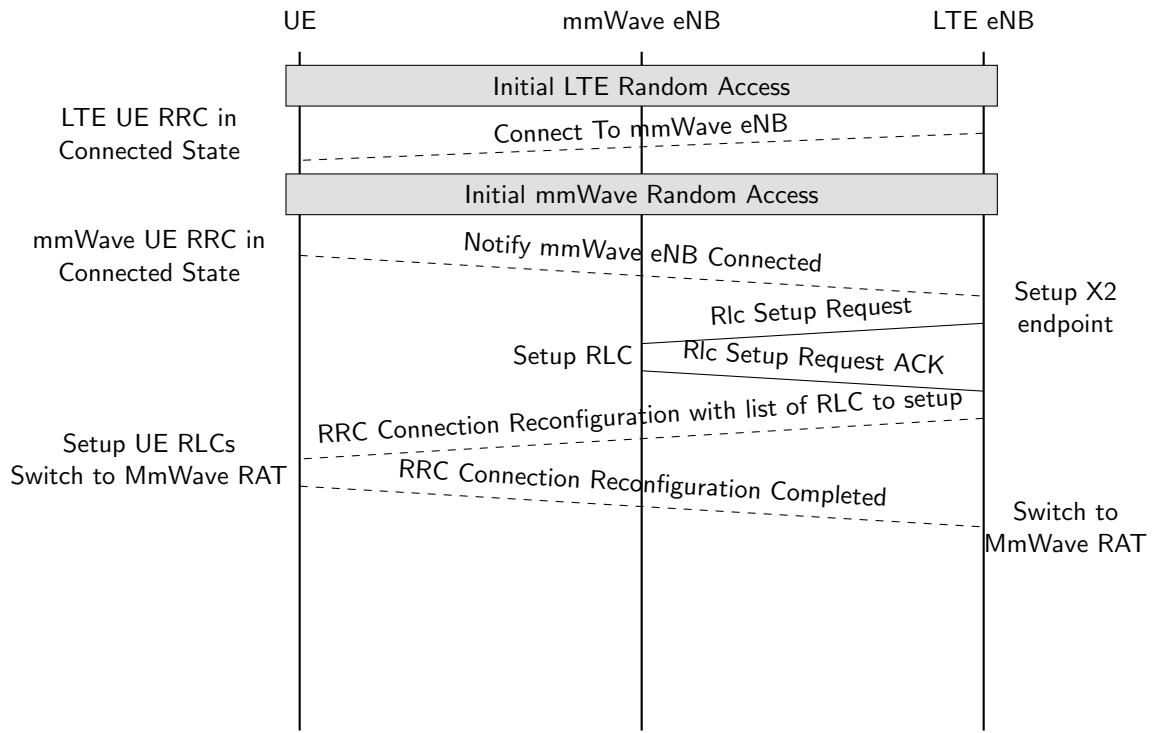
\begin{figure}[h]
\centering
  \begin{tikzpicture}[font=\sffamily\small, scale=0.9, every node/.style={scale=0.9}]
    \coordinate (UeBegin) at (0,10);
    \coordinate (UeEnd) at (0,0);
    \coordinate (mmWaveEnbBegin) at (5,10);
    \coordinate (mmWaveEnbEnd) at (5,0);
    \coordinate (lteEnbBegin) at (10,10);
    \coordinate (lteEnbEnd) at (10,0);
    
    \node [above of=UeBegin, yshift=-15pt] (UElabel) {UE};
    \node [above of=mmWaveEnbBegin,yshift=-15pt] (mmWaveEnbLabel) {mmWave eNB};
    \node [above of=lteEnbBegin,yshift=-15pt] (lteEnbLabel) {LTE eNB};

    \begin{scope}[on background layer]
      \draw [thick, black] (UeBegin) -- (UeEnd);
      \draw [thick, black] (mmWaveEnbBegin) -- (mmWaveEnbEnd);
      \draw [thick, black] (lteEnbBegin) -- (lteEnbEnd);
    \end{scope}

    \filldraw[fill=midgrey!20] (-0.2,9.8) rectangle (10.2,9.2);
    \node (initialLte) at (5,9.5) {Initial LTE Random Access};

    \node [align=center] (lteRrcConnected) at (-2.2,8.8) {LTE UE RRC in\\Connected State};
    \draw [dashed] (10,8.8) -- node[sloped, anchor=center, above] {Connect To mmWave eNB} (0, 8.4);

    \filldraw[fill=midgrey!20] (-0.2,8.2) rectangle (10.2,7.6);
    \node (initialLte) at (5,7.9) {Initial mmWave Random Access};
    
    \node [align=center] (mmRrcConnected) at (-2.2,7) {mmWave UE RRC in\\Connected State};
    \draw [dashed] (0,7) -- node[sloped, anchor=center, above] {Notify mmWave eNB Connected} (10, 6.4);

    \node [align=center] (setupRlcReq) at (11.2, 6.2) {Setup X2\\endpoint};
    \draw (10, 6.0) -- node[sloped, anchor=center, above] {Rlc Setup Request} (5, 5.7);

    \node [align=center] (setupRlc) at (4, 5.5) {Setup RLC};
    \draw (5, 5.3) -- node[sloped, anchor=center, above] {Rlc Setup Request ACK} (10, 5.0);

    \draw [dashed] (10,4.6) -- node[sloped, anchor=center, above] {RRC Connection Reconfiguration with list of RLC to setup} (0, 4.0);
    
    \node [align=center] (setupUeRlc) at (-2.2,4) {Setup UE RLCs\\Switch to MmWave RAT};
    \draw [dashed] (0,3.6) -- node[sloped, anchor=center, above] {RRC Connection Reconfiguration Completed} (10, 3.0);
    \node [align=center] (setupUeRlc) at (11.2, 3.0) {Switch to\\MmWave RAT};


  \end{tikzpicture}
  \caption{Initial Access for Dual Connected devices and mmWave RLC setup. Dashed lines are RRC messages, solid lines are X2 messages}
  \label{fig:ia}
\end{figure}

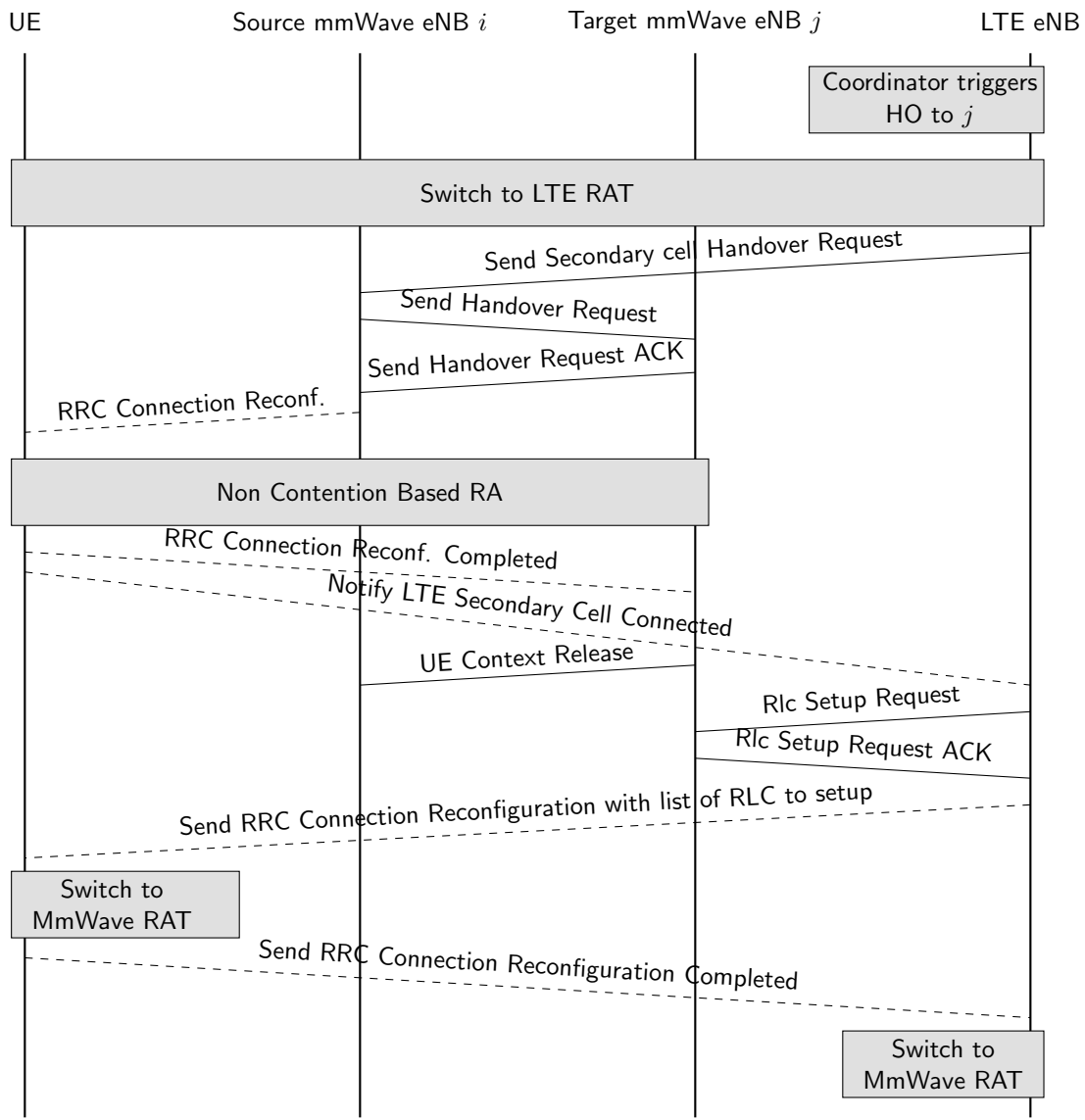
\begin{figure}[ht!]
\centering
  \begin{tikzpicture}[font=\sffamily\small, scale=0.9, every node/.style={scale=0.9}]
    \coordinate (UeBegin) at (0,10);
    \coordinate (UeEnd) at (0,-6);
    \coordinate (mmWaveEnb1Begin) at (5,10);
    \coordinate (mmWaveEnb1End) at (5,-6);
    \coordinate (mmWaveEnb2Begin) at (10,10);
    \coordinate (mmWaveEnb2End) at (10,-6);
    \coordinate (lteEnbBegin) at (15,10);
    \coordinate (lteEnbEnd) at (15,-6);
    
    \node [above of=UeBegin, yshift=-15pt] (UElabel) {UE};
    \node [above of=mmWaveEnb1Begin,yshift=-15pt] (mmWaveEnbLabel) {Source mmWave eNB $i$};
    \node [above of=mmWaveEnb2Begin,yshift=-15pt] (mmWaveEnbLabel) {Target mmWave eNB $j$};
    \node [above of=lteEnbBegin,yshift=-15pt] (lteEnbLabel) {LTE eNB};

    \begin{scope}[on background layer]
      \draw [thick, black] (UeBegin) -- (UeEnd);
      \draw [thick, black] (mmWaveEnb1Begin) -- (mmWaveEnb1End);
      \draw [thick, black] (mmWaveEnb2Begin) -- (mmWaveEnb2End);
      \draw [thick, black] (lteEnbBegin) -- (lteEnbEnd);
    \end{scope}

    \filldraw[fill=midgrey!20] (11.7,9.8) rectangle (15.2,8.8);
    \node [align=center] (triggerHO) at (13.5,9.3) {Coordinator triggers\\HO to $j$};

    \filldraw[fill=midgrey!20] (-0.2,8.4) rectangle (15.2,7.4);
    \node (switchToLte) at (7.5,7.9) {Switch to LTE RAT};

    \draw (15, 7.0) -- node[sloped, anchor=center, above] {Send Secondary cell Handover Request} (5, 6.4);
    \draw (5, 6.0) -- node[sloped, anchor=center, above] {Send Handover Request} (10, 5.7);
    \draw (10, 5.2) -- node[sloped, anchor=center, above] {Send Handover Request ACK} (5, 4.9);
    \draw [dashed] (5, 4.6) -- node[sloped, anchor=center, above] {RRC Connection Reconf.} (0, 4.3);

    \filldraw[fill=midgrey!20] (-0.2,3.9) rectangle (10.2,2.9);
    \node (initialLte) at (5,3.4) {Non Contention Based RA};

    \draw [dashed] (0, 2.5) -- node[sloped, anchor=center, above] {RRC Connection Reconf. Completed} (10, 1.9);
    \draw [dashed] (0, 2.2) -- node[sloped, anchor=center, above] {Notify LTE Secondary Cell Connected} (15, 0.5);
    \draw (10, 0.8) -- node[sloped, anchor=center, above] {UE Context Release} (5, 0.5);

    \draw (15, 0.1) -- node[sloped, anchor=center, above] {Rlc Setup Request} (10, -0.2);

    \draw (10, -0.6) -- node[sloped, anchor=center, above] {Rlc Setup Request ACK} (15, -0.9);

    \draw [dashed] (15,-1.3) -- node[sloped, anchor=center, above] {Send RRC Connection Reconfiguration with list of RLC to setup} (0, -2.1);
    
    \filldraw[fill=midgrey!20] (-0.2,-2.3) rectangle (3.2,-3.3);
    \node [align=center] (initialLte) at (1.3,-2.8) {Switch to\\MmWave RAT};
    \draw [dashed] (0,-3.6) -- node[sloped, anchor=center, above] {Send RRC Connection Reconfiguration Completed} (15, -4.5);
    \filldraw[fill=midgrey!20] (12.2,-4.7) rectangle (15.2,-5.7);
    \node [align=center] (triggerHO) at (13.7,-5.2) {Switch to\\MmWave RAT};


  \end{tikzpicture}
  \caption{Secondary cell Handover}
  \label{fig:sho}
\end{figure}

\begin{figure}[h!]
\begin{subfigure}[t]{\textwidth}
\centering
  \begin{tikzpicture}[font=\sffamily\small, scale=0.9, every node/.style={scale=0.9}]
    \coordinate (UeBegin) at (0,10);
    \coordinate (UeEnd) at (0,4.2);
    \coordinate (mmWaveEnbBegin) at (5,10);
    \coordinate (mmWaveEnbEnd) at (5,4.2);
    \coordinate (lteEnbBegin) at (10,10);
    \coordinate (lteEnbEnd) at (10,4.2);
    
    \node [above of=UeBegin, yshift=-15pt] (UElabel) {UE};
    \node [above of=mmWaveEnbBegin,yshift=-15pt] (mmWaveEnbLabel) {mmWave eNB};
    \node [above of=lteEnbBegin,yshift=-15pt] (lteEnbLabel) {LTE eNB};

    \begin{scope}[on background layer]
      \draw [thick, black] (UeBegin) -- (UeEnd);
      \draw [thick, black] (mmWaveEnbBegin) -- (mmWaveEnbEnd);
      \draw [thick, black] (lteEnbBegin) -- (lteEnbEnd);
    \end{scope}

    \filldraw[fill=midgrey!20] (6.7,9.8) rectangle (10.2,8.8);
    \node [align=center] (triggerSwitch) at (8.5,9.3) {Coordinator triggers\\switch to mMWave};

    \filldraw[fill=midgrey!20] (9.8,8.4) rectangle (13.8,7.4);
    \node [align=center] (enbSwitch) at (11.7,7.9) {Switch command to\\\texttt{McEnbPdcp}};
    \draw (10, 7.2) -- node[sloped, anchor=center, above] {Forward RLC buffer content} (5, 6.9);

    \draw [dashed] (10,7.0) -- node[sloped, anchor=center, above] {Send RRC Connection Switch$\quad\quad\quad\quad\quad\quad\quad\quad\quad\quad\quad\quad\quad$} (0, 5.8);
    \filldraw[fill=midgrey!20] (-0.2,5.4) rectangle (3.4,4.4);
    \node [align=center] (ueSwitch) at (1.5,4.9) {Switch command to\\\texttt{McUePdcp}};

  \end{tikzpicture}
  \caption{Switch from LTE RAT to mmWave RAT}
  \label{fig:lteToMm}
\end{subfigure}
\begin{subfigure}[t]{\textwidth}
\centering
  \begin{tikzpicture}[font=\sffamily\small, scale=0.9, every node/.style={scale=0.9}]
    \coordinate (UeBegin) at (0,10);
    \coordinate (UeEnd) at (0,4.2);
    \coordinate (mmWaveEnbBegin) at (5,10);
    \coordinate (mmWaveEnbEnd) at (5,4.2);
    \coordinate (lteEnbBegin) at (10,10);
    \coordinate (lteEnbEnd) at (10,4.2);
    
    \node [above of=UeBegin, yshift=-15pt] (UElabel) {UE};
    \node [above of=mmWaveEnbBegin,yshift=-15pt] (mmWaveEnbLabel) {mmWave eNB};
    \node [above of=lteEnbBegin,yshift=-15pt] (lteEnbLabel) {LTE eNB};

    \begin{scope}[on background layer]
      \draw [thick, black] (UeBegin) -- (UeEnd);
      \draw [thick, black] (mmWaveEnbBegin) -- (mmWaveEnbEnd);
      \draw [thick, black] (lteEnbBegin) -- (lteEnbEnd);
    \end{scope}

    \filldraw[fill=midgrey!20] (6.7,9.8) rectangle (10.2,8.8);
    \node [align=center] (triggerSwitch) at (8.5,9.3) {Coordinator triggers\\switch to LTE};

    \filldraw[fill=midgrey!20] (9.8,8.4) rectangle (13.8,7.4);
    \node [align=center] (enbSwitch) at (11.7,7.9) {Switch command to\\\texttt{McEnbPdcp}};
    \draw (10, 7.2) -- node[sloped, anchor=center, above] {Send Switch to LTE} (5, 6.9);
    \draw [dashed] (10,7.0) -- node[sloped, anchor=west, above] {Send RRC Connection Switch$\quad\quad\quad\quad\quad\quad\quad\quad\quad\quad\quad\quad\quad$} (0, 6.4);

    \draw (5, 6.5) -- node[sloped, anchor=center, below] {Forward RLC buffers} (10, 6.2);

    \filldraw[fill=midgrey!20] (-0.2,5.4) rectangle (3.4,4.4);
    \node [align=center] (ueSwitch) at (1.5,4.9) {Switch command to\\\texttt{McUePdcp}};

  \end{tikzpicture}
  \caption{Switch from mmWave RAT to LTE RAT}
  \label{fig:mmToLte}
\end{subfigure}
\caption{Switch RAT procedures}
\label{fig:switch}
\end{figure}
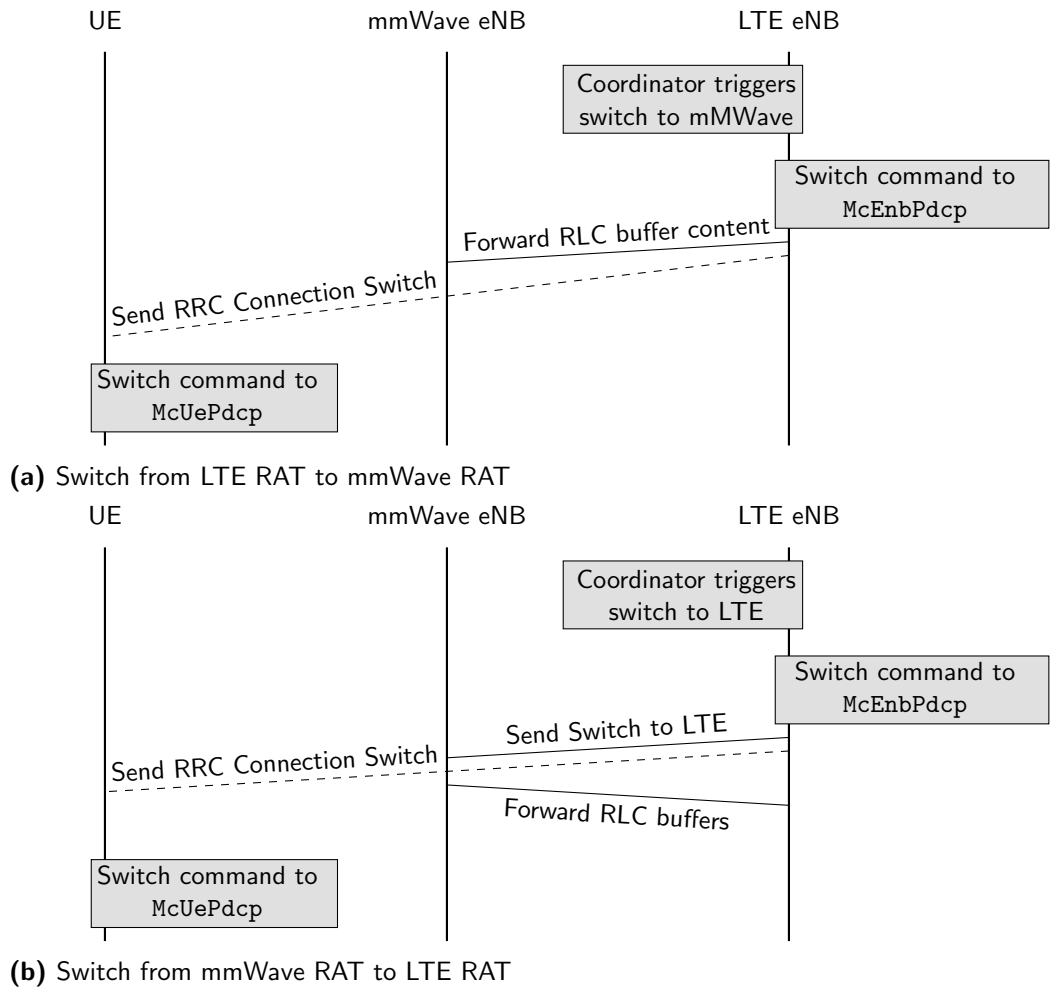

Another structure that is added to the \texttt{LteEnbRrc} is a map of X2 endpoints for Dual Connected devices, so that when a packet is received on the X2 interface it is forwarded to the correct RLC, PDCP or \texttt{UeManager}. 

Finally, the other data structures added to the \texttt{LteEnbRrc} are those needed to manage Report Tables and Complete Report Tables. They are maps that for each Dual Connected UE known to the LTE eNB store the following information by \texttt{imsi}:
\begin{itemize}
  \item the current mmWave eNB to which the UE is connected;
  \item the best mmWave eNB for a certain UE;
  \item whether the UE is using the LTE or the mmWave stack;
  \item a map of SINR of the UE in each mmWave eNB.
\end{itemize}

\subsubsection{New Signalling Procedures}
The new signalling procedures involve (i) the initial Dual Connected access; (ii) the handover for Secondary cells; (iii) the switch from the two RATs. 

In~\cite{giordani} there is a proposal for the IA for Dual Connected devices, however it does not deal with the setup of remote bearers. Therefore the IA procedure is extended in order to complete the setup of the remote RLC layer in the mmWave eNB, and of the associated RLC in the UE, and to end the IA with a switch to the mmWave RAT. The complete procedure is shown in Fig.~\ref{fig:ia}.

Fig.~\ref{fig:sho}, instead, shows the procedure for the handover of mmWave secondary cells. Firstly, the LTE RAT becomes the one in use, in order to ensure service continuity during the secondary handover. Then a handover between the two mmWave cells takes place. Notice that, since mmWave eNBs do not have Radio Bearers to set up, the Handover Request and the RRC Connection Reconfiguration PDUs have a smaller size than those used when a classic handover takes place. Once this procedure is completed, the LTE RRC is used to signal to the LTE eNB that a secondary mmWave eNB is available, so that the LTE eNB triggers the procedure to set up remote RLCs. Another difference with a classic handover is that the core network is not involved, i.e., the path switch message that a classic handover procedure has to send to the MME is not needed, since the bearer has a single endpoint in the LTE eNB. 

Finally, Fig.~\ref{fig:switch} shows the messages that are exchanged for the switch from LTE to mmWave and vice versa. 

In order to support these procedures, new X2 and RRC PDUs are defined, as well as headers, whose Information Elements are encoded following the LTE standard indications.

\section{Implementation of Hard Handover}\label{sec:hhImpl}
The X2-based handover is already implemented in the ns--3 LTE module, but it was not supported by the NYU mmWave module. Therefore the code of the LTE classes was integrated in the mmWave ones, in particular the main features that were ported are:
\begin{itemize}
  \item X2 links between mmWave eNBs, by extending the \textit{helper} of the mmWave module;
  \item Non Contention Based Random Access, that is implemented in the MAC and RRC LTE classes but was missing in the mmWave MAC.
\end{itemize}
Then also X2 links between mmWave eNBs and LTE eNBs are added. Notice that the measurement framework detailed in Sec.~\ref{sec:meas} is applied also to the handover scenario, and the handover rules are presented in Sec.~\ref{sec:selection}. The usage of different reference signals (i.e., downlink based) and of the relative handover rules that can be used are left as a future implementation. Therefore, also the handover assumes a light form of coordination, and it is modeled as an intra RAT handover, and not as an inter RAT handover. This choice was made in order to focus the comparison on the actual difference between a handover procedure and a fast switch procedure, given the same switching/handover rules and set of measurements available. 

The \texttt{McUeNetDevice} class is the basis also for devices capable of handover between LTE and mmWave eNBs. However, with respect to the dual connectivity case, a single RRC is used in the UE. This layer has interfaces to both LTE and mmWave PHY and MAC classes, but only one of the two stacks is used at a time. When the UE RRC receives a handover command to a certain target eNB, it checks whether it is an LTE or a mmWave cell. If, for example, it is connected to a mmWave cell (and thus uses the interfaces to the mmWave PHY and MAC layers) and receives a handover command to an LTE cell, it resets the mmWave PHY and MAC layers, swaps the interfaces, and starts using the LTE stack. This allows to simulate a handover among RATs while keeping a single RRC with a consistent state with respect to the eNBs to which it is connected and to the MME that controls its mobility in the core network.

\subsection{Lossless Handover and RLC Buffer Forwarding}
A feature that was missing from the LTE ns--3 module is the lossless handover, that was described in Sec.~\ref{sec:llho}. An attempt to implement it was made in~\cite{llho}. The main challenge is to reconstruct the segmented packets in the RLC AM retransmission and transmitted buffers, in order to forward to the target eNB the original PDCP SDUs. The solution adopted in~\cite{llho} is to re-use the reassembly algorithm provided by the RLC AM class. The first step is to merge the content of transmitted and retransmission buffers, by using header sequence numbers. Then the new merged buffer is given to the reassembly algorithm. Finally, the PDCP SDUs are forwarded to the target eNB before any other incoming packet is forwarded. 

The implementation of lossless handover of~\cite{llho} was ported to the mmWave module, and adapted to be used for handovers between LTE and 5G. Moreover, the same methods are used when performing fast switching.

\begin{figure}[h!]
  \centering
  \begin{tikzpicture}[font=\sffamily\small, node distance=1.5cm, scale=0.9, every node/.style={scale=0.9}]
    \node [rectangle, minimum width=9em, minimum height=2em, draw=black, align=center] (mmeApp) {\texttt{EpcMmeApplication}};
    
    \node [rectangle, minimum width=15em, minimum height=3em, draw=black, anchor=north west, align=right] at ($(mmeApp.north west) - (0.2, -0.2)$) (mmeNode) {};

    \node [left, align=center] at (mmeNode.east) (label) {MmeNode in\\core network};

    \node [rectangle, minimum width=9em, minimum height=2em, draw=black, anchor=east, align=center, below of=mmeApp, yshift=-2em] (epcs1mme) {\texttt{EpcS1Mme} at\\MME side};

    \node [align=center, below of=epcs1mme] (label) {Point to Point link\\with latency, datarate};

    \node [rectangle, minimum width=9em, minimum height=2em, draw=black, anchor=east, align=center, below of=epcs1mme, yshift=-3.5em] (epcs1enb) {\texttt{EpcS1Enb} at\\eNB side};

    \node [rectangle, minimum width=9em, minimum height=2em, draw=black, align=center, below of=epcs1enb, yshift=-2em] (enbApp) {\texttt{EpcEnbApplication}};

    \node [rectangle, minimum width=15em, minimum height=3em, draw=black, anchor=north west, align=right] at ($(enbApp.north west) - (0.2, -0.2)$) (enbNode) {};

    \node [left, align=center] at (enbNode.east) (label) {eNB node};

    \draw [sarrow] ($(mmeApp.south east) - (1, 0)$) -- node[anchor=west, align=left] {\texttt{EpcS1apSapMmeProvider}} ($(epcs1mme.north east) - (1, 0)$);

    \draw [sarrow] ($(epcs1mme.north west) + (1, 0)$) -- node[anchor=east, align=right] {\texttt{EpcS1apSapMme}} ($(mmeApp.south west) + (1, 0)$);

    \draw [dashed] ($(epcs1mme.south west) + (0.1, 0)$) -- ($(epcs1enb.north west) + (0.1, 0)$);
    \draw [dashed] ($(epcs1mme.south east) - (0.1, 0)$) -- ($(epcs1enb.north east) - (0.1, 0)$);

    \draw [sarrow] ($(epcs1enb.south east) - (1, 0)$) -- node[anchor=west, align=left] {\texttt{EpcS1apSapEnb}} ($(enbApp.north east) - (1, 0)$);

    \draw [sarrow] ($(enbApp.north west) + (1, 0)$) -- node[anchor=east, align=right] {\texttt{EpcS1apSapEnbProvider}} ($(epcs1enb.south west) + (1, 0)$);

  \end{tikzpicture}
  \caption{Relations between MME and eNB}
  \label{fig:mme}
\end{figure}
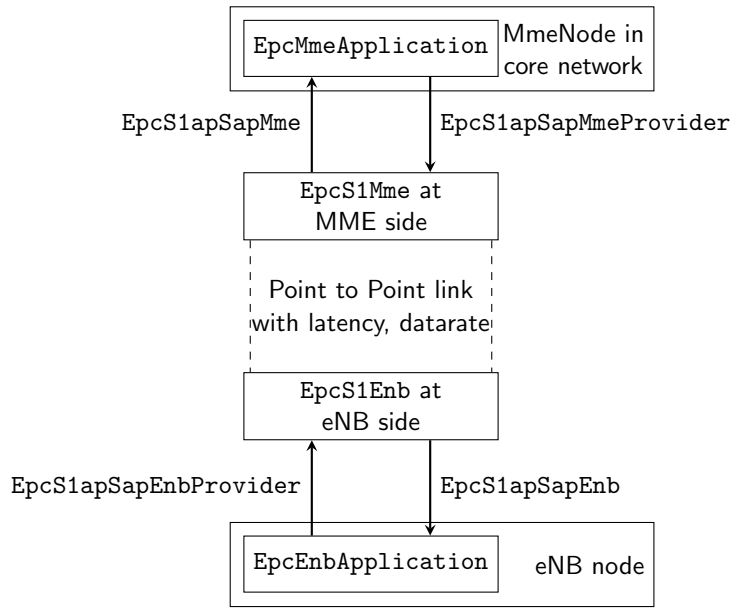 

\section{S1-AP Interface And MME Node Implementation}
As mentioned in Sec.~\ref{sec:cn}, the MME is not modeled as a real node in the EPC network. The MME is used during a handover, as shown in Fig.~\ref{fig:handoverx2}, and it receives a message from the target eNB with the path switch command. It then updates the information on the UE on handover in the MME and S-GW, so that packets are forwarded to the correct eNB. Finally, it replies with a path switch acknowledgment message. Therefore, until the MME receives the path switch, packets are forwarded from the core network to the source eNB and then to the target eNB.

In order to model the performance of handover in a more realistic way, a real interface between MME and eNBs is added to the core network implementation. In particular, the \texttt{EpcMme} class of the LTE module is modified into an application (\texttt{EpcMmeApplication}), which is installed on an ns--3 node (\texttt{MmeNode}). Then this application is interfaced with the S1-MME endpoint at the MME. This is connected with a point to point link with the other endpoint, in the eNB, which is interfaced with the \texttt{EpcEnbApplication}. 

The \texttt{EpcS1Mme} and \texttt{EpcS1Enb} classes receive SDUs from the MME and the eNB, respectively, and create PDUs that can be sent over S1 by encoding the Information Elements and adding the S1-AP header\footnote{S1-AP is the protocol which runs on top of the S1-MME link, according to LTE terminology.}. Since not all the MME-related signalling is needed in the ns--3 LTE and mmWave modules (i.e., no paging, no tracking area updates), only a subset of possible S1-AP PDUs is supported. In particular, from the eNB to the MME the messages are:
\begin{itemize}
  \item Initial UE Message, which is sent when a UE performs its first random access;
  \item Path Switch Request, which is sent at the end of a handover;
  \item E-UTRAN Radio Access Bearer (E-RAB) Release Indication;
\end{itemize}
while those from the MME to the eNB are:
\begin{itemize}
  \item Initial Context Setup Request, with information related to UE bearers;
  \item Path Switch Request ACK, to signal that the switch in the MME and S-GW happened successfully.
\end{itemize}

\section{Data Collection Framework}
The ns--3 simulator has a built-in system that allows to connect traces in different modules and to create logger classes in order to collect statistics. 

The PDCP and RLC modules already log transmission and reception of packets. The logging in the PDCP module was enhanced in order to account for the packets to and from remote RLCs. 

Moreover, for each simulation, it is possible to generate traces of:
\begin{itemize}
  \item PDCP PDUs transmission and reception, with packet size and latency;
  \item RLC PDUs transmission and reception, with packet size, latency and logical channel identity, so that it is possible to distinguish RRC PDUs and data PDUs;
  \item X2 PDUs reception for each pair of eNBs, with packet size, latency and a flag that signals whether they are control packets or data packets;
  \item the SINR measurements that are periodically taken in the mmWave eNBs;
  \item the time at which each handover starts and ends;
  \item the cell to which a UE belongs, over time;
  \item the total number of application packets sent and received.
\end{itemize}

%% file: chapters/chapter5.tex

\chapter{Simulation And Performance Analysis}\label{chap:results}

\section{Simulation Scenario}
\subsection{Simulation Assumptions}
The simulations that will be described in this Chapter use the ns--3 simulator described in Chap.~\ref{chap:ns3} with the additional features added in Chap.~\ref{chap:impl}. Therefore, UL-based reference signals for the mmWave cells will be used, with digital beamforming at the eNB side. 

\begin{figure}[t!]
	\centering
	\begin{tikzpicture}[font=\sffamily\small, scale=0.8, every node/.style={scale=0.8}]
		\centering

	    \node[anchor=south west,inner sep=0] (image) at (0,0) {\includegraphics[width=0.9\textwidth]{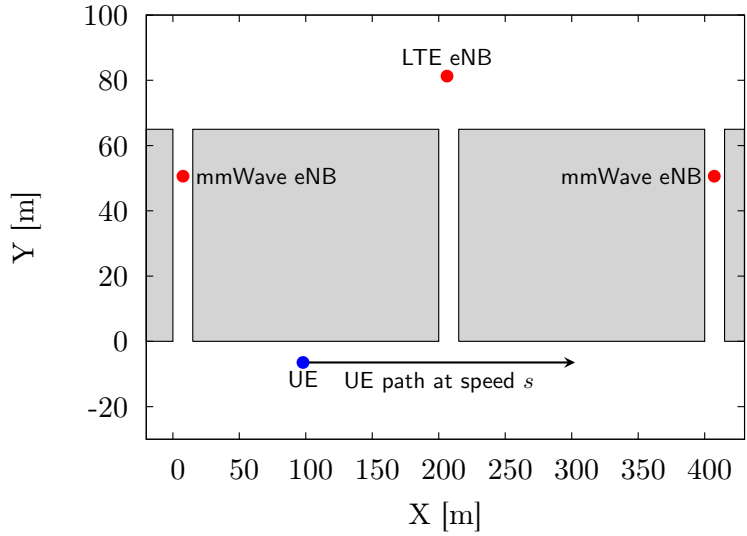}};
	    \begin{scope}[x={(image.south east)},y={(image.north west)}]
	        \filldraw[red,ultra thick] (0.23,0.645) circle (2pt);
	        \node[anchor=west] at (0.235,0.645) (mm1label) {mmWave eNB};
	        \filldraw[red,ultra thick] (0.894,0.645) circle (2pt);
	        \node[anchor=east] at (0.89,0.645) (mm2label) {mmWave eNB};
	        \filldraw[red,ultra thick] (0.56,0.825) circle (2pt);
	        \node[anchor=south] at (0.56,0.83) (ltelabel) {LTE eNB};

	        \draw[sarrow] (0.38, 0.31) -- node[anchor=north] {UE path at speed $s$} (0.72, 0.31);

	        \filldraw[blue,ultra thick] (0.38, 0.31) circle (2pt);
	        \node[anchor=north] at (0.38, 0.31) (ltelabel) {UE};
	    \end{scope}		
	\end{tikzpicture}
	\caption{Simulation scenario. The grey rectangles are buildings}
	\label{fig:scenario}
\end{figure}

The reference scenario is presented in Fig.~\ref{fig:scenario}. It is a typical urban grid, there are 4 buildings and each of them is 15 meters high. The mmWave eNBs are located in two streets along the y-axis, at a height of 10 meters. At the beginning of the simulation, the UE is at coordinates $(100, -5)$. It then moves along the x-axis at speed $s$ m/s, until it arrives in position $(300, -5)$. The goal of the simulations is indeed to test the performance of the system in a scenario where the UE is far from the mmWave eNBs, and experiences outages. The simulation duration $\tau$ is therefore depends on the UE speed $s$, and in particular is given by
\begin{equation}
	\tau = \frac{l_{path}}{s}
\end{equation}
where $l_{path} = 200$ m is the length of the path of the UE.

\subsection{Simulation Parameters and Procedures}
The parameters for the OFDM frame structure are the same of Table~\ref{table:sim_param}. Additional parameters are summarized in Table~\ref{table:all_param}.

The goal of these simulations is to assess the difference in performance between the fast switching and the hard handover setups. The simulations are performed with two different RLCs, AM and UM, and for a wide set of parameters. In particular, the \textit{for} cycle used to launch the simulation is described in Alg.~\ref{alg:sim}. A total of 5400 simulations were run.

\begin{table}[t]
	\ra{1.3}
	\begin{tabular}{@{}lll@{}}
		\toprule
		Parameter & Value & Description \\ \midrule
		mmWave BW & 1 GHz & Bandwidth of mmWave eNBs \\
		mmWave $f_c$ & 28 GHz & mmWave carrier frequency \\
		mmWave UL-DL $P_{tx}$ & 30 dBm & mmWave transmission power \\
		LTE BW & 20 MHz & Bandwidth of LTE eNBs \\
		LTE $f_c$ & 2.1 GHz & LTE downlink carrier frequency \\
		LTE DL $P_{tx}$ & 30 dBm & LTE transmission power for downlink \\
		LTE UL $P_{tx}$ & 23 dBm & LTE transmission power for uplink \\
		$F$ & 5 dB & noise figure \\
		$\Delta_{LTE}$ & -5 dB & Threshold for switch/handover to LTE \\
		$\Delta_{hys}$ & 3 dB & Hysteresis for handover\\
		mmWave eNB antenna & 16x16 ULA & \\
		mmWave UE antenna & 8x8 ULA & \\
		$s$ & $\{2,4,8,16\}$ m/s & UE speed \\
		$B_{RLC}$ & $\{1, 10, 100\}$ MB & RLC buffer size \\
		$\lambda$ & $\{20,40,80,160\}\,\mu$s & UDP packet inter-arrival time \\
		$D_{X2}$ & $\{0.1, 1, 10\}$ ms & One-way delay on X2 links \\
		$D_{MME}$ & 10 ms & One-way MME Delay \\
		$N$ & 10 & Iterations per set of parameters \\
		\bottomrule
	\end{tabular}
	\caption{Simulation parameters}
	\label{table:all_param}
\end{table}

\begin{algorithm}[ht]
  \sffamily
  \setstretch{1.2}
  \caption{Simulation campaign}\label{alg:sim}
  \begin{algorithmic}[1]
    \For {UE speed $s \in \{2,4,8,16\}$ m/s}
    	\For{RLC AM or UM}
    		\For{$\lambda \in \{20, 40, 80, 160\}\,\mu$s}
    			\For{$B_{RLC} \in \{1, 10, 100\}$ MB}
    				\For{$D_{X2} \in \{0.1, 1, 10\}$ ms}
    					Run $N$ simulations with these parameters
    				\EndFor

    			\EndFor
    		\EndFor
    	\EndFor
    \EndFor
  \end{algorithmic}
\end{algorithm}

The LTE bandwidth of the eNBs is the highest that can be used in an LTE system. The transmission powers are typical values in LTE deployments, while mmWave UEs and eNBs are expected to use a transmission power in the 20-30 dBm range~\cite{rappaport1}. The noise figure of 5 dB accounts for the noise at the receiver side.

The value of the delay to the MME node ($D_{MME}$) is chosen in order to model both the propagation delay to a node which is usually centralized and far from the access network, and the processing delays of the MME server. 

The one-way delay of the X2 interface, instead, varies in $\{0.1, 1, 10\}$ ms in order to understand the dependence of the system performance on this value. However, for practical applications, the typical delay is around 1 ms. The Next Generation Mobile Networks Alliance requires that the round-trip delay must be smaller than 10 ms (i.e., a one-way delay of 5 ms), and recommends a round-trip value smaller than 5 ms~\cite{x2lat}. 

The parameter $N$ was chosen as a trade-off between the time required to perform the simulations, which are computationally very expensive, due to the level of detail of the simulator, and the need for small enough confidence intervals. These are however not shown in the figures, in order to make them easier to read. 

The main metrics that will be collected are related to the UDP packet losses, the latency at the RLC layer, the PDCP and the RRC throughput, and finally the X2 link traffic. For each of these metrics, the dependence on simulation parameters will be investigated. Only downlink traffic is considered.

\section{Main Results}

\subsection{Packet Losses}\label{sec:losses}
The first element to consider in this performance analysis is the number of packets lost, i.e., the difference between sent and received packets, averaged over the $N$ different iterations for each set of parameters.

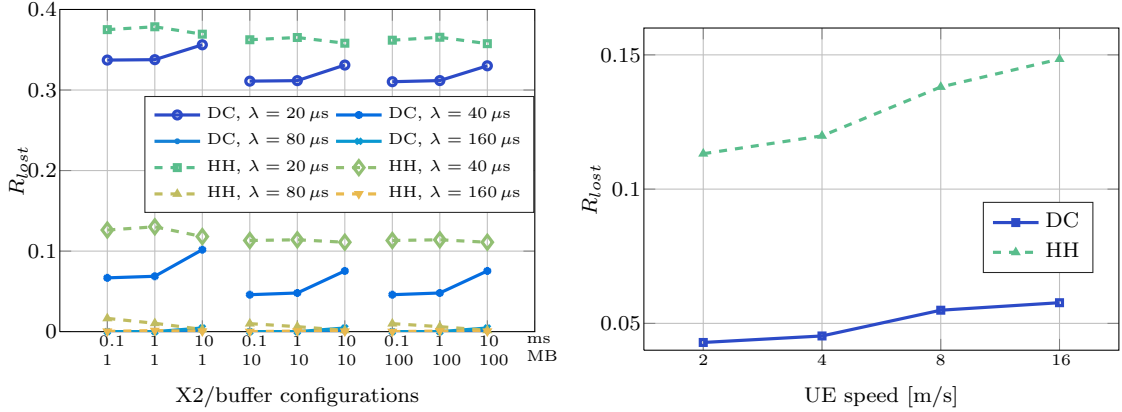
\begin{figure}[!t]
\centering
\begin{subfigure}[t]{0.45\textwidth}
	\captionsetup[sub]{justification=centering,font=tiny}
	\centering
	\setlength\fwidth{\textwidth}
	\setlength\fheight{0.65\textwidth}
	\input{./figures/perf/lost_2.tex}
	\caption{UE speed $s=2$ m/s, $\lambda \in \{20, 40, 80, 160\} \,\mu$s}
	\label{fig:lost_2}
\end{subfigure}
\hspace{2em}%
\begin{subfigure}[t]{0.45\textwidth}
	\captionsetup[sub]{justification=centering,font=tiny}
	\centering
	\setlength\fwidth{\textwidth}
	\setlength\fheight{0.65\textwidth}
	\input{./figures/perf/lost_speed_x2_100_b_10.tex}
	\caption{$R_{lost}$ as a function of the UE speed, for $\lambda = 40\,\mu$s, $B_{RLC}=10$MB, $D_{X2}=0.1$ms}
	\label{fig:lost_speed}
\end{subfigure}
\caption{UDP packet losses for simulations with RLC AM}
\label{fig:lost}
\end{figure}

\begin{figure}[!t]
\centering
\begin{subfigure}[t]{0.45\textwidth}
	\captionsetup[sub]{justification=centering,font=tiny}
	\centering
	\setlength\fwidth{\textwidth}
	\setlength\fheight{0.65\textwidth}
	\input{./figures/perf/lost_2_80160.tex}
	\caption{RLC AM}
	\label{fig:lost_am_2_80160}
\end{subfigure}
\hfill
\begin{subfigure}[t]{0.45\textwidth}
	\captionsetup[sub]{justification=centering,font=tiny}
	\centering
	\setlength\fwidth{\textwidth}
	\setlength\fheight{0.65\textwidth}
	\input{./figures/perf/lost_um_2.tex}
	\caption{RLC UM}
	\label{fig:lost_um_2_80160}
\end{subfigure}
\caption{UDP packet losses for UE speed $s= 2$ m/s, $\lambda = 80\,\mu$s}
\label{fig:lost_amum}
\end{figure}
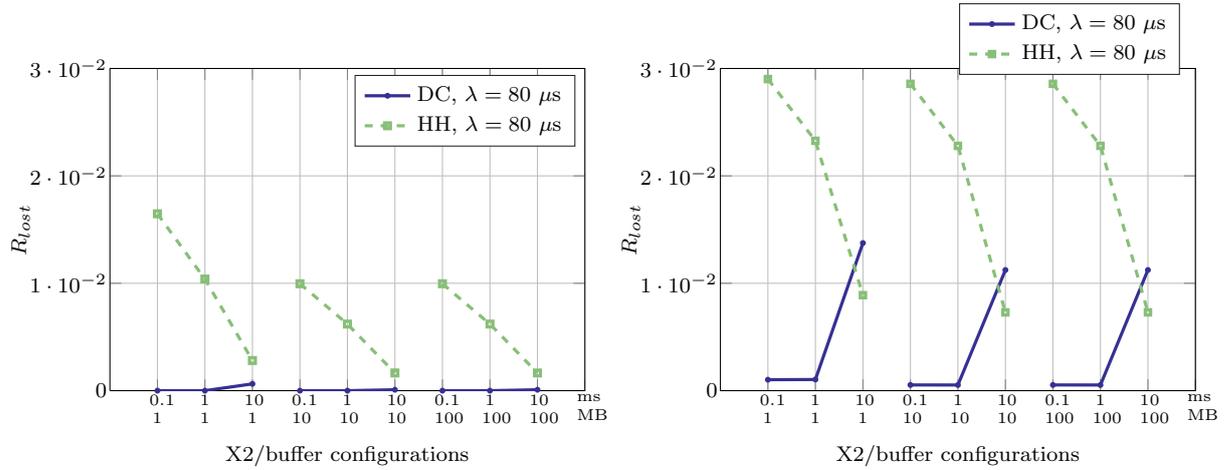

In Figs.~\ref{fig:lost} and~\ref{fig:lost_amum} the metric considered is the ratio of lost packets over the total sent packets. Since the UDP source constantly pushes packets to the system, with inter-arrival time $\lambda$, it can be computed as
\begin{equation}
	R_{lost} = 1 - \frac{\lambda}{\tau} r
\end{equation}
where $r$ is the number of received packets, and $\tau$ the duration of the simulation. 

In the x-axis of Figs.~\ref{fig:lost} and~\ref{fig:lost_amum}, there are different pairs $(D_{X2}, B_{RLC})$, where $D_{X2}$ is the latency of X2 links between eNBs, and $B_{RLC}$ is the transmission buffer size of RLC entities.

In Fig.~\ref{fig:lost_2} the first thing to observe is that at very low packet inter-arrival intervals ($\lambda = 20 \,\mu$s) the mobile network is not able to successfully deliver to the UE as many packets as the eNBs receive from the core network, and this causes system instability, with very high packet losses. It must be said that (i) the SNR of mmWave links is generally low in the simulation scenario, thus low MCS must be used, and therefore the rate offered on the mobile network is limited; (ii) the RLC entities have a limited buffer size $B_{RLC}$, i.e., packets are discarded if the buffer is full\footnote{However, the retransmission buffer is not limited in size (because of the particular ns--3 implementation), therefore once packets are sent a first time, they are moved to the retransmission buffer which can be filled without limits. This explains why there is no difference in packet losses for $B_{RLC}=10$ MB and $B_{RLC}=100$ MB.}. This behavior for $\lambda=20\,\mu$s is independent of the UE speed $s$, therefore the trends for $s \in \{4,8,16\}$ m/s are redundant and are not shown. 

While simulations of an unstable system cannot be used to derive statistically sound results, it is interesting to observe the general behavior at very low $\lambda$. This allows to understand the application rate that can be supported with very small losses by the setup under analysis, and compare the fast switching and hard handover solutions.

Therefore, the second observation that can be made is that with the fast switching solution fewer packets are lost. Notice that, since with many parameter combinations the number of lost packets is 0, it is not possible to compute a loss probability. However, since the number of packets sent in a simulation for $s = 2$m/s and $\lambda=80\,\mu$s is in the order of $10^7$, then the confidence interval at 95\% for the loss probability for $D_{X2} < 1$ ms will be $[0, \, 3\cdot10^{-7}]$, for the fast switching solution. 

The reason behind this behavior is the following. In the simulated scenario, the UE experiences a lot of handovers and/or switches, because of the simple handover/switch algorithm and of the high variability of the mmWave channel. The design of the LTE procedure for the lossless handover with RLC AM tries to minimize the packet losses, but in an extreme scenario like the simulated one this is not enough. There are two elements that contribute to the losses:
\begin{itemize}
	\item The first, which depends weakly on the buffer size $B_{RLC}$, is the fact that some packets, which are segmented in the RLC AM retransmission buffer, cannot be re-composed as the original PDCP SDU. Therefore they are lost;
	\item The second, which depends on the buffer size, is that during handover, the target RLC AM transmission buffer receives both the packets sent by the UDP source at rate $\lambda$, and the packets that were in the source RLC buffers. If the source RLC buffers are full, then the target buffer may overflow and discard packets.
\end{itemize}

Both these phenomena are stressed by the fact that the handover procedure takes more time than the switching procedure. Indeed, it involves 3 messages on the X2 interface and a message to the core network before the handover is completed at the target eNB. Moreover, until the UE has completed the Non Contention Based RA procedure with the target eNB, packets cannot be sent to the UE and must be buffered at the RLC layer. This worsens the overflow behavior of the RLC buffer. Instead, with fast switching, the UE does not need to perform random access, since it is already connected, so as soon as packets get to the buffer of the eNB that is the target of the switch, they are transmitted to the UE.

In Fig.~\ref{fig:lost_am_2_80160} there is the trend of $R_{lost}$ for $s = 2$m/s and $\lambda =80\,\mu$s and the behavior that was just described can be seen in detail. When the RLC AM buffer size $B_{RLC}$ increases from 1 to 10 MB, the value of $R_{lost}$ decreases, but it remains constant for 10 and 100 MB. It can also be seen that the packet losses with the hard handover setup decrease as the X2 latency $D_{X2}$ increases. This can be explained by the fact that once the handover is completed, with a greater X2 latency $D_{X2}$ it takes more time to trigger the next handover, in case the channel suddenly changed during the previous handover operation, therefore RLC buffers have a smaller chance to overflow. However, since the update on the channel is reported with delay $D_{X2}$ to the LTE eNB and a handover command would take another $D_{X2}$ to be received by the mmWave eNB, the UE may be connected to an eNB with a low SINR link. Therefore the packets in the buffer cannot be transmitted (no transmission opportunity is issued by the MAC layer) or are transmitted with errors and need to be retransmitted. This is why there are still packet losses, and also the performance of the fast switching setup worsens. 

This behavior is exacerbated when RLC UM is used, as shown in Fig.~\ref{fig:lost_um_2_80160}. Indeed, since there are no retransmissions, the packet losses for the fast switching solution with a high X2 latency are higher than those with hard handover, which benefits from the fact that fewer handovers happen, thus the RLC buffers are reset less often. For $D_{X2} \in \{0.1, 1\}$ ms, i.e., for less extreme X2 latencies, fast switching performs better than hard handover also with RLC UM, and as expected packet losses are higher with RLC UM than with RLC AM.

In Fig.~\ref{fig:lost_speed}, finally, there is a comparison between $R_{lost}$ for fast switching and hard handover as the UE speed increases, for $\lambda=40\,\mu$s, and it can be seen that the ratio of lost packets increases with the UE speed in both systems.

\subsection{Latency}
The latency $L$ is measured for each packet, from the time at which the PDCP PDU enters the RLC buffer of the eNB to when it is successfully received at the PDCP layer in the UE. Therefore, it is the latency of only the correctly received packets. 

The choice of measuring this type of latency was made in order to exclude the first $D_{X2}$ delay in the fast switching setup, which is given by the forwarding of the packet from the LTE eNB to the mmWave RLC. Indeed, the coordinator may not be placed in the LTE eNB, unlike assumed in these simulations, but, for example, it may be deployed in the core network, or even in the mmWave eNB. It may also be the case that the X2 link is not a point to point link, but a more complex multi-hop network, where the delays from the coordinator to the LTE eNB and from the coordinator to the mmWave eNB are similar. Therefore, a measure of the latency which includes also the first X2 delay would be deployment-dependent, and would limit the validity of the observations to the particular simulation setup. 

\begin{figure}[!t]
\centering
\begin{subfigure}[t]{0.45\textwidth}
	\captionsetup[sub]{justification=centering,font=tiny}
	\centering
	\setlength\fwidth{\textwidth}
	\setlength\fheight{0.65\textwidth}
	\input{./figures/perf/lat_2_80160.tex}
	\caption{RLC AM}
	\label{fig:lat_2_am}
\end{subfigure}
\hfill%
\begin{subfigure}[t]{0.45\textwidth}
	\captionsetup[sub]{justification=centering,font=tiny}
	\centering
	\setlength\fwidth{\textwidth}
	\setlength\fheight{0.65\textwidth}
	\input{./figures/perf/lat_um_801602.tex}
	\caption{RLC UM}
	\label{fig:lat_2_um}
\end{subfigure}
\caption{Latency $L$ for different $D_{X2}$ and $B_{RLC}$, UE speed $s = 2$~m/s}
\label{fig:lat_2}
\end{figure}
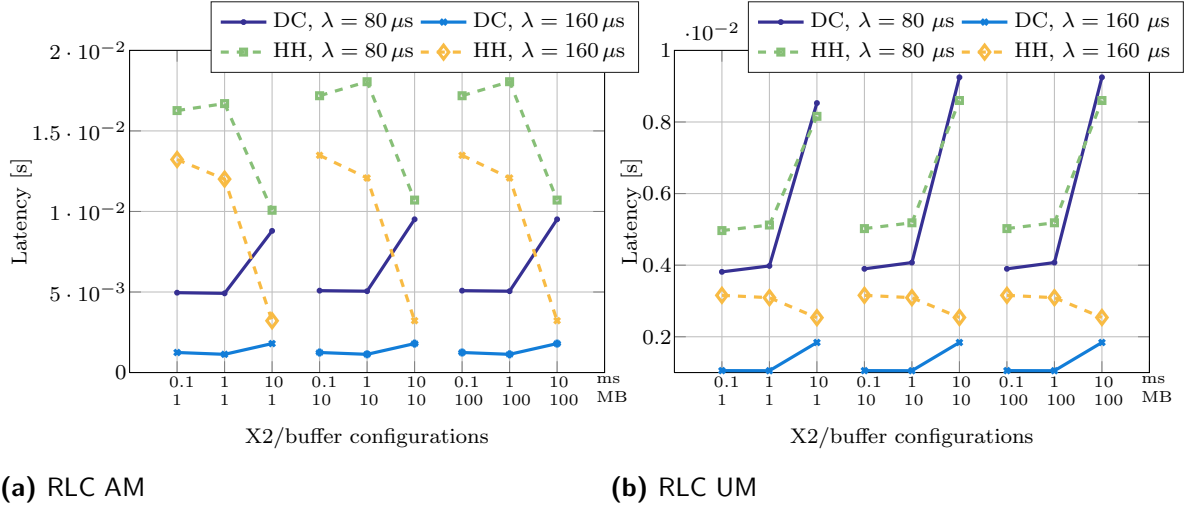

Moreover, the main behavior that is of interest for this analysis is given by the time that packets spend in the RLC buffers (at transmitter and receiver side), before successful reception, and the additional latency that occurs when a switch or handover happens and the packet is forwarded to the target eNB or RAT. This kind of delay has a large impact on cellular networks, which are designed to limit packet losses, at a price of higher buffering latencies~\cite{bufferbloat}. Therefore it will be of interest to evaluate how fast switching and hard handover compare with respect to this metric, for both RLC AM, and RLC UM, where the buffering delay issue should be less relevant (but there are more packet losses, as shown in Sec.~\ref{sec:losses}).

\begin{figure}[!t]
\centering
	\captionsetup[sub]{justification=centering,font=tiny}
	\setlength\fwidth{0.7\textwidth}
	\setlength\fheight{0.45\textwidth}
	\input{./figures/perf/latDiff_a_80160_b_10_x_1.tex}
	\caption{Difference between fast switching and hard handover latency, $B_{RLC}=10$~MB, $D_{X2}=1$~ms}
\label{fig:lat_diff}
\end{figure}
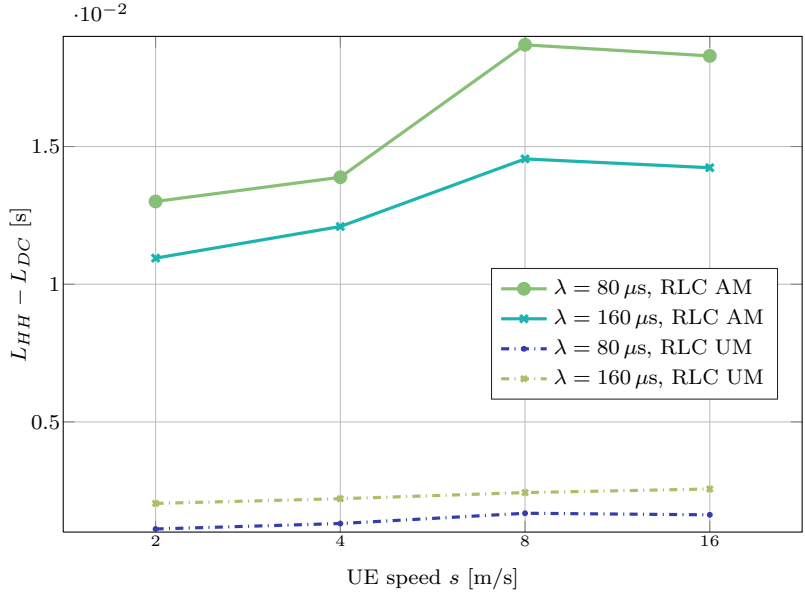

Fig.~\ref{fig:lat_2} shows the mean packet latency over $N$ simulations for UE speed $s=2$~m/s and different buffer and X2 latency values. It can be seen that fast switching outperforms hard handover, both with RLC AM and UM, when the X2 latency is 0.1 or 1 ms. However, for RLC AM, the latency of hard handover decreases for $D_{X2}=10$~ms, while the fast switching slightly increases. This is due to the fact that for $D_{X2}=10$~ms less frequent and longer handovers are performed, therefore there is a smaller chance that an RLC PDU already in the buffer is forwarded between source and target eNBs. Instead, the increase in the fast switching latency can be explained by the fact that, since the updates on mmWave SINR are reported with a larger delay to the LTE eNB, the overall performance of a dual-connected setup becomes worse, also from the point of view of the packet losses and throughput, as seen in Sec.~\ref{sec:losses}. For example, the UE may experience an outage in the mmWave link, but given the X2 delay, the LTE eNB becomes aware of this after $D_{X2}=10$~ms. During this interval, the UE may not receive, or may receive with errors, the packets sent from the mmWave eNB, which must then be retransmitted, and this further increases the latency. 

With RLC UM the latency is smaller than with RLC AM, because no retransmissions are performed. However, the difference between RLC AM and UM latency for fast switching is smaller than for hard handover, i.e., the latency of hard handover with RLC UM decreases more than the latency for fast switching. In particular, for $D_{X2} = 10$ ms and $\lambda = 80\,\,\mu$s, the latency of hard handover is slightly smaller than that of fast switching. The reasons are the same as for the respective behavior for RLC AM. 

In Fig.~\ref{fig:lat_diff} the dependence on the UE speed $s$ is shown. In particular, the quantity plotted is
\begin{equation}
	L_{HH} - L_{DC}
\end{equation}
i.e., the difference in latency between the two setups, as a function of the UE speed, for two different $\lambda \in \{80, 160\}\,\mu$s, $B_{RLC}=10$~MB and $D_{X2}=1$~ms. It can be seen that, for RLC AM, the difference, which is in the order of 10 ms, is higher for higher $\lambda$. Moreover, the difference increases as the UE speed increases from 2 to 8 m/s, as expected, but slightly decreases when further increasing the UE speed $s$ from 8 to 16 m/s. However, notice that this is the latency of packets which are actually received, and, as shown in Sec.~\ref{sec:losses}, the packet losses are higher at $s=16$~m/s. The same trend is observed also for RLC UM, but the difference is nearly one order of magnitude smaller and the dependence on the UE speed is weaker. 

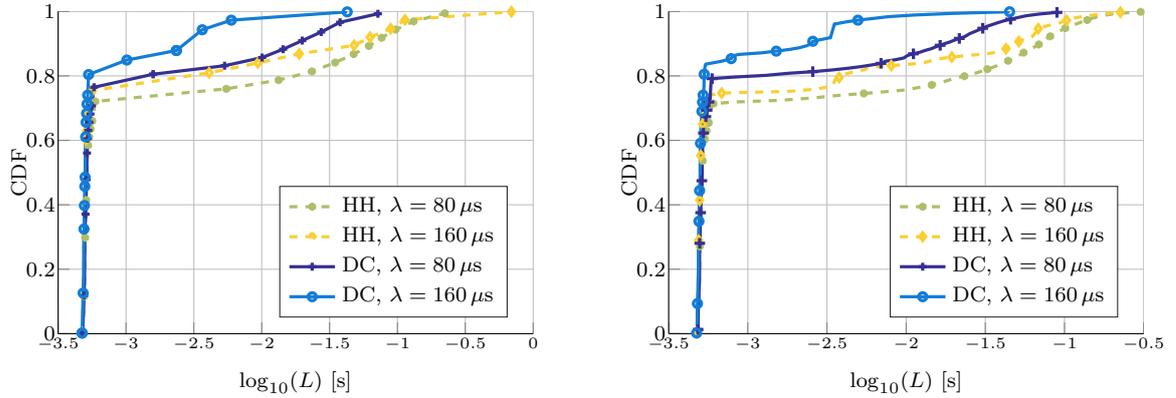
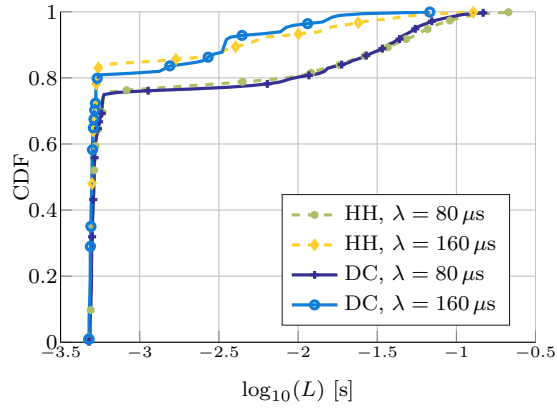
\begin{figure}[!t]
\centering
\begin{subfigure}[t]{0.45\textwidth}
	\captionsetup[sub]{justification=centering,font=tiny}
	\centering
	\setlength\fwidth{\textwidth}
	\setlength\fheight{0.65\textwidth}
	\input{./figures/perf/latCDF_a_80160_b_10_x_100.tex}
	\caption{$D_{X2} = 0.1$~ms}
	\label{fig:lat_cdf_01}
\end{subfigure}
\hfill
\begin{subfigure}[t]{0.45\textwidth}
	\captionsetup[sub]{justification=centering,font=tiny}
	\centering
	\setlength\fwidth{\textwidth}
	\setlength\fheight{0.65\textwidth}
	\input{./figures/perf/latCDF_a_80160_b_10_x_1000.tex}
	\caption{$D_{X2} = 1$~ms}
	\label{fig:lat_cdf_1}
\end{subfigure}
\begin{subfigure}[b]{\textwidth}
	\captionsetup[sub]{justification=centering,font=tiny}
	\centering
	\setlength\fwidth{0.45\textwidth}
	\setlength\fheight{0.3\textwidth}
	\input{./figures/perf/latCDF_a_80160_b_10_x_10000.tex}
	\caption{$D_{X2} = 10$~ms}
	\label{fig:lat_cdf_10}
\end{subfigure}
\caption{CDF of packet latency $L$ for UE speed $s=2$~m/s, $B_{RLC}=10$~MB, RLC AM. The x-axis is in logarithmic scale}
\label{fig:cdf_lat}
\end{figure}

Fig.~\ref{fig:cdf_lat} shows why the fast switching setup for RLC AM has a latency which is smaller than that of the hard handover system. The metric reported in Fig.~\ref{fig:cdf_lat} is the CDF computed on all the latencies of the received packets, for each set of $N$ simulations with $s=2$~m/s and $B_{RLC}=10$~MB. Moreover, two different $\lambda \in \{80, 160\}\,\mu$s are shown. Notice that the x-axis is in logarithmic scale. 

The first observation is that most of the packets (up to 80\% in some cases) are sent with a very small latency (in the order of $10^{-3.4}$~s, i.e., in less than a millisecond). This is actually the latency of the mmWave radio access network, which by design is the one that should be used most of the time, since it should provide a higher throughput. Therefore, the mean values shown in Fig.~\ref{fig:lat_2} are very different from the median values, which are similar for the fast switching and hard handover setups. 

The second observation is that for $D_{X2}=0.1$~ms and $D_{X2}=1$~ms the fast switching option manages to send nearly 5 and 10\% more packets with the mmWave interface latency. Moreover, the latency value for which the CDF for fast switching reaches 1 is smaller than the respective value for hard handover. This explains why the fast switching setup has a lower mean latency. The reason for this behavior is that a switch is much faster than a handover, therefore the UE experiences (i) no service interruptions; (ii) a mmWave channel with a good SINR (higher than $\Delta_{LTE}$) most of the time, so that more packets can be served by the mmWave interface without additional delays. 
This phenomenon is instead less relevant for $D_{X2}=10$~ms, and this explains why the difference in the mean values is smaller.

\subsection{PDCP Throughput}\label{sec:th}
The throughput over time at the PDCP layer is measured by sampling the logs of received PDCP PDUs every $T_s = 5$ ms and summing the received packet sizes to obtain the total number of bytes received $B(t)$. Then the throughput $S(t)$ is computed in bit/s as
\begin{equation}\label{eq:th}
	S(t) = \frac{B(t)\times 8}{T_s}.
\end{equation}
Then, in order to get the mean throughput $S_{PDCP}$ for a simulation, these samples can be averaged over the total simulation time, and finally over the $N$ simulations to obtain $\hat{S}_{PDCP}$. 

Notice that the PDCP throughput is mainly made up of the transmission of new incoming packets, but it may also account for retransmission of already transmitted packets. Indeed, in the RLC AM setup, if a packet was transmitted, but not already ACKed, it is stored in the RLC AM retransmission buffer. Then, when a handover (switch) happens, the retransmission buffer is forwarded to the target eNB (RAT) and transmitted again. Therefore, if at the first time it was received successfully, it is wastefully retransmitted. 

The PDCP throughput is mainly a measure of the rate that the radio network can offer, given a certain application rate. 

\begin{figure}[!t]
\centering
\begin{subfigure}[t]{0.4\textwidth}
	\captionsetup[sub]{justification=centering,font=tiny}
	\centering
	\setlength\fwidth{\textwidth}
	\setlength\fheight{0.65\textwidth}
	\input{./figures/perf/th_2_2040.tex}
	\caption{$\hat{S}_{PDCP}$, UE speed ${s=2}$~m/s, $\lambda \in \{20, 40, 80, 160\}\,\mu$s}
	\label{fig:pdcp_th}
\end{subfigure}
\hfill
\begin{subfigure}[t]{0.4\textwidth}
	\captionsetup[sub]{justification=centering,font=tiny}
	\centering
	\setlength\fwidth{\textwidth}
	\setlength\fheight{0.65\textwidth}
	\input{./figures/perf/th_a_80_b_10_x2.tex}
	\caption{$\hat{S}_{PDCP}$ as a function of $D_{X2}$, for ${\lambda = 80\,\mu}$s, ${B_{RLC}=10}$~MB and different UE speed $s$}
	\label{fig:pdcp_th_speed}
\end{subfigure}
\caption{PDCP throughput $\hat{S}_{PDCP}$}
\label{fig:th_an}
\end{figure}
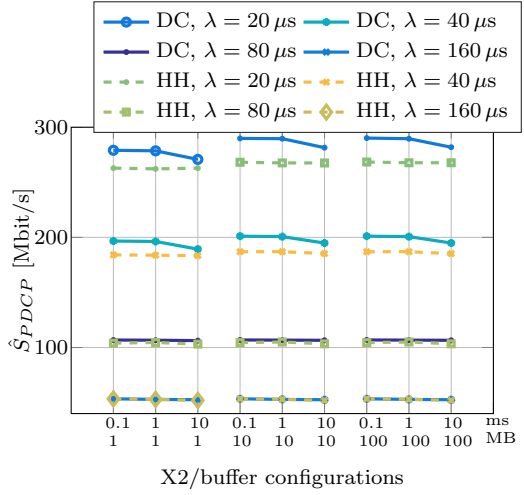
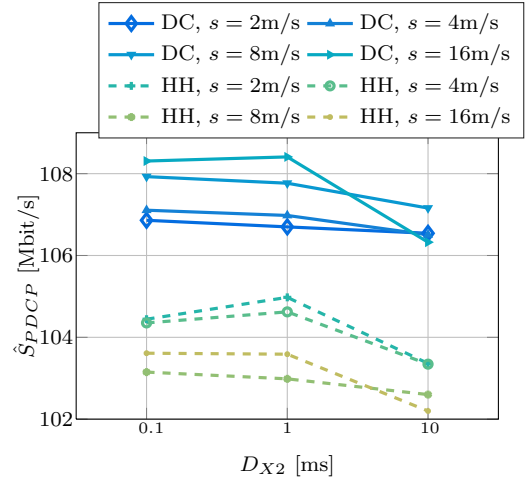

Fig.~\ref{fig:pdcp_th} shows the PDCP throughput $\hat{S}_{PDCP}$ for UE speed $s=2$ m/s and different combinations of $D_{X2}$ and $B_{RLC}$. It can be observed that the throughput achievable with the fast switching dual connectivity solution is higher than the one with hard handover. Moreover, the difference in throughput increases as the application rate increases, in accordance with the results on packet losses described in the previous section.

As for the relation with the UE speed, there are different behaviors with respect to the hard handover and fast switching setup. Fig.~\ref{fig:pdcp_th_speed} shows the PDCP throughput for different speeds, different $D_{X2}$, $\lambda = 80\,\mu$s and $B_{RLC}=10$ MB. Notice that since the size of PDCP SDUs is 1042 bytes, then the rate at which new packets arrive at PDCP is $S_{incoming} = 104.2$ Mbit/s. 
If the PDCP throughput $\hat{S}_{PDCP}$ is higher than $S_{incoming}$, it means that the contribution given by unneeded retransmissions (that increase the throughput) is more significant than that of packet losses (that decrease the throughput), and vice versa if $\hat{S}_{PDCP} < S_{incoming}$.

The throughput generally decreases as $D_{X2}$ increases from 1 to 10 ms. The reason is the same one that explains the behavior of the packet losses in Fig.~\ref{fig:lost_2}: there are fewer handovers/switches. Therefore, with the fast switching, where the switch to the best channel happens in a very short time, the number of unnecessary retransmissions decreases as the number of switches decreases. The same holds for the handover case, but since the service interruption is longer and the channel that the UE uses is not the optimal one for a longer time, then there is a loss in performance also due to the non optimal choice of the channel.

The fast switching option, at lower speed, experiences fewer unneeded retransmissions, thus the throughput is smaller with respect to the one for higher UE speeds, with the exception of $s=16$ m/s and $D_{X2}=10$ ms, where besides the increase given by retransmission there is also a loss due to the combination of a faster UE and less timely SINR estimation updates. The handover performance instead decreases as the UE speed increases.


\subsection{RRC Traffic}\label{sec:rrcth}
The RRC traffic is measured at the RLC layer by analyzing the received RLC PDUs logs and accounting only for packets of signalling radio bearers. Then, Eq.~\eqref{eq:th} is applied to get the instantaneous RRC throughput, which is then averaged over the duration of a simulation.

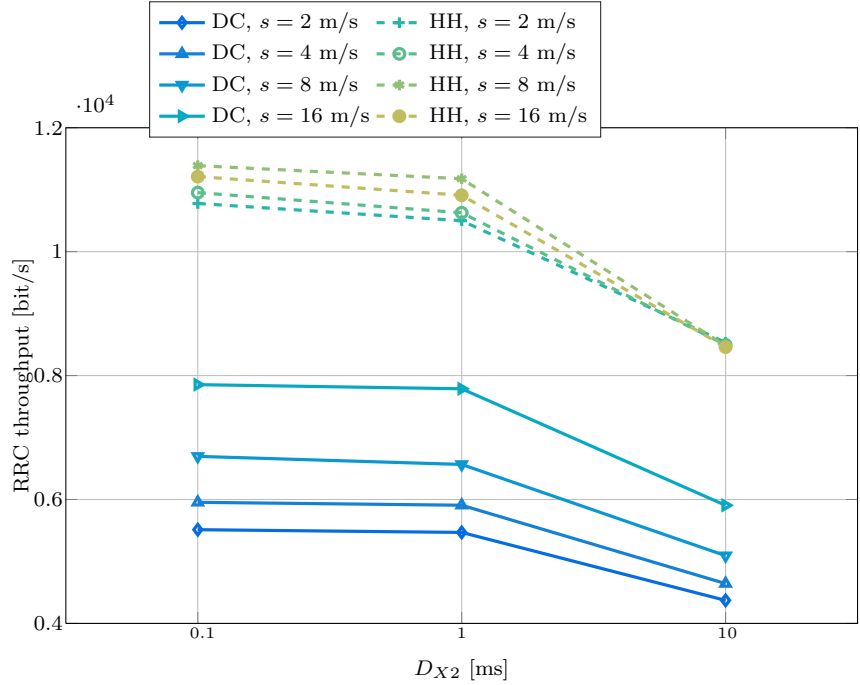
\begin{figure}[!t]
\centering
\captionsetup[sub]{justification=centering,font=tiny}
\centering
\setlength\fwidth{0.75\textwidth}
\setlength\fheight{0.45\textwidth}
\input{./figures/perf/rrc_80_B_10_speed.tex}
\caption{RRC traffic as a function of the UE speed and X2 latency}
\label{fig:rrc}
\end{figure}

The RRC traffic is an indication of how many control operations are done by the UE-eNB pairs. Moreover, it is dependent also on the RRC PDU size. For example, a switch message contains 1 byte for each of the bearers that should be switched, while an RRC connection reconfiguration message (which triggers the handover) carries several data structures, for a minimum of 59 bytes for a single bearer reconfiguration. 

Fig.~\ref{fig:rrc} shows the RRC throughput for different $D_{X2}$ delays and different UE speed $s$. Notice that the RRC traffic is independent of the buffer size $B_{RLC}$, since even 1 MB is enough to buffer the RRC PDUs, and of the UDP packet inter-arrival time, $\lambda$.

It can be seen that fast switching has an RRC traffic which is 4 to 5 Kbit/s lower than for hard handover. A lower RRC traffic is better, since it allows to allocate more resources to data transmission. Moreover, in this simulation a single UE is accounted for, but the number of devices that an LTE or mmWave eNB has to serve may be large, thus the RRC traffic could cause a large overhead. 

The other trends that can be observed are:
\begin{itemize}
	\item the RRC traffic increases with the UE speed. This is probably due to the fact that at a higher speed, more retransmissions of the messages are required. Moreover, at higher speeds the channel changes more frequently, therefore there are more handovers and switches;
	\item the RRC traffic slightly decreases as $D_{X2}$ increases from 0.1 to 1 ms. Instead, when $D_{X2} = 10$~ms, the RRC traffic decreases. This shows that with such a high X2 delay, handover and switch procedures last longer, thus fewer messages are exchanged. Moreover, in the hard handover case, the difference between the RRC throughput at different speeds is minimized when $D_{X2}=10$~ms. This is due to the fact that with less timely updates there is no actual difference between how the channel is seen at the coordinator for different UE speeds. 
\end{itemize}

\subsection{X2 Traffic}
Another metric that must be considered when analyzing a dual connected system is the X2 link traffic. Indeed, if the coordinator is placed in the LTE eNB, the X2 link has to support the forwarding of data packets to the mmWave remote RLCs. If, instead, it is placed in the core network, the same considerations made for the X2 link will also be valid for the link connecting the coordinator to the LTE and mmWave eNBs.

In this simulation campaign only one UE is used, therefore it is not meaningful to show the ratio at which X2 links are used, since it also depends on the particular choice of the datarate for the X2 link. Instead, the metric chosen is
\begin{equation}\label{eq:x}
	X = \frac{\hat{S}_{X2}}{\hat{S}_{PDCP}}
\end{equation}
where $\hat{S}_{PDCP}$ is the PDCP throughput, as described in Eq.~\eqref{eq:th}, and 
\begin{equation}
	\hat{S}_{X2} = \sum_{i=1}^3 \hat{S}_{X2, i}
\end{equation}
where $\hat{S}_{X2, i}$ is the mean throughput of X2 link $i$ across the $N$ iterations. In these simulations there are 3 X2 links, one for each mmWave eNB to the LTE eNB and one between the two mmWave eNBs.

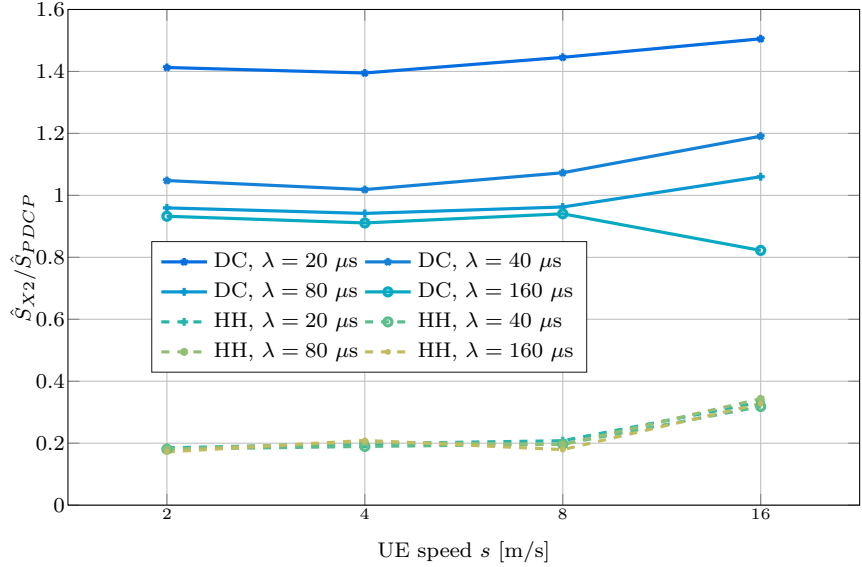
\begin{figure}[!t]
\centering
\captionsetup[sub]{justification=centering,font=tiny}
\centering
\setlength\fwidth{0.75\textwidth}
\setlength\fheight{0.45\textwidth}
\input{./figures/perf/x2pdcpRatio_x_1_b_10.tex}
\caption{Metric $X$ (see Eq.~\eqref{eq:x}) for different UE speed $s$ and $\lambda$, for $B_{RLC} = 10$~MB and $D_{X2}=1$~ms}
\label{fig:x2}
\end{figure}

This metric aims to show how much the X2 link is used, given a certain PDCP throughput, i.e., given the rate at which data packets are sent over the radio access network. This is shown in Fig.~\ref{fig:x2} for different $\lambda$ and different UE speed $s$, and $B_{RLC} = 10$~MB, $D_{X2}=1$~ms. Hard handover has a ratio $X$ close to 0.2, in general much smaller than the ratio for fast switching, and independent of the rate $\lambda$. This means that, given a certain application rate that the system must support, the X2 link must be dimensioned to offer just 20\% of it, per UE. Notice also that in this case there are many handovers, therefore this is an extreme scenario with respect to the utilization of X2 links.

The most interesting comparison is however with the ratio $X$ of the fast switching solution, which is close to 1 for $\lambda \in \{40, 80, 160\}\,\mu$s. This means that X2 links have to forward many more data packets, as expected, and a certain amount of control information, which explains why the ratio is in some cases greater than 1. For $\lambda = 20\,\mu$s, instead, the ratio $X$ is not meaningful, since the rate at which packets are sent to the RAN is higher than the rate at which they are forwarded to the UE, i.e., packets reach the LTE eNB, are forwarded to the mmWave eNB and then either discarded (because the transmission buffer overflows), or not transmitted successfully and moved to the retransmission buffer.

\section{Comments And Further Analysis}
It can be seen that, in general, the fast switching option performs better than the hard handover setup. 

The main benefit of the fast switching setup is the short time it takes to change radio access network. This is shown by several metrics. The latency is in general smaller for the fast switching setup, because the handover is a long procedure and packets have to be buffered in the target eNB before they can be sent. When the buffer is too small (1 MB), this also causes buffer overflow and packet losses. 

Fast switching performs better with lower X2 one-way latencies, while it appears that the hard handover prefers higher X2 latencies. This however depends on the fact that the simulated scenario is extreme with respect to the mmWave channel quality and to the number of handovers performed, therefore what causes the reduction of latency and packet losses is the smaller number of handovers that are performed as the X2 latency increases. This does not mean that having a higher X2 latency is better: indeed, as shown in Sec.~\ref{sec:th}, the PDCP throughput decreases because the updates on channel quality are not timely.

The gain in performance is consistent also at different UE speeds, therefore the fast switching solution may be a good candidate for its robustness to mobility.

Another advantage of the dual connectivity solution is that the control signalling related to the user plane (i.e., setup of Data Radio Bearers, switching) is performed on the LTE connection. This may cause an increase in the traffic of the LTE eNB, however, as shown in Sec.~\ref{sec:rrcth}, the overall\footnote{The analysis in Sec.~\ref{sec:rrcth} aggregates the RRC traffic of the LTE eNB and of the mmWave eNBs} RRC traffic is smaller with the fast switching solution. Moreover, the LTE eNB is used for user plane traffic only when the SINR of all the mmWave eNBs is smaller than $\Delta_{LTE}$, therefore the UEs use the mmWave eNBs most of the time. This allows the LTE eNB to handle the load of many more UEs than it would be able to manage if the LTE user plane were always used. 

The computational load on the LTE eNB, however, is expected to increase, in particular if the coordinator is co-located on the LTE eNB. This aspect was not investigated in the simulations, because the simulation framework in ns--3 does not model CPU or memory loads. However, the LTE eNB has to collect the report tables and find the optimal mmWave association for each UE, and it has to encode and send RRC messages to the UE. Moreover, if the coordinator is co-located, and since the PDCP encrypts the PDUs it receives from higher layers, the LTE eNB has to encrypt all the traffic of dual connected UEs under its coverage. Therefore, when deploying the next generation 5G networks, it will be necessary to consider if an update of computational elements in the LTE eNBs is required. 

Another drawback of the dual-connected fast switching solution is that the X2 links are heavily stressed, and the deployment of this solution must be carefully planned with respect to the datarate of X2 links, otherwise they may be the bottleneck of the system.

%% file: figures/perf/lost_2.tex
%
%
\definecolor{mycolor1}{rgb}{0.17073,0.29194,0.77925}%
\definecolor{mycolor2}{rgb}{0.01651,0.42660,0.87863}%
\definecolor{mycolor3}{rgb}{0.07794,0.50399,0.83837}%
\definecolor{mycolor4}{rgb}{0.03434,0.59658,0.81985}%
\definecolor{mycolor5}{rgb}{0.34817,0.74243,0.54727}%
\definecolor{mycolor6}{rgb}{0.57086,0.74852,0.44939}%
\definecolor{mycolor7}{rgb}{0.75249,0.73840,0.37681}%
\definecolor{mycolor8}{rgb}{0.91393,0.72579,0.30628}%
\begin{tikzpicture}
\tikzstyle{every node}=[font=\scriptsize]
\pgfplotsset{every x tick label/.append style={font=\tiny, yshift=0.5ex}}

\begin{axis}[%
width=0.951\fwidth,
height=\fheight,
at={(0\fwidth,0\fheight)},
scale only axis,
xmin=0,
xmax=10,
xlabel={X2/buffer configurations},
xtick={1,2,3,4,5,6,7,8,9,10},
xticklabels={$0.1$ $1$, $1$ $1$, $10$ $1$, $0.1$ $10$, $1$ $10$, $10$ $10$, $0.1$ $100$, $1$ $100$, $10$ $100$, ms MB},
xticklabel style={text width=6, align=center}, xmajorgrids,
ymin=0,
ymax=0.4,
ylabel={$R_{lost}$},
ymajorgrids, ylabel shift = -5 pt, yticklabel shift = -3 pt,
axis background/.style={fill=white},
legend columns = 2,
legend style={legend cell align=left,align=left,draw=white!15!black,at={(0.95\fwidth,0.55\fheight)},anchor=east,}
]
\addplot [color=mycolor1,solid,line width=1.2pt,mark size=1.5pt,mark=o,mark options={solid}]
  table[row sep=crcr]{%
1	0.33717112\\
2	0.337677444444444\\
3	0.35614442\\
};
\addplot [color=mycolor1,solid,line width=1.2pt,mark size=1.5pt,mark=o,mark options={solid}, forget plot]
  table[row sep=crcr]{%
4	0.31103512\\
5	0.311630133333333\\
6	0.3310024\\
};
\addplot [color=mycolor1,solid,line width=1.2pt,mark size=1.5pt,mark=o,mark options={solid}, forget plot]
  table[row sep=crcr]{%
7	0.31031348\\
8	0.311714511111111\\
9	0.3300724\\
};
\addlegendentry{\tiny{DC, $\lambda = 20 \,\mu$s}};

\addplot [color=mycolor2,solid,line width=1.2pt,mark size=1.5pt,mark=asterisk,mark options={solid}]
  table[row sep=crcr]{%
1	0.0668058222222222\\
2	0.068748\\
3	0.10176635\\
};
\addplot [color=mycolor2,solid,line width=1.2pt,mark size=1.5pt,mark=asterisk,mark options={solid}, forget plot]
  table[row sep=crcr]{%
4	0.0458853777777778\\
5	0.0480104444444444\\
6	0.0753855428571429\\
};
\addplot [color=mycolor2,solid,line width=1.2pt,mark size=1.5pt,mark=asterisk,mark options={solid}, forget plot]
  table[row sep=crcr]{%
7	0.0458852888888889\\
8	0.0480104\\
9	0.0753854857142857\\
};
\addlegendentry{\tiny{DC, $\lambda = 40 \,\mu$s}};

\addplot [color=mycolor3,solid,line width=1.2pt,mark size=0.5pt,mark=*,mark options={solid}]
  table[row sep=crcr]{%
1	0\\
2	0\\
3	0.00063216\\
};
\addplot [color=mycolor3,solid,line width=1.2pt,mark size=0.5pt,mark=*,mark options={solid}, forget plot]
  table[row sep=crcr]{%
4	0\\
5	0\\
6	8.536e-05\\
};
\addplot [color=mycolor3,solid,line width=1.2pt,mark size=0.5pt,mark=*,mark options={solid}, forget plot]
  table[row sep=crcr]{%
7	0\\
8	0\\
9	8.536e-05\\
};
\addlegendentry{\tiny{DC, $\lambda = 80 \,\mu$s}};

\addplot [color=mycolor4,solid,line width=1.2pt,mark size=1.5pt,mark=x,mark options={solid}]
  table[row sep=crcr]{%
1	0\\
2	0\\
3	0.00456944\\
};
\addplot [color=mycolor4,solid,line width=1.2pt,mark size=1.5pt,mark=x,mark options={solid}, forget plot]
  table[row sep=crcr]{%
4	0\\
5	0\\
6	0.00456944\\
};
\addplot [color=mycolor4,solid,line width=1.2pt,mark size=1.5pt,mark=x,mark options={solid}, forget plot]
  table[row sep=crcr]{%
7	0\\
8	0\\
9	0.00456944\\
};
\addlegendentry{\tiny{DC, $\lambda = 160 \,\mu$s}};

\addplot [color=mycolor5,dashed,line width=1.2pt,mark size=1.1pt,mark=square,mark options={solid}]
  table[row sep=crcr]{%
1	0.37497158\\
2	0.378507155555556\\
3	0.36932098\\
};
\addplot [color=mycolor5,dashed,line width=1.2pt,mark size=1.1pt,mark=square,mark options={solid}, forget plot]
  table[row sep=crcr]{%
4	0.36241924\\
5	0.365202177777778\\
6	0.358113088888889\\
};
\addplot [color=mycolor5,dashed,line width=1.2pt,mark size=1.1pt,mark=square,mark options={solid}, forget plot]
  table[row sep=crcr]{%
7	0.36185154\\
8	0.365501266666667\\
9	0.357543622222222\\
};
\addlegendentry{\tiny{HH, $\lambda = 20 \,\mu$s}};

\addplot [color=mycolor6,dashed,line width=1.2pt,mark size=2.6pt,mark=diamond,mark options={solid}]
  table[row sep=crcr]{%
1	0.126176755555556\\
2	0.1301798\\
3	0.11803395\\
};
\addplot [color=mycolor6,dashed,line width=1.2pt,mark size=2.6pt,mark=diamond,mark options={solid}, forget plot]
  table[row sep=crcr]{%
4	0.1131856\\
5	0.114069911111111\\
6	0.111019542857143\\
};
\addplot [color=mycolor6,dashed,line width=1.2pt,mark size=2.6pt,mark=diamond,mark options={solid}, forget plot]
  table[row sep=crcr]{%
7	0.1131856\\
8	0.114069911111111\\
9	0.111019542857143\\
};
\addlegendentry{\tiny{HH, $\lambda = 40 \,\mu$s}};

\addplot [color=mycolor7,dashed,line width=1.2pt,mark size=1.0pt,mark=triangle,mark options={solid}]
  table[row sep=crcr]{%
1	0.01646928\\
2	0.01039416\\
3	0.00280704\\
};
\addplot [color=mycolor7,dashed,line width=1.2pt,mark size=1.0pt,mark=triangle,mark options={solid}, forget plot]
  table[row sep=crcr]{%
4	0.00995544\\
5	0.00620112\\
6	0.00164776\\
};
\addplot [color=mycolor7,dashed,line width=1.2pt,mark size=1.0pt,mark=triangle,mark options={solid}, forget plot]
  table[row sep=crcr]{%
7	0.00995544\\
8	0.00620112\\
9	0.00164776\\
};
\addlegendentry{\tiny{HH, $\lambda = 80 \,\mu$s}};

\addplot [color=mycolor8,dashed,line width=1.2pt,mark size=1.0pt,mark=triangle,mark options={solid,rotate=180}]
  table[row sep=crcr]{%
1	0.00043488\\
2	0.00130144\\
3	0.00052144\\
};
\addplot [color=mycolor8,dashed,line width=1.2pt,mark size=1.0pt,mark=triangle,mark options={solid,rotate=180}, forget plot]
  table[row sep=crcr]{%
4	8.032e-05\\
5	0.0007288\\
6	0.00048032\\
};
\addplot [color=mycolor8,dashed,line width=1.2pt,mark size=1.0pt,mark=triangle,mark options={solid,rotate=180}, forget plot]
  table[row sep=crcr]{%
7	8.032e-05\\
8	0.0007288\\
9	0.00048032\\
};
\addlegendentry{\tiny{HH, $\lambda = 160 \,\mu$s}};

\end{axis}
\end{tikzpicture}%

%% file: figures/perf/lost_speed_x2_100_b_10.tex
%
%
\definecolor{mycolor1}{rgb}{0.17073,0.29194,0.77925}%
\definecolor{mycolor2}{rgb}{0.34817,0.74243,0.54727}%
\begin{tikzpicture}
\tikzstyle{every node}=[font=\scriptsize]
\pgfplotsset{every x tick label/.append style={font=\tiny, yshift=0.5ex}}

\begin{axis}[%
width=0.951\fwidth,
height=\fheight,
at={(0\fwidth,0\fheight)},
scale only axis,
scaled ticks=false, tick label style={/pgf/number format/fixed},
xmin=0.5,
xmax=4.5,
xlabel={UE speed [m/s]},
xtick={1,2,3,4},
xticklabels={$2$, $4$, $8$, $16$},
xticklabel style={text width=6, align=center}, xmajorgrids,
xmajorgrids,
ymin=0.04,
ymax=0.16,
ylabel={$R_{lost}$},
ymajorgrids, ylabel shift = -5 pt, yticklabel shift = -3 pt,
axis background/.style={fill=white},
legend style={legend cell align=left,align=left,draw=white!15!black,at={(0.9\fwidth,0.35\fheight)},anchor=east}
]
\addplot [color=mycolor1,solid,line width=1.2pt,mark size=1.1pt,mark=square,mark options={solid}]
  table[row sep=crcr]{%
1	0.0428853777777778\\
2	0.0453078222222222\\
3	0.0549\\
4	0.05769344\\
};
\addlegendentry{DC};

\addplot [color=mycolor2,dashed,line width=1.2pt,mark size=1.0pt,mark=triangle,mark options={solid}]
  table[row sep=crcr]{%
1	0.1131856\\
2	0.119830666666667\\
3	0.13803872\\
4	0.14843328\\
};
\addlegendentry{HH};

\end{axis}
\end{tikzpicture}%

%% file: figures/perf/lost_2_80160.tex
%
%
\definecolor{mycolor3}{rgb}{0.53003,0.74911,0.46611}%
\definecolor{mycolor2}{rgb}{0.97390,0.73140,0.26665}%
\definecolor{mycolor1}{rgb}{0.21162,0.18978,0.57768}%
\definecolor{mycolor4}{rgb}{0.07227,0.48867,0.84670}%
\begin{tikzpicture}
\tikzstyle{every node}=[font=\scriptsize]
\pgfplotsset{every x tick label/.append style={font=\tiny, yshift=0.5ex}}
\begin{axis}[%
width=0.951\fwidth,
height=\fheight,
at={(0\fwidth,0\fheight)},
scale only axis,
xmin=0,
xmax=10,
xlabel={X2/buffer configurations},
xtick={1,2,3,4,5,6,7,8,9,10},
xticklabels={$0.1$ $1$, $1$ $1$, $10$ $1$, $0.1$ $10$, $1$ $10$, $10$ $10$, $0.1$ $100$, $1$ $100$, $10$ $100$, ms MB},
xticklabel style={text width=6, align=center}, xmajorgrids,
ymin=0,
ymax=0.03,
ylabel={$R_{lost}$},
ymajorgrids, ylabel shift = -5 pt, yticklabel shift = -3 pt,
axis background/.style={fill=white},
legend style={legend cell align=left,align=left,draw=white!15!black}
]
\addplot [color=mycolor1,solid,line width=1.2pt,mark size=0.5pt,mark=*,mark options={solid}]
  table[row sep=crcr]{%
1	0\\
2	0\\
3	0.00063216\\
};
\addplot [color=mycolor1,solid,line width=1.2pt,mark size=0.5pt,mark=*,mark options={solid}, forget plot]
  table[row sep=crcr]{%
4	0\\
5	0\\
6	8.536e-05\\
};
\addplot [color=mycolor1,solid,line width=1.2pt,mark size=0.5pt,mark=*,mark options={solid}, forget plot]
  table[row sep=crcr]{%
7	0\\
8	0\\
9	8.536e-05\\
};
\addlegendentry{DC, $\lambda = 80 \,\,\mu$s};


\addplot [color=mycolor3,dashed,line width=1.2pt,mark size=1.1pt,mark=square,mark options={solid}]
  table[row sep=crcr]{%
1	0.01646928\\
2	0.01039416\\
3	0.00280704\\
};
\addplot [color=mycolor3,dashed,line width=1.2pt,mark size=1.1pt,mark=square,mark options={solid}, forget plot]
  table[row sep=crcr]{%
4	0.00995544\\
5	0.00620112\\
6	0.00164776\\
};
\addplot [color=mycolor3,dashed,line width=1.2pt,mark size=1.1pt,mark=square,mark options={solid}, forget plot]
  table[row sep=crcr]{%
7	0.00995544\\
8	0.00620112\\
9	0.00164776\\
};
\addlegendentry{HH, $\lambda = 80 \,\,\mu$s};


\end{axis}
\end{tikzpicture}%

%% file: figures/perf/lost_um_2.tex
%
%
\definecolor{mycolor1}{rgb}{0.21162,0.18978,0.57768}%
\definecolor{mycolor2}{rgb}{0.53003,0.74911,0.46611}%
\begin{tikzpicture}
\tikzstyle{every node}=[font=\scriptsize]
\pgfplotsset{every x tick label/.append style={font=\tiny, yshift=0.5ex}}

\begin{axis}[%
width=0.951\fwidth,
height=\fheight,
at={(0\fwidth,0\fheight)},
scale only axis,
xmin=0,
xmax=10,
xlabel={X2/buffer configurations},
xtick={1,2,3,4,5,6,7,8,9,10},
xticklabels={$0.1$ $1$, $1$ $1$, $10$ $1$, $0.1$ $10$, $1$ $10$, $10$ $10$, $0.1$ $100$, $1$ $100$, $10$ $100$, ms MB},
xticklabel style={text width=6, align=center}, xmajorgrids,
ymin=0,
ymax=0.03,
ylabel={$R_{lost}$},
ymajorgrids, ylabel shift = -5 pt, yticklabel shift = -3 pt,
axis background/.style={fill=white},
legend style={legend cell align=left,align=left,draw=white!15!black,at={(0.92\fwidth,0.98\fheight)},anchor=south east}
]
\addplot [color=mycolor1,solid,line width=1.2pt,mark size=0.5pt,mark=*,mark options={solid}]
  table[row sep=crcr]{%
1	0.00101957333333333\\
2	0.00103909333333333\\
3	0.0137476266666667\\
};
\addplot [color=mycolor1,solid,line width=1.2pt,mark size=0.5pt,mark=*,mark options={solid}, forget plot]
  table[row sep=crcr]{%
4	0.000532693333333333\\
5	0.000527946666666667\\
6	0.01123664\\
};
\addplot [color=mycolor1,solid,line width=1.2pt,mark size=0.5pt,mark=*,mark options={solid}, forget plot]
  table[row sep=crcr]{%
7	0.000532693333333333\\
8	0.000527946666666667\\
9	0.01123664\\
};
\addlegendentry{DC, $\lambda = 80 \, \,\mu$s};

\addplot [color=mycolor2,dashed,line width=1.2pt,mark size=1.0pt,mark=square,mark options={solid}]
  table[row sep=crcr]{%
1	0.0290145066666667\\
2	0.0232667733333333\\
3	0.00889514666666667\\
};
\addplot [color=mycolor2,dashed,line width=1.2pt,mark size=1.0pt,mark=square,mark options={solid}, forget plot]
  table[row sep=crcr]{%
4	0.028572\\
5	0.02279616\\
6	0.00729072\\
};
\addplot [color=mycolor2,dashed,line width=1.2pt,mark size=1.0pt,mark=square,mark options={solid}, forget plot]
  table[row sep=crcr]{%
7	0.028572\\
8	0.02279616\\
9	0.00729072\\
};
\addlegendentry{HH, $\lambda = 80 \, \,\mu$s};

\end{axis}
\end{tikzpicture}%

%% file: figures/perf/lat_2_80160.tex
%
%
\definecolor{mycolor3}{rgb}{0.53003,0.74911,0.46611}%
\definecolor{mycolor4}{rgb}{0.97390,0.73140,0.26665}%
\definecolor{mycolor1}{rgb}{0.21162,0.18978,0.57768}%
\definecolor{mycolor2}{rgb}{0.07227,0.48867,0.84670}%
\begin{tikzpicture}
\tikzstyle{every node}=[font=\scriptsize]
\pgfplotsset{every x tick label/.append style={font=\tiny, yshift=0.5ex}}
\begin{axis}[%
width=0.951\fwidth,
height=\fheight,
at={(0\fwidth,0\fheight)},
scale only axis,
xmin=0,
xmax=10,
xlabel={X2/buffer configurations},
xtick={1,2,3,4,5,6,7,8,9,10},
xticklabels={$0.1$ $1$, $1$ $1$, $10$ $1$, $0.1$ $10$, $1$ $10$, $10$ $10$, $0.1$ $100$, $1$ $100$, $10$ $100$, ms MB},
xticklabel style={text width=6, align=center}, xmajorgrids,
ymin=0,
ymax=0.02,
ylabel={Latency [s]},
ymajorgrids, ylabel shift = -5 pt, yticklabel shift = -3 pt,
axis background/.style={fill=white},
legend columns=2,
legend style={legend cell align=left,align=left,draw=white!15!black,at={(1.02\fwidth,0.93\fheight)},anchor=south east}
]
\addplot [color=mycolor1,solid,line width=1.2pt,mark size=0.5pt,mark=*,mark options={solid}]
  table[row sep=crcr]{%
1	0.00495849990507489\\
2	0.0049164562162578\\
3	0.00879526921531602\\
};
\addplot [color=mycolor1,solid,line width=1.2pt,mark size=0.5pt,mark=*,mark options={solid}, forget plot]
  table[row sep=crcr]{%
4	0.00508514681164273\\
5	0.00505032370873645\\
6	0.00951282037443823\\
};
\addplot [color=mycolor1,solid,line width=1.2pt,mark size=0.5pt,mark=*,mark options={solid}, forget plot]
  table[row sep=crcr]{%
7	0.00508514681164273\\
8	0.00505032370873645\\
9	0.00951282037443823\\
};
\addlegendentry{DC, $\lambda = 80 \,\mu$s};

\addplot [color=mycolor2,solid,line width=1.2pt,mark size=1.5pt,mark=x,mark options={solid}]
  table[row sep=crcr]{%
1	0.00124510184744398\\
2	0.00113183338876704\\
3	0.00179892757222714\\
};
\addplot [color=mycolor2,solid,line width=1.2pt,mark size=1.5pt,mark=asterisk,mark options={solid}, forget plot]
  table[row sep=crcr]{%
4	0.00124510184744398\\
5	0.00113183338876704\\
6	0.00179892757222714\\
};
\addplot [color=mycolor2,solid,line width=1.2pt,mark size=1.5pt,mark=asterisk,mark options={solid}, forget plot]
  table[row sep=crcr]{%
7	0.00124510184744398\\
8	0.00113183338876704\\
9	0.00179892757222714\\
};
\addlegendentry{DC, $\lambda = 160 \,\mu$s};

\addplot [color=mycolor3,dashed,line width=1.2pt,mark size=1.1pt,mark=square,mark options={solid}]
  table[row sep=crcr]{%
1	0.0162627269598343\\
2	0.016696331379154\\
3	0.0100738172875013\\
};
\addplot [color=mycolor3,dashed,line width=1.2pt,mark size=1.1pt,mark=square,mark options={solid}, forget plot]
  table[row sep=crcr]{%
4	0.017184224742292\\
5	0.0180554996852624\\
6	0.0107070659300479\\
};
\addplot [color=mycolor3,dashed,line width=1.2pt,mark size=1.1pt,mark=square,mark options={solid}, forget plot]
  table[row sep=crcr]{%
7	0.017184224742292\\
8	0.0180554996852624\\
9	0.0107070659300479\\
};
\addlegendentry{HH, $\lambda = 80 \,\mu$s};

\addplot [color=mycolor4,dashed,line width=1.2pt,mark size=2.6pt,mark=diamond,mark options={solid}]
  table[row sep=crcr]{%
1	0.013225522952361\\
2	0.0120080800075429\\
3	0.00321559751455657\\
};
\addplot [color=mycolor4,dashed,line width=1.2pt,mark size=1.5pt,mark=x,mark options={solid}, forget plot]
  table[row sep=crcr]{%
4	0.0134854275792661\\
5	0.012079701913355\\
6	0.0032224125746326\\
};
\addplot [color=mycolor4,dashed,line width=1.2pt,mark size=1.5pt,mark=x,mark options={solid}, forget plot]
  table[row sep=crcr]{%
7	0.0134854275792661\\
8	0.012079701913355\\
9	0.0032224125746326\\
};
\addlegendentry{HH, $\lambda = 160 \,\mu$s};

\end{axis}
\end{tikzpicture}%

%% file: figures/perf/lat_um_801602.tex
%
%
\definecolor{mycolor3}{rgb}{0.53003,0.74911,0.46611}%
\definecolor{mycolor4}{rgb}{0.97390,0.73140,0.26665}%
\definecolor{mycolor1}{rgb}{0.21162,0.18978,0.57768}%
\definecolor{mycolor2}{rgb}{0.07227,0.48867,0.84670}%
\begin{tikzpicture}
\tikzstyle{every node}=[font=\scriptsize]
\pgfplotsset{every x tick label/.append style={font=\tiny, yshift=0.5ex}}

\begin{axis}[%
width=0.951\fwidth,
height=\fheight,
at={(0\fwidth,0\fheight)},
scale only axis,
xmin=0,
xmax=10,
xlabel={X2/buffer configurations},
xtick={1,2,3,4,5,6,7,8,9,10},
xticklabels={$0.1$ $1$, $1$ $1$, $10$ $1$, $0.1$ $10$, $1$ $10$, $10$ $10$, $0.1$ $100$, $1$ $100$, $10$ $100$, ms MB},
xticklabel style={text width=6, align=center}, xmajorgrids,
ymin=0.001,
ymax=0.01,
ylabel={Latency [s]},
ymajorgrids, ylabel shift = -5 pt, yticklabel shift = -3 pt,
axis background/.style={fill=white},
legend columns=2,
legend style={legend cell align=left,align=left,draw=white!15!black,at={(1.02\fwidth,0.93\fheight)},anchor=south east}
]
\addplot [color=mycolor1,solid,line width=1.2pt,mark size=0.5pt,mark=*,mark options={solid}]
  table[row sep=crcr]{%
1	0.00381322575081439\\
2	0.00397862862347171\\
3	0.00852922990347389\\
};
\addplot [color=mycolor1,solid,line width=1.2pt,mark size=0.5pt,mark=*,mark options={solid}, forget plot]
  table[row sep=crcr]{%
4	0.00389776974733884\\
5	0.00407261037209189\\
6	0.00924819525038717\\
};
\addplot [color=mycolor1,solid,line width=1.2pt,mark size=0.5pt,mark=*,mark options={solid}, forget plot]
  table[row sep=crcr]{%
7	0.00389776974733884\\
8	0.00407261037209189\\
9	0.00924819525038717\\
};
\addlegendentry{DC, $\lambda = 80 \, \,\mu$s};

\addplot [color=mycolor2,solid,line width=1.2pt,mark size=1.5pt,mark=x,mark options={solid}]
  table[row sep=crcr]{%
1	0.00105390191532909\\
2	0.00105136796532035\\
3	0.00184049307107087\\
};
\addplot [color=mycolor2,solid,line width=1.2pt,mark size=1.5pt,mark=x,mark options={solid}, forget plot]
  table[row sep=crcr]{%
4	0.00105390191532909\\
5	0.00105136796532035\\
6	0.00184049307107087\\
};
\addplot [color=mycolor2,solid,line width=1.2pt,mark size=1.5pt,mark=x,mark options={solid}, forget plot]
  table[row sep=crcr]{%
7	0.00105390191532909\\
8	0.00105136796532035\\
9	0.00184049307107087\\
};
\addlegendentry{DC, $\lambda = 160 \, \,\mu$s};

\addplot [color=mycolor3,dashed,line width=1.2pt,mark size=1.1pt,mark=square,mark options={solid}]
  table[row sep=crcr]{%
1	0.00496646757615594\\
2	0.00512148911481464\\
3	0.00815785136990895\\
};
\addplot [color=mycolor3,dashed,line width=1.2pt,mark size=1.1pt,mark=square,mark options={solid}, forget plot]
  table[row sep=crcr]{%
4	0.00502183451679203\\
5	0.00518425929947802\\
6	0.00860014667980409\\
};
\addplot [color=mycolor3,dashed,line width=1.2pt,mark size=1.1pt,mark=square,mark options={solid}, forget plot]
  table[row sep=crcr]{%
7	0.00502183451679203\\
8	0.00518425929947802\\
9	0.00860014667980409\\
};
\addlegendentry{HH, $\lambda = 80 \, \,\mu$s};

\addplot [color=mycolor4,dashed,line width=1.2pt,mark size=2.6pt,mark=diamond,mark options={solid}]
  table[row sep=crcr]{%
1	0.00315738712045772\\
2	0.00309177532337623\\
3	0.00253332751929033\\
};
\addplot [color=mycolor4,dashed,line width=1.2pt,mark size=2.6pt,mark=diamond,mark options={solid}, forget plot]
  table[row sep=crcr]{%
4	0.00315738712045772\\
5	0.00309177532337623\\
6	0.00253875125083396\\
};
\addplot [color=mycolor4,dashed,line width=1.2pt,mark size=2.6pt,mark=diamond,mark options={solid}, forget plot]
  table[row sep=crcr]{%
7	0.00315738712045772\\
8	0.00309177532337623\\
9	0.00253875125083396\\
};
\addlegendentry{HH, $\lambda = 160 \, \,\mu$s};

\end{axis}
\end{tikzpicture}%

%% file: figures/perf/latDiff_a_80160_b_10_x_1.tex
%
%
\definecolor{mycolor1}{rgb}{0.53003,0.74911,0.46611}%
\definecolor{mycolor2}{rgb}{0.11329,0.7015,0.685871}%
\definecolor{mycolor1a}{rgb}{0.2081,0.2386,0.67708}%
\definecolor{mycolor2a}{rgb}{0.683419,0.7434762,0.404433}%
\begin{tikzpicture}
\tikzstyle{every node}=[font=\scriptsize]
\pgfplotsset{every x tick label/.append style={font=\tiny, yshift=0.5ex}}

\begin{axis}[%
width=0.951\fwidth,
height=\fheight,
at={(0\fwidth,0\fheight)},
scale only axis,
xmin=0.5,
xmax=4.5,
xlabel={UE speed $s$ [m/s]},
xmajorgrids,
xtick={1,2,3,4},
xticklabels={$2$, $4$, $8$, $16$},
ymin=0.001,
ymax=0.019,
ylabel={$L_{HH} - L_{DC}$ [s]},
ymajorgrids, ylabel shift = -5 pt, yticklabel shift = -3 pt,
axis background/.style={fill=white},
legend style={legend cell align=left,align=left,draw=white!15!black,at={(0.92\fwidth,0.4\fheight)},anchor=east}
]
\addplot [color=mycolor1,solid,line width=1.2pt,mark=*,mark options={solid}]
  table[row sep=crcr]{%
1	0.013005175976526\\
2	0.0138851951288641\\
3	0.0186929970411252\\
4	0.0182907320894984\\
};
\addlegendentry{$\lambda = 80\,\mu$s, RLC AM};

\addplot [color=mycolor2,solid,line width=1.2pt,mark=x,mark options={solid}]
  table[row sep=crcr]{%
1	0.0109478685245879\\
2	0.0120956249574184\\
3	0.01455067466465\\
4	0.0142311704114832\\
};
\addlegendentry{$\lambda = 160\,\mu$s, RLC AM};

\addplot [color=mycolor1a,dashdotted,line width=1.2pt,mark size=0.5pt,mark=*,mark options={solid}]
  table[row sep=crcr]{%
1	0.00111164892738613\\
2	0.00130997292818425\\
3	0.00168291544194881\\
4	0.00162639711769885\\
};
\addlegendentry{$\lambda = 80\,\mu$s, RLC UM};

\addplot [color=mycolor2a,dashdotted,line width=1.2pt,mark size=1.5pt,mark=x,mark options={solid}]
  table[row sep=crcr]{%
1	0.00204040735805588\\
2	0.00221094983507366\\
3	0.00243398928194831\\
4	0.00256451267380676\\
};
\addlegendentry{$\lambda = 160\,\mu$s, RLC UM};

\end{axis}
\end{tikzpicture}%

%% file: figures/perf/latCDF_a_80160_b_10_x_100.tex
%
%
\definecolor{mycolor1}{rgb}{0.21162,0.18978,0.57768}%
\definecolor{mycolor2}{rgb}{0.68342,0.74348,0.40443}%
\definecolor{mycolor3}{rgb}{0.07227,0.48867,0.84670}%
\definecolor{mycolor4}{rgb}{0.98800,0.80660,0.17937}%
\begin{tikzpicture}
\tikzstyle{every node}=[font=\scriptsize]
\pgfplotsset{every x tick label/.append style={font=\tiny, yshift=0.5ex}}

\begin{axis}[%
width=0.951\fwidth,
height=\fheight,
at={(0\fwidth,0\fheight)},
scale only axis,
unbounded coords=jump,
xmin=-3.5,
xmax=0,
xlabel={$\log_{10}(L)$ [s]},
xmajorgrids,
ymin=0,
ymax=1,
ylabel={CDF},
ymajorgrids, ylabel shift = -5 pt, yticklabel shift = -3 pt,
axis background/.style={fill=white},
axis x line*=bottom,
axis y line*=left,
legend style={legend cell align=left,align=left,draw=white!15!black,at={(0.9\fwidth,0.05\fheight)},anchor=south east}
]

\addplot [color=mycolor2,dashed,line width=1.2pt,forget plot]
  table[row sep=crcr]{%
-3.31926515557629 0.00226654578422458\\
-3.31081057851158 0.0826030019139717\\
-3.30823289590381 0.115341996574996\\
-3.30624233154265 0.14349753198348\\
-3.30447394368328 0.171401228971495\\
-3.3027859502566  0.269064168429538\\
-3.3008664491337  0.297723380678958\\
-3.29901015345474 0.333484436385617\\
-3.29715560582813 0.360985191900879\\
-3.29530807173392 0.387730432154731\\
-3.29326305110928 0.414526040092677\\
-3.2910956516838  0.441926060239751\\
-3.28804896318486 0.468217991336756\\
-3.28428367953245 0.543165105268457\\
-3.27899364798869 0.584365870857257\\
-3.27074146467376 0.609650448272389\\
-3.26088157140875 0.634985393371615\\
-3.24838854897009 0.660219603102653\\
-3.22553125999782 0.720660824015316\\
-2.26095569564525 0.759897249924454\\
-1.87580394605778 0.786995063966964\\
-1.62964242600444 0.813891407273099\\
-1.45922402569256 0.841392162788362\\
-1.32248795186532 0.867583358517183\\
-1.21086665901373 0.892918303616407\\
-1.11649251404146 0.918303616399724\\
-1.01328103156118 0.943789664551228\\
-0.880988867437579  0.969074241966356\\
-0.651193122929571  0.994308451697391\\
};
\addplot [color=mycolor2,line width=1.2pt,only marks,mark=asterisk,mark size=1.5,mark options={solid},forget plot]
  table[row sep=crcr]{%
-3.31926515557629 0.00226654578422458\\
-3.30823289590381 0.115341996574996\\
-3.3008664491337  0.297723380678958\\
-3.29326305110928 0.414526040092677\\
-3.27899364798869 0.584365870857257\\
-3.27074146467376 0.609650448272389\\
-3.26088157140875 0.634985393371615\\
-3.24838854897009 0.660219603102653\\
-3.22553125999782 0.720660824015316\\
-2.26095569564525 0.759897249924454\\
-1.87580394605778 0.786995063966964\\
-1.62964242600444 0.813891407273099\\
-1.45922402569256 0.841392162788362\\
-1.32248795186532 0.867583358517183\\
-1.21086665901373 0.892918303616407\\
-1.11649251404146 0.918303616399724\\
-1.01328103156118 0.943789664551228\\
-0.880988867437579  0.969074241966356\\
-0.651193122929571  0.994308451697391\\
};
\addplot [color=mycolor2,dashed,line width=1.2pt,mark=asterisk,mark size=1.5,mark options={solid}]
  table[row sep=crcr]{%
-3.31926515557629 0.00226654578422458\\
};
\addlegendentry{HH, $\lambda = 80 \,\mu$s};

\addplot [color=mycolor4,dashed,line width=1.2pt,forget plot]
  table[row sep=crcr]{%
-3.32274111668722 0.00261938343743717\\
-3.31734866924602 0.0907213379004628\\
-3.31520157672288 0.123010276042712\\
-3.31343889213735 0.154745113842427\\
-3.31187415924902 0.184616159580887\\
-3.31035605887296 0.28903888776948\\
-3.30878381106813 0.318053596614932\\
-3.30725247815363 0.349234334072111\\
-3.30556111133683 0.389784404593968\\
-3.30375987718802 0.419655450332429\\
-3.30186731176276 0.448821277453118\\
-3.29977659791337 0.530928873665088\\
-3.29714812551347 0.559943582510545\\
-3.29367242570215 0.602609308885724\\
-3.28944198884834 0.6300624622204\\
-3.2846550542225  0.657213379004603\\
-3.27789872685102 0.68290348579485\\
-3.24633775922211 0.754483175498661\\
-2.38729747845076 0.809691718718495\\
-2.02845377061742 0.840267983074734\\
-1.72315207062775 0.868476727785598\\
-1.31934401404871 0.894720934918384\\
-1.20172035365817 0.919957686882924\\
-1.05003048479914 0.946353012290947\\
-0.944041878124999  0.973906911142451\\
-0.160139586824741  0.999294781382228\\
-inf  0\\
-inf  0\\
-inf  0\\
};
\addplot [color=mycolor4,line width=1.2pt,only marks,mark size=1.5,mark=diamond,forget plot]
  table[row sep=crcr]{%
-3.32274111668722 0.00261938343743717\\
-3.29367242570215 0.602609308885724\\
-3.28944198884834 0.6300624622204\\
-3.27789872685102 0.68290348579485\\
-3.24633775922211 0.754483175498661\\
-2.38729747845076 0.809691718718495\\
-2.02845377061742 0.840267983074734\\
-1.72315207062775 0.868476727785598\\
-1.31934401404871 0.894720934918384\\
-1.20172035365817 0.919957686882924\\
-1.05003048479914 0.946353012290947\\
-0.944041878124999  0.973906911142451\\
-0.160139586824741  0.999294781382228\\
};
\addplot [color=mycolor4,dashed,line width=1.2pt,mark size=1.5,mark=diamond]
  table[row sep=crcr]{%
-3.32274111668722 0.00261938343743717\\
};
\addlegendentry{HH, $\lambda = 160 \,\mu$s};

\addplot [color=mycolor1,solid,line width=1.2pt,forget plot]
  table[row sep=crcr]{%
-3.31992234727061	0.00211172004625682\\
-3.31147103428686	0.085172708532355\\
-3.30906902805243	0.118105485444216\\
-3.30713416598028	0.146865101312284\\
-3.30534373454726	0.174820252400826\\
-3.30366693080227	0.278445371813564\\
-3.30195708098337	0.306249685755945\\
-3.30002784222378	0.343104228467997\\
-3.2983066246876	0.370757705264214\\
-3.29642693955051	0.397807833475787\\
-3.29455404048891	0.424506008346317\\
-3.29247045532881	0.451153904168126\\
-3.29012934312387	0.478053195233536\\
-3.28733260375809	0.560611393232431\\
-3.28366364679597	0.604907235155104\\
-3.27811753445276	0.631002061440993\\
-3.2701771113186	0.656342701996076\\
-3.26195995812091	0.681582784453719\\
-3.25213925652039	0.706873145960081\\
-3.23531020602772	0.764895168183423\\
-2.80029244648459	0.805218965257182\\
-2.27353477611897	0.831615465835392\\
-1.99725829314582	0.857006385439194\\
-1.84226097905469	0.88284981648148\\
-1.70143508004105	0.90964854944945\\
-1.56188241438522	0.936145608125098\\
-1.42265309011642	0.966463874503497\\
-1.14381774351165	0.992860375081704\\
};
\addplot [color=mycolor1,line width=1.2pt,only marks,mark=+,mark size=1.5,mark options={solid},forget plot]
  table[row sep=crcr]{%
-3.31992234727061	0.00211172004625682\\
-3.30906902805243	0.118105485444216\\
-3.2983066246876	0.370757705264214\\
-3.29012934312387	0.478053195233536\\
-3.28733260375809	0.560611393232431\\
-3.28366364679597	0.604907235155104\\
-3.27811753445276	0.631002061440993\\
-3.2701771113186	0.656342701996076\\
-3.26195995812091	0.681582784453719\\
-3.25213925652039	0.706873145960081\\
-3.23531020602772	0.764895168183423\\
-2.80029244648459	0.805218965257182\\
-2.80029244648459	0.805218965257182\\
-2.27353477611897	0.831615465835392\\
-1.99725829314582	0.857006385439194\\
-1.84226097905469	0.88284981648148\\
-1.70143508004105	0.90964854944945\\
-1.56188241438522	0.936145608125098\\
-1.42265309011642	0.966463874503497\\
-1.14381774351165	0.992860375081704\\
};
\addplot [color=mycolor1,solid,line width=1.2pt,mark=+,mark size=1.5,mark options={solid}]
  table[row sep=crcr]{%
-3.31992234727061	0.00211172004625682\\
};
\addlegendentry{DC, $\lambda = 80 \,\mu$s};

\addplot [color=mycolor3,solid,line width=1.2pt,forget plot]
  table[row sep=crcr]{%
-3.32351462121305 0.00191069991954951\\
-3.31800861097888 0.0925181013676604\\
-3.3159363330115  0.126005631536607\\
-3.31424319968799 0.15833668543846\\
-3.31264449376371 0.18790225261465\\
-3.31120437196323 0.297013274336294\\
-3.30972517247784 0.324869267900255\\
-3.30827938429564 0.356898632341127\\
-3.30669442181228 0.397676991150461\\
-3.30520576917198 0.428298471440085\\
-3.30359980462291 0.45705953338699\\
-3.3017501837444  0.485770313757065\\
-3.29974114227563 0.581355591311364\\
-3.29721308886368 0.610870876910717\\
-3.29421222793502 0.655973451327446\\
-3.29071428876359 0.683678600160909\\
-3.28696999341139 0.712389380530977\\
-3.28241534893481 0.740144810941269\\
-3.27562201029281 0.804253821399836\\
-2.99359220937524 0.849306114239741\\
-2.62891242824738 0.878771118262268\\
-2.43812988094413 0.943282381335478\\
-2.22425361039074 0.972998793242156\\
-1.37020256959657 0.998742960579244\\
-inf  0\\
-inf  0\\
-inf  0\\
-inf  0\\
};
\addplot [color=mycolor3,line width=1.2pt,only marks,mark=o,mark size=1.5,mark options={solid},forget plot]
  table[row sep=crcr]{%
-3.32351462121305 0.00191069991954951\\
-3.3159363330115  0.126005631536607\\
-3.30972517247784 0.324869267900255\\
-3.30669442181228 0.397676991150461\\
-3.30359980462291 0.45705953338699\\
-3.3017501837444  0.485770313757065\\
-3.29721308886368 0.610870876910717\\
-3.29421222793502 0.655973451327446\\
-3.29071428876359 0.683678600160909\\
-3.28696999341139 0.712389380530977\\
-3.28241534893481 0.740144810941269\\
-3.27562201029281 0.804253821399836\\
-2.99359220937524 0.849306114239741\\
-2.62891242824738 0.878771118262268\\
-2.62891242824738 0.878771118262268\\
-2.43812988094413 0.943282381335478\\
-2.22425361039074 0.972998793242156\\
-1.37020256959657 0.998742960579244\\
};
\addplot [color=mycolor3,solid,line width=1.2pt,mark=o,mark size=1.5,mark options={solid}]
  table[row sep=crcr]{%
-3.32351462121305 0.00191069991954951\\
};
\addlegendentry{DC, $\lambda = 160 \,\mu$s};

\end{axis}
\end{tikzpicture}%

%% file: figures/perf/latCDF_a_80160_b_10_x_1000.tex
%
%
\definecolor{mycolor1}{rgb}{0.21162,0.18978,0.57768}%
\definecolor{mycolor2}{rgb}{0.68342,0.74348,0.40443}%
\definecolor{mycolor3}{rgb}{0.07227,0.48867,0.84670}%
\definecolor{mycolor4}{rgb}{0.98800,0.80660,0.17937}%
\begin{tikzpicture}
\tikzstyle{every node}=[font=\scriptsize]
\pgfplotsset{every x tick label/.append style={font=\tiny, yshift=0.5ex}}

\begin{axis}[%
width=0.951\fwidth,
height=\fheight,
at={(0\fwidth,0\fheight)},
scale only axis,
unbounded coords=jump,
xmin=-3.5,
xmax=-0.5,
xlabel={$\log_{10}(L)$ [s]}, xmajorgrids,
ymin=0,
ymax=1,
ylabel={CDF},
ymajorgrids, ylabel shift = -5 pt, yticklabel shift = -3 pt,
axis background/.style={fill=white},
axis x line*=bottom,
axis y line*=left,
legend style={legend cell align=left,align=left,draw=white!15!black,at={(0.9\fwidth,0.05\fheight)},anchor=south east}
]
\addplot [color=mycolor2,dashed,line width=1.2pt,forget plot]
  table[row sep=crcr]{%
-3.31927118486104	0.00211832349826024\\
-3.31507562548314	0.00726282342260609\\
-3.31322386397163	0.0123568870731839\\
-3.31231074860569	0.0711655822867822\\
-3.31140228185215	0.078327533161852\\
-3.31070211116921	0.0839763958238791\\
-3.3101799371028	0.0906844202350355\\
-3.30969695134773	0.0965350279921344\\
-3.30922288600519	0.103495233772131\\
-3.30864274149963	0.109598022898071\\
-3.30825957385777	0.116961718868213\\
-3.30778187966196	0.123114944267921\\
-3.30736382016649	0.128662934382412\\
-3.30699169763524	0.134765723508352\\
-3.30660412721708	0.140212841075307\\
-3.30619935034778	0.145861703737334\\
-3.30581181160982	0.151460130125593\\
-3.30549785075385	0.157008120240084\\
-3.30509948999843	0.162606546628343\\
-3.30468422766725	0.168104100469067\\
-3.30438452538386	0.173399909214717\\
-3.30410272048859	0.179401825793121\\
-3.30374770062069	0.251626569829032\\
-3.30329248122004	0.257124123669756\\
-3.3030094906803	0.265900035305405\\
-3.30264393939536	0.271246280324824\\
-3.30221915929915	0.276592525344243\\
-3.30182632499508	0.281837897816126\\
-3.30149675510918	0.288192868310908\\
-3.30113407204024	0.293791294699167\\
-3.30074324752986	0.299793211277571\\
-3.30033012974386	0.305089020023222\\
-3.29996464836727	0.310334392495105\\
-3.29961853469144	0.324355676602636\\
-3.29925599485014	0.329550612800751\\
-3.29886324469939	0.334795985272634\\
-3.29849915878012	0.340192666565822\\
-3.29816318037004	0.345589347859009\\
-3.297834852452	0.350834720330892\\
-3.29745755377423	0.356584455540456\\
-3.29710564063069	0.361981136833643\\
-3.29677759160129	0.367226509305526\\
-3.2964080462201	0.372421445503641\\
-3.2960504655273	0.377868563070596\\
-3.29566067305582	0.383214808090015\\
-3.29521139571561	0.38840974428813\\
-3.29477275248388	0.393806425581318\\
-3.29433986722149	0.399051798053201\\
-3.29394242363516	0.404549351893924\\
-3.2935592293757	0.409845160639575\\
-3.29305953374131	0.41504009683769\\
-3.29263599069395	0.420184596762037\\
-3.29224619873909	0.42568215060276\\
-3.29175833243799	0.431129268169716\\
-3.29131216644962	0.437232057295656\\
-3.29084383908689	0.442275684672467\\
-3.29032633285029	0.447319312049277\\
-3.28972970888678	0.452615120794928\\
-3.28912820243583	0.457709184445507\\
-3.28850242163682	0.462904120643622\\
-3.28788063564827	0.468099056841737\\
-3.28716826950562	0.515912644373885\\
-3.28645698192199	0.525192918747213\\
-3.2856939577408	0.53028698239779\\
-3.28486751207787	0.536440207797497\\
-3.28394107022895	0.542240379280827\\
-3.28312989812647	0.562414888788062\\
-3.28210836595987	0.567710697533712\\
-3.28118476847541	0.572855197458057\\
-3.27971010692089	0.577949261108634\\
-3.27833597855671	0.58309376103298\\
-3.27677205320034	0.588137388409789\\
-3.27510689430979	0.593181015786598\\
-3.27318653154578	0.598325515710943\\
-3.27142469500514	0.603369143087752\\
-3.26935072599051	0.608463206738329\\
-3.2674681114623	0.613658142936442\\
-3.26526380121348	0.618701770313251\\
-3.26313317807489	0.62374539769006\\
-3.26073178397144	0.628839461340637\\
-3.25860729356449	0.633933524991215\\
-3.25590673496227	0.639027588641792\\
-3.25335095747315	0.644172088566138\\
-3.25082023456826	0.649266152216715\\
-3.24639220514458	0.654309779593524\\
-3.24160791735328	0.659353406970334\\
-3.23681014895692	0.664447470620911\\
-3.23290937356082	0.6924396025622\\
-3.22985933197976	0.705956523932048\\
-3.21679004891301	0.714026327734942\\
-3.08230402525382	0.719069955111752\\
-2.80090998735763	0.724113582488561\\
-2.59599996117331	0.731376405911167\\
-2.42249435061433	0.739849699904206\\
-2.26765950583685	0.745548998840001\\
-2.13423210642274	0.750844807585651\\
-2.01747036903169	0.755938871236227\\
-1.95295297552411	0.76113380743434\\
-1.89319904478578	0.76668179754883\\
-1.83842183986146	0.772078478842016\\
-1.79755044248534	0.777525596408969\\
-1.75545214514038	0.783124022797227\\
-1.71987888926773	0.788520704090412\\
-1.68470732049156	0.794119130478669\\
-1.63213968806968	0.799213194129246\\
-1.59044413094406	0.804710747969968\\
-1.55241839507189	0.80990568416808\\
-1.51986185446377	0.815251929187497\\
-1.48683219254553	0.820951228123291\\
-1.45518514463234	0.82624703686894\\
-1.42848905215014	0.831441973067052\\
-1.40456869558205	0.836838654360237\\
-1.37843123534357	0.841983154284582\\
-1.35611909416117	0.847077217935159\\
-1.3321454760445	0.852120845311967\\
-1.31132703354854	0.857164472688775\\
-1.29293232923735	0.862208100065584\\
-1.27362681864782	0.86730216371616\\
-1.25792523858175	0.872396227366737\\
-1.23673288019694	0.877439854743545\\
-1.21512365923264	0.882483482120354\\
-1.19858251923636	0.887527109497163\\
-1.17923368504814	0.892570736873972\\
-1.16180621615325	0.897614364250781\\
-1.14530612440493	0.902708427901358\\
-1.1320234372466	0.907752055278167\\
-1.11978075667558	0.912846118928744\\
-1.10719477110344	0.917889746305553\\
-1.08991679406058	0.922933373682362\\
-1.07256348652648	0.927977001059171\\
-1.05436502628226	0.933121500983516\\
-1.033028139731	0.938165128360325\\
-1.01460499849025	0.943208755737134\\
-0.994954210563026	0.948252383113942\\
-0.96762417406592	0.953296010490751\\
-0.942877970822521	0.95833963786756\\
-0.914082055837792	0.963383265244368\\
-0.887963363352344	0.968426892621177\\
-0.856554977970557	0.973470519997986\\
-0.826077632849926	0.978615019922331\\
-0.791561310990506	0.983709083572908\\
-0.741627214704893	0.988752710949716\\
-0.665939482018739	0.993796338326525\\
-0.518845122444278	0.998839965703334\\
};
\addplot [color=mycolor2,line width=1.2pt,only marks,mark=asterisk,mark size=1.5,mark options={solid},forget plot]
  table[row sep=crcr]{%
-3.30264393939536	0.271246280324824\\
-3.28486751207787	0.536440207797497\\
-3.27142469500514	0.603369143087752\\
-3.26073178397144	0.628839461340637\\
-3.24639220514458	0.654309779593524\\
-3.21679004891301	0.714026327734942\\
-2.26765950583685	0.745548998840001\\
-1.83842183986146	0.772078478842016\\
-1.63213968806968	0.799213194129246\\
-1.48683219254553	0.820951228123291\\
-1.35611909416117	0.847077217935159\\
-1.25792523858175	0.872396227366737\\
-1.16180621615325	0.897614364250781\\
-1.08991679406058	0.922933373682362\\
-0.994954210563026	0.948252383113942\\
-0.856554977970557	0.973470519997986\\
-0.518845122444278	0.998839965703334\\
};
\addplot [color=mycolor2,dashed,line width=1.2pt,mark=asterisk,mark size=1.5,mark options={solid}]
  table[row sep=crcr]{%
-3.31927118486104	0.00211832349826024\\
};
\addlegendentry{HH, $\lambda = 80 \,\mu$s};

\addplot [color=mycolor4,dashed,line width=1.2pt,forget plot]
  table[row sep=crcr]{%
-3.322531564483	0.00357826831972574\\
-3.32031032895638	0.00987803648825736\\
-3.31913957315161	0.0164297953835302\\
-3.31845638804917	0.0775123475456105\\
-3.31776966139459	0.0851224674931961\\
-3.31728253881688	0.0915734300977717\\
-3.31673400721969	0.0978731982663024\\
-3.31621963375642	0.104777744179012\\
-3.315826213579	0.111279104928936\\
-3.3154754850209	0.118536437859084\\
-3.31509501339098	0.124836206027615\\
-3.31473514419008	0.131135974196146\\
-3.31439926065811	0.139199677451865\\
-3.31401036069099	0.145197056748307\\
-3.31367286421968	0.15084164902731\\
-3.31333531945023	0.156788630178403\\
-3.31300689015737	0.162836407620193\\
-3.3126828298439	0.168984981352679\\
-3.31238552825713	0.175032758794469\\
-3.31211504450838	0.18072774921882\\
-3.31181527909247	0.186725128515261\\
-3.31155735942474	0.192672109666354\\
-3.31131003637627	0.258945670799304\\
-3.31096827974995	0.275425864328181\\
-3.31067474498081	0.283892752746686\\
-3.31031555696202	0.289738937607082\\
-3.30999311727891	0.295383529886086\\
-3.30963532553868	0.301128918455785\\
-3.30927075920392	0.30672311258944\\
-3.30894823747841	0.312770890031229\\
-3.30864222168163	0.319574639653242\\
-3.30830341531851	0.325824009676424\\
-3.30802045821615	0.332224574135651\\
-3.30770156860285	0.338423546013485\\
-3.30738037285712	0.344269730873881\\
-3.30707538584647	0.359590767059749\\
-3.3067516562793	0.365940933373628\\
-3.30644363184825	0.372240701542158\\
-3.30608232269004	0.37818768269325\\
-3.30573612549321	0.384638645297825\\
-3.30545104590001	0.390636024594266\\
-3.30511312729403	0.3968349964721\\
-3.30478287827524	0.402882773913889\\
-3.30441550954721	0.408728958774285\\
-3.30405621121347	0.41432315290794\\
-3.30366064012068	0.420572522931122\\
-3.30321318232361	0.42636830964617\\
-3.30280520736275	0.432164096361218\\
-3.30245760241268	0.437707892349525\\
-3.30208223424893	0.443402882773876\\
-3.30168713459055	0.449249067634272\\
-3.30127782234372	0.454843261767926\\
-3.30093632026127	0.460487854046929\\
-3.300480237427	0.503023888720855\\
-3.30004896074733	0.524291906057819\\
-3.29954422812222	0.529533313174037\\
-3.29898660544149	0.534925914726301\\
-3.29848160841837	0.541427275476226\\
-3.29781708573336	0.546719080737793\\
-3.29726827745794	0.553119645197021\\
-3.2966120492312	0.558814635621373\\
-3.29599079380202	0.580133051103685\\
-3.29535547751351	0.585424856365252\\
-3.29472455443295	0.59066626348147\\
-3.29394753830171	0.59585727245234\\
-3.29320652737222	0.601401068440647\\
-3.29246486705566	0.606944864428955\\
-3.29160114872464	0.612186271545173\\
-3.29067237395683	0.617427678661391\\
-3.28978452975753	0.622769882068306\\
-3.28871630353496	0.628263279911265\\
-3.28779077248461	0.633857474044921\\
-3.2868455981627	0.639350871887881\\
-3.28590343912264	0.644894667876188\\
-3.28495495958487	0.650085676847059\\
-3.28387585199523	0.655276685817929\\
-3.28253647510944	0.660518092934147\\
-3.28118462685692	0.665658703759669\\
-3.27942659313619	0.670900110875887\\
-3.27782376016704	0.676191916137453\\
-3.27599649356587	0.69186573934076\\
-3.27373601795117	0.715704062090484\\
-3.27115755444116	0.732033061183319\\
-3.26592064391005	0.737425662735582\\
-3.16523567937485	0.746850115915705\\
-2.7458528091664	0.751940328595878\\
-2.56215983476389	0.759197661526027\\
-2.46399307499523	0.769478883177071\\
-2.44811191428772	0.788630178409409\\
-2.42309571564928	0.795232335450031\\
-2.39800635948166	0.804354399758066\\
-2.34440250200988	0.811107751234733\\
-2.26218483766563	0.819171454490454\\
-2.24113368505386	0.826479185565951\\
-2.09417148756675	0.831821388972867\\
-1.95676459571488	0.837113194234434\\
-1.8794318566697	0.842304203205304\\
-1.83917075104433	0.847596008466872\\
-1.79356454156315	0.85313980445518\\
-1.71150223321428	0.858381211571399\\
-1.60978544159174	0.863521822396921\\
-1.48997668694252	0.868612035077095\\
-1.42894184836829	0.873702247757269\\
-1.3958325611798	0.878842858582792\\
-1.36940379536536	0.884185061989707\\
-1.35905809497465	0.890484830158239\\
-1.32714755834209	0.895575042838413\\
-1.30612083512063	0.900917246245328\\
-1.2882152161476	0.906763431105725\\
-1.26644217382237	0.911904041931247\\
-1.23206328183876	0.916994254611421\\
-1.20399445481641	0.922336458018336\\
-1.1885821465699	0.930248966838012\\
-1.165480081219	0.946678762221544\\
-1.13978497652683	0.951920169337763\\
-1.09548562059852	0.957111178308633\\
-1.05043821990683	0.962251789134155\\
-1.01504733878981	0.967291603668981\\
-0.98682963479172	0.972381816349155\\
-0.959971830248338	0.977421630883981\\
-0.907528835044638	0.982461445418807\\
-0.83543363778284	0.987501259953632\\
-0.766616276517705	0.992591472633806\\
-0.646762963230689	0.997631287168632\\
-inf	0\\
-inf	0\\
-inf	0\\
-inf	0\\
-inf	0\\
-inf	0\\
-inf	0\\
-inf	0\\
-inf	0\\
-inf	0\\
-inf	0\\
-inf	0\\
-inf	0\\
-inf	0\\
-inf	0\\
-inf	0\\
-inf	0\\
-inf	0\\
-inf	0\\
-inf	0\\
};
\addplot [color=mycolor4,line width=1.2pt,only marks,mark=diamond,mark size=1.5,mark options={solid},forget plot]
  table[row sep=crcr]{%
-3.322531564483	0.00357826831972574\\
-3.31031555696202	0.289738937607082\\
-3.30405621121347	0.41432315290794\\
-3.29726827745794	0.553119645197021\\
-3.28495495958487	0.650085676847059\\
-3.27782376016704	0.676191916137453\\
-3.16523567937485	0.746850115915705\\
-2.42309571564928	0.795232335450031\\
-2.09417148756675	0.831821388972867\\
-1.71150223321428	0.858381211571399\\
-1.36940379536536	0.884185061989707\\
-1.2882152161476	0.906763431105725\\
-1.165480081219	0.946678762221544\\
-0.98682963479172	0.972381816349155\\
-0.646762963230689	0.997631287168632\\
};
\addplot [color=mycolor4,dashed,line width=1.2pt,mark=diamond,mark size=1.5,mark options={solid}]
  table[row sep=crcr]{%
-3.322531564483	0.00357826831972574\\
};
\addlegendentry{HH, $\lambda = 160 \,\mu$s};

\addplot [color=mycolor1,solid,line width=1.2pt,forget plot]
  table[row sep=crcr]{%
-3.31954627302956	0.00231283624113843\\
-3.31588339978907	0.00759213635677991\\
-3.31421565427416	0.0127708783749811\\
-3.31322337800458	0.0727035044496968\\
-3.31232464112297	0.0795917341243902\\
-3.31160133790949	0.085675499019559\\
-3.31104472866645	0.0927648448891353\\
-3.3104684129231	0.0991502840766258\\
-3.30996535861956	0.106289908994923\\
-3.30944656290958	0.114233998692746\\
-3.30901625370908	0.119915531198151\\
-3.30866126945923	0.125798179898438\\
-3.3082446238279	0.131680828598724\\
-3.30784865948055	0.137362361104129\\
-3.30745089211725	0.142993614560815\\
-3.30705885344762	0.148423751822618\\
-3.30670343734255	0.154205842425464\\
-3.30636181798323	0.159585700638547\\
-3.30602429093671	0.165216954095232\\
-3.30558926814876	0.171200160892959\\
-3.30519164804954	0.176831414349645\\
-3.30488067622981	0.183116295439695\\
-3.30448332919017	0.260294635225505\\
-3.30414972193324	0.26728342299764\\
-3.30384949292388	0.2750766755493\\
-3.30357206080982	0.280506812811103\\
-3.30319602059232	0.285886671024185\\
-3.30285938097934	0.291266529237268\\
-3.30247882570157	0.296948061742672\\
-3.3020533559199	0.303685454271206\\
-3.3016873128731	0.309417265825332\\
-3.30140480409319	0.315953542158984\\
-3.30100550909423	0.321584795615668\\
-3.30060236019495	0.335964603549702\\
-3.30017783619708	0.341495298908946\\
-3.29986060234577	0.34702599426819\\
-3.2994188577078	0.352707526773595\\
-3.29906808732672	0.35843933832772\\
-3.29872290833059	0.363970033686963\\
-3.29837557241704	0.369852682387249\\
-3.29798920858729	0.375584493941373\\
-3.29759877597441	0.380964352154456\\
-3.29722610685316	0.386394489416259\\
-3.29683503803241	0.391975463824223\\
-3.29647776586378	0.397656996329628\\
-3.29610034091929	0.403187691688872\\
-3.29569556763462	0.4091708984866\\
-3.29526747240302	0.414500477650962\\
-3.29484490612693	0.420031173010205\\
-3.29448701095055	0.425411031223288\\
-3.29415836548335	0.430891447533812\\
-3.29375802128029	0.436120468600733\\
-3.29330539885652	0.441299210618933\\
-3.29285412051217	0.447734928855143\\
-3.29242554569671	0.453014228970784\\
-3.2919603962553	0.458544924330027\\
-3.29145240974359	0.46397506159183\\
-3.29098523485202	0.469455477902353\\
-3.29054938295579	0.474634219920553\\
-3.2900153173033	0.480315752425957\\
-3.28931067082406	0.485645331590319\\
-3.28873337974851	0.4907737945598\\
-3.28816030887036	0.546684096736883\\
-3.2876470666429	0.557896324601532\\
-3.28698370659784	0.562974508522293\\
-3.28627600531561	0.568253808637935\\
-3.28565186277672	0.574588968776706\\
-3.28496635175799	0.580622454623154\\
-3.28423579830844	0.602192166524206\\
-3.2833781004877	0.607220071396246\\
-3.2825610553076	0.612398813414447\\
-3.28132522571712	0.617627834481369\\
-3.28020787610321	0.62270601840213\\
-3.27896593460407	0.627935039469052\\
-3.27758383202734	0.632962944341092\\
-3.27636511708154	0.637990849213132\\
-3.27488220080258	0.643119312182614\\
-3.27324956300264	0.648147217054654\\
-3.27135520394618	0.653275680024135\\
-3.26970412227982	0.658303584896175\\
-3.26803758318739	0.663381768816936\\
-3.26654656868545	0.668409673688976\\
-3.26502907125788	0.673487857609737\\
-3.26307665255377	0.678515762481777\\
-3.26139778612968	0.683593946402538\\
-3.25960778987748	0.688621851274578\\
-3.25793312196488	0.693700035195339\\
-3.25608531671091	0.6987782191161\\
-3.25394574213292	0.70380612398814\\
-3.25116243180701	0.708884307908901\\
-3.24840590237784	0.713912212780941\\
-3.24551757974726	0.718940117652981\\
-3.24219207883179	0.724018301573742\\
-3.2376818919512	0.729046206445782\\
-3.23411345582158	0.766302981547596\\
-3.23175750255222	0.782995625722768\\
-3.22558231292816	0.791593343053956\\
-3.07991020135269	0.796621247925996\\
-2.88587123096969	0.801699431846756\\
-2.74685555656376	0.808034591985527\\
-2.58963547047102	0.813163054955007\\
-2.46669259179105	0.818492634119369\\
-2.37015578758884	0.82362109708885\\
-2.28741066073784	0.828749560058331\\
-2.22023132327921	0.833777464930371\\
-2.15874409764565	0.838956206948572\\
-2.1316251845135	0.845442204233503\\
-2.06433699534828	0.850520388154264\\
-2.01465181515472	0.855699130172465\\
-1.99143011237496	0.863341545577966\\
-1.95407499720592	0.868620845693608\\
-1.91174027876745	0.873749308663088\\
-1.88094456665427	0.87897832973001\\
-1.84159742032479	0.884307908894371\\
-1.81674351539196	0.889637488058733\\
-1.78530862502266	0.894916788174375\\
-1.75359722155327	0.90069887877722\\
-1.72011650724931	0.905978178892862\\
-1.69444403512554	0.911156920911062\\
-1.66409831928508	0.916285383880542\\
-1.63067380442866	0.921816079239786\\
-1.59879002456143	0.931017145155618\\
-1.57126640246431	0.936648398612302\\
-1.544600220932	0.942179093971546\\
-1.51639198358825	0.947910905525671\\
-1.48859953484755	0.953994670420839\\
-1.45425158381512	0.959625923877523\\
-1.42242146428461	0.964955503041885\\
-1.38380389121792	0.97068731459601\\
-1.34042766591286	0.975966614711651\\
-1.2950548141444	0.981095077681132\\
-1.2398947804098	0.986826889235257\\
-1.15829937336468	0.992055910302178\\
-1.04727323048506	0.997134094222938\\
-inf	0\\
-inf	0\\
-inf	0\\
};
\addplot [color=mycolor1,line width=1.2pt,only marks,mark=+,mark options={solid},forget plot]
  table[row sep=crcr]{%
-3.31421565427416	0.0127708783749811\\
-3.30357206080982	0.280506812811103\\
-3.29798920858729	0.375584493941373\\
-3.29054938295579	0.474634219920553\\
-3.28020787610321	0.62270601840213\\
-3.26502907125788	0.673487857609737\\
-3.25793312196488	0.693700035195339\\
-3.24551757974726	0.718940117652981\\
-3.22558231292816	0.791593343053956\\
-2.58963547047102	0.813163054955007\\
-2.15874409764565	0.838956206948572\\
-1.95407499720592	0.868620845693608\\
-1.78530862502266	0.894916788174375\\
-1.66409831928508	0.916285383880542\\
-1.51639198358825	0.947910905525671\\
-1.34042766591286	0.975966614711651\\
-1.04727323048506	0.997134094222938\\
-inf	0\\
};
\addplot [color=mycolor1,solid,line width=1.2pt,mark=+,mark size=1.5,mark options={solid}]
  table[row sep=crcr]{%
-3.31421565427416	0.0127708783749811\\
};
\addlegendentry{DC, $\lambda = 80 \,\mu$s};

\addplot [color=mycolor3,solid,line width=1.2pt,forget plot]
  table[row sep=crcr]{%
-3.323439195777	0.00201126307320998\\
-3.32110414985333	0.0100563153660498\\
-3.31994795806027	0.0165426387771527\\
-3.31929730473302	0.0788415124698325\\
-3.31866658311314	0.0865848753016915\\
-3.31809446630699	0.093523732904266\\
-3.31759096499757	0.0997083668543868\\
-3.31708383816322	0.105893000804508\\
-3.31670268730913	0.112127916331458\\
-3.31635786086442	0.119368463395014\\
-3.31597631234333	0.126810136765891\\
-3.31557576628228	0.132793644408691\\
-3.31524669473579	0.138978278358812\\
-3.31495000953627	0.145263475462594\\
-3.31462182421845	0.153409090909095\\
-3.31430354562056	0.159040627514084\\
-3.31399714329384	0.165174979887374\\
-3.31374388755992	0.171108205953344\\
-3.31335376124742	0.177091713596144\\
-3.31309164463582	0.183175784392605\\
-3.31279594936419	0.189259855189066\\
-3.31248942346152	0.195293644408695\\
-3.31221967084954	0.223049074818993\\
-3.31191538404877	0.282481898632348\\
-3.31161176087973	0.293393000804511\\
-3.31126897273748	0.299778761061953\\
-3.31094362472298	0.30515888978279\\
-3.31059169330242	0.31089098954144\\
-3.31028179628554	0.316723652453749\\
-3.30997994603568	0.322355189058739\\
-3.30970841230577	0.328439259855199\\
-3.30941329883837	0.335076427996793\\
-3.30912367196494	0.341914722445707\\
-3.30882290058668	0.348853580048282\\
-3.30853457343306	0.354786806114252\\
-3.30825730715347	0.361122284794866\\
-3.30795487201856	0.367508045052308\\
-3.30759683040413	0.383397023330666\\
-3.30733481452403	0.389380530973468\\
-3.30701887486101	0.395062349155286\\
-3.30664434149721	0.401749798873709\\
-3.30631912940158	0.408034995977491\\
-3.30605253034826	0.41381737731297\\
-3.30570501045042	0.420152855993583\\
-3.30539849050427	0.426086082059552\\
-3.30504934173252	0.432320997586503\\
-3.30474158175177	0.438304505229304\\
-3.30448192495399	0.444137168141612\\
-3.30408980838176	0.449818986323431\\
-3.30373288573691	0.45550080450525\\
-3.30335109614134	0.460931214802917\\
-3.30299859390158	0.466713596138396\\
-3.30257759094705	0.472998793242177\\
-3.30228131753836	0.478730893000825\\
-3.30191832953911	0.484161303298492\\
-3.30155987499676	0.489943684633971\\
-3.3011414566114	0.543543845535016\\
-3.3007626413739	0.568584070796479\\
-3.30028551842658	0.574416733708788\\
-3.29985307249293	0.579947707160114\\
-3.29937579321917	0.585478680611441\\
-3.29891455191505	0.590909090909108\\
-3.29843431737288	0.598149637972663\\
-3.29785024605451	0.605088495575237\\
-3.29733408565363	0.610720032180224\\
-3.29679564061131	0.633447304907496\\
-3.29618024860672	0.638927996781992\\
-3.29564934976248	0.644509251810149\\
-3.29497909065155	0.650291633145627\\
-3.29438995979334	0.655872888173784\\
-3.29375906345179	0.661504424778771\\
-3.29301504172008	0.667236524537419\\
-3.29231238465854	0.672817779565576\\
-3.29149856378324	0.678700724054714\\
-3.29082530070947	0.6843322606597\\
-3.29010636273776	0.690014078841518\\
-3.28928618549721	0.695746178600165\\
-3.28852727514534	0.701377715205152\\
-3.28771136777231	0.70705953338697\\
-3.28671086341409	0.712942477876108\\
-3.28598903763201	0.718674577634756\\
-3.28518072621883	0.724306114239743\\
-3.28426279680289	0.729736524537409\\
-3.28333534751633	0.735066371681414\\
-3.2821994508736	0.740396218825419\\
-3.28117062198189	0.745826629123085\\
-3.28009149559878	0.751156476267091\\
-3.27872143762637	0.756637168141588\\
-3.27717929616805	0.762017296862425\\
-3.27553910052452	0.80510860820595\\
-3.27326117484303	0.810438455349957\\
-3.27051751144781	0.829847144006433\\
-3.26526013946918	0.836484312148026\\
-3.16877584619614	0.84523330651649\\
-3.10499627102174	0.853529766693481\\
-3.09697592683161	0.86132341110217\\
-3.06055391383537	0.866502413515686\\
-2.91267856346946	0.871631134352372\\
-2.82018826277343	0.876910699919548\\
-2.73436580875107	0.882290828640385\\
-2.67317345588768	0.887570394207561\\
-2.62561884107196	0.893956154465003\\
-2.5991592205324	0.900492759452935\\
-2.5892345520243	0.90748189863234\\
-2.53522740305781	0.913063153660497\\
-2.47776152425978	0.918644408688655\\
-2.45306301833492	0.960830651649235\\
-2.38222611029148	0.967316975060337\\
-2.30439320952251	0.973049074818986\\
-2.20839260362292	0.978328640386162\\
-2.11250755874319	0.983809332260659\\
-1.95150825863855	0.988938053097345\\
-1.70744803531015	0.99396621078037\\
-1.34659976560556	0.999094931617056\\
-inf	0\\
-inf	0\\
-inf	0\\
-inf	0\\
-inf	0\\
-inf	0\\
-inf	0\\
-inf	0\\
-inf	0\\
-inf	0\\
-inf	0\\
-inf	0\\
-inf	0\\
-inf	0\\
-inf	0\\
-inf	0\\
-inf	0\\
-inf	0\\
-inf	0\\
-inf	0\\
-inf	0\\
-inf	0\\
-inf	0\\
};
\addplot [color=mycolor3,line width=1.2pt,only marks,mark=o,mark size=1.5,mark options={solid},forget plot]
  table[row sep=crcr]{%
-3.31809446630699	0.093523732904266\\
-3.30882290058668	0.348853580048282\\
-3.30448192495399	0.444137168141612\\
-3.29891455191505	0.590909090909108\\
-3.29010636273776	0.690014078841518\\
-3.28598903763201	0.718674577634756\\
-3.2821994508736	0.740396218825419\\
-3.27553910052452	0.80510860820595\\
-3.10499627102174	0.853529766693481\\
-2.82018826277343	0.876910699919548\\
-2.5892345520243	0.90748189863234\\
-2.30439320952251	0.973049074818986\\
-1.34659976560556	0.999094931617056\\
};
\addplot [color=mycolor3,solid,line width=1.2pt,mark size=1.5,mark=o,mark options={solid}]
  table[row sep=crcr]{%
-3.323439195777	0.00201126307320998\\
};
\addlegendentry{DC, $\lambda = 160 \,\mu$s};

\end{axis}
\end{tikzpicture}%

%% file: figures/perf/latCDF_a_80160_b_10_x_10000.tex
%
%
\definecolor{mycolor1}{rgb}{0.21162,0.18978,0.57768}%
\definecolor{mycolor2}{rgb}{0.68342,0.74348,0.40443}%
\definecolor{mycolor3}{rgb}{0.07227,0.48867,0.84670}%
\definecolor{mycolor4}{rgb}{0.98800,0.80660,0.17937}%
\begin{tikzpicture}
\tikzstyle{every node}=[font=\scriptsize]
\pgfplotsset{every x tick label/.append style={font=\tiny, yshift=0.5ex}}

\begin{axis}[%
width=0.951\fwidth,
height=\fheight,
at={(0\fwidth,0\fheight)},
scale only axis,
unbounded coords=jump,
xmin=-3.5,
xmax=-0.5,
xlabel={$\log_{10}(L)$ [s]},
xmajorgrids,
ymin=0,
ymax=1,
ylabel={CDF},
ymajorgrids, ylabel shift = -5 pt, yticklabel shift = -3 pt,
axis background/.style={fill=white},
axis x line*=bottom,
axis y line*=left,
legend style={legend cell align=left,align=left,draw=white!15!black,at={(0.9\fwidth,0.05\fheight)},anchor=south east}
]

\addplot [color=mycolor2,dashed,line width=1.2pt,forget plot]
  table[row sep=crcr]{%
-3.31927118486104	0.0023149313069295\\
-3.31501132329053	0.00754868904433537\\
-3.31316646644702	0.0128327713753718\\
-3.31233412973672	0.0720144934829673\\
-3.31155625686168	0.0784560414674663\\
-3.31079875266377	0.0843440189220489\\
-3.31023514213082	0.0907352423129197\\
-3.3096481833075	0.0974284132655646\\
-3.30911843903091	0.103417039907405\\
-3.30862418604529	0.109657289517388\\
-3.30814970738412	0.116803381812693\\
-3.30768658447668	0.12243973629913\\
-3.30728463285124	0.128478687534599\\
-3.30689589951126	0.133762769865634\\
-3.30646495668367	0.139550098132959\\
-3.30607920559441	0.145337426400284\\
-3.30571986972036	0.15102410548035\\
-3.30542776307504	0.156559810779531\\
-3.30507031219931	0.162447788234112\\
-3.30471091276562	0.167631221377889\\
-3.30444279085766	0.172965628302554\\
-3.30412601062227	0.178803281163507\\
-3.30376425877692	0.252881082985257\\
-3.3033889464557	0.258114840722663\\
-3.30307047622495	0.264908660862564\\
-3.30265134555335	0.272407025313271\\
-3.30225736995959	0.277691107644306\\
-3.30189917587122	0.282874540788083\\
-3.30158914305725	0.28896381661718\\
-3.30126042047236	0.294600171103618\\
-3.30091016081893	0.300085551809169\\
-3.30050754456187	0.305822555482865\\
-3.30011661753283	0.311458909969302\\
-3.29975818862921	0.316742992300337\\
-3.29943883720232	0.331186150671832\\
-3.29905018548526	0.336419908409238\\
-3.2986638847482	0.342005938302047\\
-3.29831491711569	0.347541643601227\\
-3.297964346244	0.353027024306777\\
-3.29761674255814	0.358311106637812\\
-3.29725170384376	0.36389713653062\\
-3.29685907034629	0.369130894268025\\
-3.29651938296166	0.374716924160834\\
-3.29612874432664	0.380001006491869\\
-3.29575966825097	0.385234764229275\\
-3.2953978595457	0.390569171153939\\
-3.29494326002307	0.395853253484974\\
-3.29454881660877	0.40108701122238\\
-3.2941892543548	0.406622716521559\\
-3.29380917251069	0.411856474258966\\
-3.29345093074757	0.417241205777258\\
-3.29301726623097	0.422374314327406\\
-3.29260359275193	0.427557747471183\\
-3.29209369086781	0.432690856021332\\
-3.29162016855727	0.438025262945997\\
-3.29111608119598	0.444215187962351\\
-3.29061431496429	0.449348296512498\\
-3.29008134758835	0.454481405062645\\
-3.28944556014427	0.459664838206423\\
-3.28884542362532	0.464697297569313\\
-3.28827130524778	0.469780081525831\\
-3.28767041834789	0.520507271903769\\
-3.287000568074	0.526344924764723\\
-3.28621564059643	0.535504000805184\\
-3.28561415274333	0.540536460168075\\
-3.28482851846552	0.546374113029028\\
-3.28402594884549	0.551809169140951\\
-3.28336093596023	0.572844849277834\\
-3.28247786801618	0.578078607015241\\
-3.28137264133818	0.58321171556539\\
-3.28015046810906	0.588395148709168\\
-3.27880727055919	0.593528257259317\\
-3.27747242683233	0.598560716622208\\
-3.27601024933492	0.603744149765986\\
-3.27453677925652	0.608877258316135\\
-3.2726550168532	0.613909717679026\\
-3.27075926405481	0.618992501635547\\
-3.26894347890266	0.624024960998438\\
-3.26719280852645	0.629208394142216\\
-3.26528837553924	0.634240853505108\\
-3.26350992332067	0.639273312867999\\
-3.26177727760707	0.644305772230891\\
-3.25966357302946	0.649338231593782\\
-3.25794069598915	0.654471340143931\\
-3.25613015267768	0.659554124100452\\
-3.25394190493766	0.664687232650601\\
-3.2515296256623	0.669770016607121\\
-3.2488393700575	0.674802475970013\\
-3.24623546205608	0.679834935332904\\
-3.24279002272079	0.684968043883053\\
-3.2384022666556	0.690000503245945\\
-3.2354911318428	0.695737506919641\\
-3.23192550363464	0.731367319208906\\
-3.2282463316313	0.743948467616133\\
-3.22367228743775	0.75240299934579\\
-3.17798517599722	0.757586432489567\\
-3.08290482361845	0.762618891852457\\
-2.94079318369977	0.767651351215348\\
-2.78743862173605	0.772683810578238\\
-2.62999455599901	0.777816919128387\\
-2.48208621855151	0.782849378491277\\
-2.35060654292842	0.787932162447797\\
-2.22056108302608	0.792964621810687\\
-2.10746213420836	0.798047405767207\\
-2.03075175088502	0.803130189723727\\
-1.96353869697398	0.808816868803793\\
-1.91837087255684	0.815560364350066\\
-1.87249371457101	0.822857430426258\\
-1.83349052438952	0.828594434099953\\
-1.78237451482305	0.834381762367277\\
-1.73812900977628	0.839414221730168\\
-1.70596837281513	0.845100900810234\\
-1.67978155341584	0.850636606109414\\
-1.64609845151238	0.856121986814965\\
-1.61040982591958	0.861557042926887\\
-1.5785390604589	0.866841125257922\\
-1.54296444345241	0.871923909214441\\
-1.51024594568815	0.877006693170961\\
-1.48565694870109	0.882139801721109\\
-1.45916000638972	0.887222585677628\\
-1.43304829697079	0.892305369634147\\
-1.40915197502258	0.897388153590666\\
-1.38773680325197	0.902621911328072\\
-1.36470437471912	0.907704695284591\\
-1.33895061404895	0.912737154647481\\
-1.31705848133262	0.917870263197629\\
-1.28566220255107	0.923003371747777\\
-1.25815601531371	0.928035831110667\\
-1.23160183389711	0.933068290473558\\
-1.21365047570031	0.941422173015956\\
-1.18193616291598	0.94766242262594\\
-1.16128748643813	0.953047154144232\\
-1.13451787486121	0.958180262694381\\
-1.11021605813821	0.963212722057271\\
-1.07506889167698	0.968245181420162\\
-1.04109245139396	0.973529263751196\\
-1.00532691835309	0.978561723114087\\
-0.964334119857579	0.983644507070606\\
-0.908251640207635	0.988676966433497\\
-0.844627721417335	0.993709425796387\\
-0.671632684618316	0.998741885159277\\
-inf	0\\
-inf	0\\
-inf	0\\
};
\addplot [color=mycolor2,line width=1.2pt,only marks,mark=asterisk,mark size=1.5,mark options={solid},forget plot]
  table[row sep=crcr]{%
-3.3096481833075	0.0974284132655646\\
-3.29831491711569	0.347541643601227\\
-3.28767041834789	0.520507271903769\\
-3.27880727055919	0.593528257259317\\
-3.26177727760707	0.644305772230891\\
-3.2354911318428	0.695737506919641\\
-3.08290482361845	0.762618891852457\\
-2.35060654292842	0.787932162447797\\
-1.91837087255684	0.815560364350066\\
-1.73812900977628	0.839414221730168\\
-1.5785390604589	0.866841125257922\\
-1.43304829697079	0.892305369634147\\
-1.31705848133262	0.917870263197629\\
-1.18193616291598	0.94766242262594\\
-1.04109245139396	0.973529263751196\\
-0.671632684618316	0.998741885159277\\
};
\addplot [color=mycolor2,dashed,line width=1.2pt,mark=asterisk,mark size=1.5,mark options={solid}]
  table[row sep=crcr]{%
-3.31316646644702	0.0128327713753718\\
};
\addlegendentry{HH, $\lambda = 80 \,\mu$s};
\addplot [color=mycolor4,dashed,line width=1.2pt,forget plot]
  table[row sep=crcr]{%
-3.32248633610878	0.00337089957737979\\
-3.32035178126331	0.00955926745824121\\
-3.31924576090417	0.0165023143489637\\
-3.31856917596292	0.0776816260817066\\
-3.31793766360657	0.0849768565103644\\
-3.31741434719049	0.0918192795331054\\
-3.31686891252678	0.0982592070839206\\
-3.31638407098395	0.104246327228819\\
-3.31596803368014	0.111239686053532\\
-3.31556535563307	0.117981485208292\\
-3.31523609346504	0.125025155966996\\
-3.31482313560965	0.131666331253774\\
-3.31448241197095	0.139766552626284\\
-3.31410002685844	0.146156168243108\\
-3.3137713164659	0.151942040652044\\
-3.31344253545009	0.157526665325016\\
-3.31311735840009	0.163312537733951\\
-3.31282116600267	0.169199034010868\\
-3.31252637810028	0.175085530287785\\
-3.31222027451428	0.181022338498693\\
-3.31194609208263	0.187462266049508\\
-3.31163204258355	0.193046890722481\\
-3.31138551358292	0.24315757697726\\
-3.31112140072828	0.275960957939224\\
-3.31081248799914	0.286375528275308\\
-3.31053283896444	0.292262024552225\\
-3.31023840340038	0.297846649225198\\
-3.30991788165213	0.303682833568124\\
-3.3095399237278	0.309217146307106\\
-3.30923240125993	0.315053330650032\\
-3.30890671456854	0.321543570134838\\
-3.30863691672485	0.327832561883681\\
-3.30833326059056	0.33457436103844\\
-3.30804753176588	0.340360233447376\\
-3.30771831964041	0.346498289394246\\
-3.30742105397914	0.352787281143089\\
-3.30714553807066	0.362497484403303\\
-3.30682884183919	0.374421412759109\\
-3.30650702675947	0.380358220970016\\
-3.30618115851166	0.386445964982896\\
-3.30588786149907	0.392533708995776\\
-3.30559821684615	0.398772388810628\\
-3.30527558610404	0.404860132823508\\
-3.30499095217267	0.41104850070437\\
-3.30468396259409	0.416683437311333\\
-3.30439793906718	0.422469309720269\\
-3.30405352855073	0.428305494063195\\
-3.30372517657481	0.434443550010065\\
-3.3033521339912	0.439977862749047\\
-3.30298312828328	0.445461863554038\\
-3.30262646801711	0.451247735962974\\
-3.30223387894371	0.457385791909844\\
-3.30182537109697	0.463221976252771\\
-3.30150285447672	0.468705977057762\\
-3.30109526961833	0.474340913664725\\
-3.30075834826974	0.480026162205679\\
-3.30038593591705	0.528023747232847\\
-3.29995205705142	0.548500704367079\\
-3.29954865305214	0.553783457436107\\
-3.29910152061209	0.559015898571144\\
-3.29860495175341	0.565254578385997\\
-3.29814732139269	0.570738579190988\\
-3.29767239204703	0.576574763533914\\
-3.29715824692256	0.582461259810831\\
-3.29668938932473	0.587744012879859\\
-3.29627046906115	0.610837190581609\\
-3.29580371013298	0.61697524652848\\
-3.29527239197516	0.622912054739388\\
-3.29457004392506	0.628597303280342\\
-3.29391976439966	0.634030992151342\\
-3.2933006456589	0.639816864560278\\
-3.29273212293583	0.645049305695315\\
-3.2919510320694	0.650482994566315\\
-3.29124522879987	0.655916683437315\\
-3.29053607659576	0.661048500704371\\
-3.28982172882321	0.666431877641381\\
-3.28901175045382	0.671664318776418\\
-3.28829441013861	0.677098007647418\\
-3.28749488290108	0.682582008452409\\
-3.28677513681993	0.687864761521437\\
-3.286010035203	0.693348762326428\\
-3.28518985545174	0.698832763131419\\
-3.28431680319557	0.704115516200447\\
-3.283370935221	0.709247333467503\\
-3.28235565337829	0.714328838800568\\
-3.28137270604737	0.719460656067624\\
-3.28022690780838	0.724642785268671\\
-3.27901467925924	0.729724290601736\\
-3.27785577040043	0.735057355604754\\
-3.27624486864135	0.740390420607773\\
-3.27512402104046	0.781897766150135\\
-3.27314452934957	0.787029583417191\\
-3.27115588580897	0.808613403099219\\
-3.26856412503789	0.813745220366275\\
-3.26520109623221	0.8201348359831\\
-3.2619409588052	0.830096598913266\\
-3.25606134191556	0.835127792312341\\
-3.22095047787464	0.840410545381369\\
-3.08009007458243	0.845492050714434\\
-2.90050273519382	0.850925739585434\\
-2.77202444201497	0.856107868786481\\
-2.64521981230704	0.861239686053537\\
-2.55029319231317	0.866673374924537\\
-2.50335573043249	0.872459247333472\\
-2.45328932667044	0.885590662105056\\
-2.39451029713061	0.894395250553435\\
-2.3132970839502	0.903652646407732\\
-2.21459967415097	0.922821493258204\\
-2.08992712626728	0.927852686657278\\
-1.99748840087804	0.932934191990343\\
-1.88287752819188	0.938166633125381\\
-1.82379200070054	0.946065606761927\\
-1.78448499122576	0.954568323606362\\
-1.69472571556857	0.96055544375126\\
-1.61975425522092	0.966894747434093\\
-1.54243707120769	0.97348561078688\\
-1.43086458092091	0.978567116119945\\
-1.28547880497822	0.988377943248139\\
-1.13885132491909	0.993459448581204\\
-0.89314324260488	0.998490641980278\\
-inf	0\\
};
\addplot [color=mycolor4,line width=1.2pt,only marks,mark=diamond,mark size=1.5,mark options={solid},forget plot]
  table[row sep=crcr]{%
-3.32035178126331	0.00955926745824121\\
-3.30075834826974	0.480026162205679\\
-3.2933006456589	0.639816864560278\\
-3.28982172882321	0.666431877641381\\
-3.28137270604737	0.719460656067624\\
-3.27512402104046	0.781897766150135\\
-3.2619409588052	0.830096598913266\\
-2.77202444201497	0.856107868786481\\
-2.39451029713061	0.894395250553435\\
-1.99748840087804	0.932934191990343\\
-1.61975425522092	0.966894747434093\\
-0.89314324260488	0.998490641980278\\
};
\addplot [color=mycolor4,dashed,line width=1.2pt,mark=diamond,mark size=1.5,mark options={solid}]
  table[row sep=crcr]{%
-3.32035178126331	0.00955926745824121\\
};
\addlegendentry{HH, $\lambda = 160 \,\mu$s};

\addplot [color=mycolor1,solid,line width=1.2pt,forget plot]
  table[row sep=crcr]{%
-3.3201137594368	0.00238215914850493\\
-3.3157794652981	0.0074505828687278\\
-3.31402247038278	0.0126710593005577\\
-3.31318175837289	0.0720729853015712\\
-3.31232464112297	0.0784591991890519\\
-3.31148151180404	0.0844906234161167\\
-3.31092746135954	0.0912823112012151\\
-3.31042092149496	0.0971616827166732\\
-3.3099419801988	0.103142422706536\\
-3.30942185221335	0.109680689305623\\
-3.30901059160683	0.117536746071968\\
-3.30858297653795	0.123162696401415\\
-3.30823286764413	0.128687278256458\\
-3.30786948725487	0.134515965534714\\
-3.30754049262587	0.140141915864161\\
-3.30715265788048	0.145565129244799\\
-3.30678085879311	0.151039026862639\\
-3.30640278911567	0.156462240243277\\
-3.30602901450528	0.162392295995938\\
-3.30567803586024	0.167815509376576\\
-3.30535900614794	0.173188038520012\\
-3.30500318264942	0.178712620375054\\
-3.30469608439405	0.249619868220974\\
-3.30432113904003	0.260010136847431\\
-3.30394618833476	0.266903193106934\\
-3.30360248019663	0.274961986822088\\
-3.3032666430278	0.280283831728322\\
-3.30290627596416	0.285554992397353\\
-3.30255267868124	0.290775468829182\\
-3.3022078415568	0.296654840344641\\
-3.30190974173862	0.302280790674088\\
-3.30151587112317	0.307957425240737\\
-3.30113026325027	0.31348200709578\\
-3.30070455658746	0.319006588950822\\
-3.30034120217152	0.333299543841851\\
-3.29998192247492	0.338773441459691\\
-3.29964241655214	0.344196654840329\\
-3.29923922965115	0.349923973644181\\
-3.29888307267148	0.355549923973628\\
-3.29847326710832	0.360922453117064\\
-3.29812502632996	0.366599087683713\\
-3.29776117449315	0.37222503801316\\
-3.29739211460222	0.377648251393798\\
-3.29705906085981	0.383122149011638\\
-3.29658515876125	0.38839330968067\\
-3.29618828556623	0.393765838824106\\
-3.29572522062107	0.399189052204744\\
-3.29527778563445	0.404409528636573\\
-3.29491872218598	0.409832742017211\\
-3.29449467720386	0.415205271160647\\
-3.29405704133314	0.420628484541285\\
-3.29359645677634	0.426153066396328\\
-3.29318311176481	0.431525595539764\\
-3.29271208780417	0.436746071971593\\
-3.29223505840472	0.441966548403422\\
-3.29172351021443	0.447997972630487\\
-3.29128561258463	0.453421186011125\\
-3.29076193369633	0.458692346680157\\
-3.29016210905614	0.463862138874784\\
-3.28963111831593	0.469031931069411\\
-3.28901425096514	0.474100354789633\\
-3.28846447932118	0.479624936644676\\
-3.28779077620567	0.530714647744526\\
-3.2871125681202	0.541155600608186\\
-3.28642368843549	0.546325392802814\\
-3.28560293150305	0.552559553978688\\
-3.28484905695223	0.558438925494147\\
-3.28403846840474	0.579320831221466\\
-3.2829491729266	0.584490623416094\\
-3.28210067577725	0.589761784085125\\
-3.28115589990858	0.595032944754157\\
-3.27969060388891	0.60010136847438\\
-3.27842352048254	0.605220476431805\\
-3.27667602133455	0.610440952863634\\
-3.27507758034258	0.615560060821059\\
-3.27346309438549	0.620729853015687\\
-3.27184714066474	0.625848960973112\\
-3.26996239117951	0.630968068930537\\
-3.26815916900951	0.636087176887962\\
-3.2663921458679	0.641206284845387\\
-3.26430605130565	0.646376077040014\\
-3.26222118506462	0.651444500760237\\
-3.26042221305936	0.656512924480459\\
-3.2582875446371	0.661784085149491\\
-3.25643311911215	0.667055245818522\\
-3.25409293585328	0.672174353775947\\
-3.251308359422	0.677293461733372\\
-3.24829646819708	0.682361885453594\\
-3.24457328698291	0.687430309173817\\
-3.24040421036437	0.692549417131242\\
-3.23539668352066	0.727318803851974\\
-3.23117057314494	0.741256969082588\\
-3.22551831023167	0.749619868220956\\
-3.14084914937795	0.756006082108437\\
-2.94674241925081	0.761125190065863\\
-2.6735417931869	0.766294982260491\\
-2.44428210452131	0.771515458692322\\
-2.29838237220114	0.776583882412545\\
-2.1941225478149	0.781753674607173\\
-2.11961804272364	0.786822098327397\\
-2.0813356633245	0.792143943233631\\
-2.04685753426277	0.797263051191057\\
-1.998618827328	0.802382159148483\\
-1.93151973443748	0.807704004054718\\
-1.86893643891316	0.813025848960953\\
-1.83364645236318	0.819412062848434\\
-1.81242638020333	0.82929548910287\\
-1.76825558011196	0.834718702483509\\
-1.72430011289608	0.840141915864148\\
-1.68261882445917	0.845615813481989\\
-1.6590895018064	0.85129244804864\\
-1.62620026272772	0.85651292448047\\
-1.59532834749345	0.8617334009123\\
-1.56921913627348	0.867511403953354\\
-1.54430380042648	0.872883933096791\\
-1.5186161959769	0.878155093765824\\
-1.49158310714098	0.883476938672058\\
-1.46766391168208	0.888545362392282\\
-1.44156606343977	0.89371515458691\\
-1.42606605739326	0.901672579827661\\
-1.40042573469011	0.906741003547885\\
-1.38113467181931	0.912468322351738\\
-1.35895837162961	0.917536746071962\\
-1.33776578170751	0.927318803851994\\
-1.31653966115397	0.932488596046622\\
-1.29677784220984	0.937557019766846\\
-1.27810683428985	0.942878864673081\\
-1.25984173553494	0.948403446528124\\
-1.24027829370892	0.955043081601617\\
-1.21475554706313	0.960111505321841\\
-1.18548328775885	0.965179929042064\\
-1.15534244006399	0.971312721743535\\
-1.11216447659325	0.976381145463758\\
-1.06517009077657	0.981601621895589\\
-1.01163377666182	0.986670045615812\\
-0.959523901462386	0.991789153573238\\
-0.830825434068899	0.996857577293461\\
};
\addplot [color=mycolor1,line width=1.2pt,only marks,mark=+,mark size=1.5,mark options={solid},forget plot]
  table[row sep=crcr]{%
-3.3201137594368	0.00238215914850493\\
-3.30070455658746	0.319006588950822\\
-3.29318311176481	0.431525595539764\\
-3.28484905695223	0.558438925494147\\
-3.26430605130565	0.646376077040014\\
-3.25643311911215	0.667055245818522\\
-3.24040421036437	0.692549417131242\\
-2.94674241925081	0.761125190065863\\
-2.1941225478149	0.781753674607173\\
-1.93151973443748	0.807704004054718\\
-1.72430011289608	0.840141915864148\\
-1.56921913627348	0.867511403953354\\
-1.46766391168208	0.888545362392282\\
-1.35895837162961	0.917536746071962\\
-1.25984173553494	0.948403446528124\\
-1.15534244006399	0.971312721743535\\
-0.830825434068899	0.996857577293461\\
};
\addplot [color=mycolor1,solid,line width=1.2pt,mark=+,mark size=1.5,mark options={solid}]
  table[row sep=crcr]{%
-3.3201137594368	0.00238215914850493\\
};
\addlegendentry{DC, $\lambda = 80 \,\mu$s};

\addplot [color=mycolor3,solid,line width=1.2pt,forget plot]
  table[row sep=crcr]{%
-3.32352691087624	0.00191484001007802\\
-3.32113074130221	0.00932224741748588\\
-3.32003005410392	0.0156714537666931\\
-3.31943169777175	0.0779541446208115\\
-3.3187890158151	0.0849080372889898\\
-3.31818040078887	0.091912320483749\\
-3.31761541275427	0.0984126984126982\\
-3.31713006100574	0.104711514235324\\
-3.31673247653814	0.11156462585034\\
-3.31637400833064	0.118568909045099\\
-3.31601537616478	0.125573192239859\\
-3.31560239664191	0.134139581758629\\
-3.31529899906713	0.140136054421769\\
-3.31498818610483	0.146283698664652\\
-3.31463144490102	0.152431342907534\\
-3.31429805221071	0.158075081884606\\
-3.3139937343193	0.163970773494584\\
-3.31366271181823	0.17031997984379\\
-3.31336336412861	0.176568405139834\\
-3.31308712051463	0.182312925170069\\
-3.31276528703649	0.18841017888637\\
-3.31246348298014	0.194658604182414\\
-3.31221329255642	0.224086671705721\\
-3.31190807633694	0.281380700428321\\
-3.31160491468423	0.289846308893928\\
-3.311344897486	0.295691609977325\\
-3.31104354991181	0.301587301587303\\
-3.3107178933574	0.307130259511213\\
-3.31038664259399	0.312975560594609\\
-3.31006543897348	0.318669690098262\\
-3.30975356200307	0.324716553287983\\
-3.30946234570612	0.330763416477703\\
-3.30915749874709	0.337969261778786\\
-3.30885766462331	0.344419249181154\\
-3.30857384962019	0.350667674477197\\
-3.30826403421778	0.365986394557822\\
-3.30795128168202	0.372688334593094\\
-3.30765038597977	0.378835978835977\\
-3.30734294959015	0.385084404132022\\
-3.30700005315828	0.391131267321741\\
-3.30668044107131	0.397228521038043\\
-3.30635942309752	0.403224993701182\\
-3.30607256358485	0.409372637944064\\
-3.30576759265928	0.41526832955404\\
-3.30544483611268	0.421264802217179\\
-3.30510132855584	0.427059712773994\\
-3.30475536734862	0.432703451751067\\
-3.30440123424892	0.438498362307882\\
-3.30406456110729	0.444041320231791\\
-3.30376381340245	0.449584278155701\\
-3.30334893861195	0.455076845553029\\
-3.30303606156154	0.460720584530101\\
-3.30266286737874	0.466515495086915\\
-3.30233313962756	0.471856890904501\\
-3.30193212644062	0.477551020408153\\
-3.30156403839094	0.483295540438386\\
-3.3012350310516	0.502191987906262\\
-3.30075199189114	0.555102040816316\\
-3.30026771647113	0.560443436633903\\
-3.29994237085197	0.566943814562853\\
-3.2994684250124	0.572083648274116\\
-3.29901165214726	0.577374653565122\\
-3.29848075474477	0.582615268329546\\
-3.29798778608086	0.588863693625591\\
-3.29747234587612	0.611337868480719\\
-3.29674663947253	0.616628873771725\\
-3.29610322706242	0.622020660115893\\
-3.29537754840776	0.627462836986642\\
-3.29479405823548	0.63285462333081\\
-3.29412824998338	0.638095238095235\\
-3.29344228381635	0.643487024439403\\
-3.29282794772663	0.648778029730409\\
-3.29203917542494	0.654270597127739\\
-3.29117534949458	0.659511211892164\\
-3.29032707686267	0.66475182665659\\
-3.28954388702411	0.670093222474177\\
-3.28878220227901	0.675535399344926\\
-3.28796285144149	0.680977576215675\\
-3.28723688385497	0.686570924666167\\
-3.28629248234013	0.691861929957174\\
-3.28528256763122	0.697354497354504\\
-3.28429836881038	0.702595112118929\\
-3.2833889727605	0.707886117409935\\
-3.28219947131431	0.712975560594617\\
-3.28098262890947	0.718216175359042\\
-3.27953671308113	0.723456790123468\\
-3.27787151439035	0.728495842781569\\
-3.27646400638592	0.768909045099532\\
-3.27423583684637	0.774149659863956\\
-3.27191822373188	0.793550012597642\\
-3.26865434743431	0.798689846308904\\
-3.2537668911423	0.808969513731428\\
-3.04479479460458	0.814260519022433\\
-2.8754572783364	0.819803476946344\\
-2.83116620616264	0.830435878054935\\
-2.80998920312644	0.836281179138331\\
-2.72063773355643	0.841622574955917\\
-2.64342576699992	0.847064751826665\\
-2.58643998287184	0.853262786596128\\
-2.56805290382034	0.862635424540194\\
-2.53314906803608	0.871705719324775\\
-2.48046278495922	0.877802973041076\\
-2.45267063768805	0.916099773242635\\
-2.42800835806989	0.923204837490556\\
-2.35374643639717	0.928495842781561\\
-2.2198010841899	0.933585285966242\\
-2.11190282287294	0.939329805996476\\
-2.07546710424419	0.952834467120184\\
-2.02769246667512	0.959082892416228\\
-1.93971525954419	0.964121945074328\\
-1.84477103922331	0.969211388259009\\
-1.78610703781486	0.98337112622827\\
-1.73836025212416	0.988964474678761\\
-1.59348575252494	0.994205089443185\\
-1.16786213025245	0.999244142101285\\
};
\addplot [color=mycolor3,line width=1.2pt,only marks,mark=o,mark size=1.5,mark options={solid},forget plot]
  table[row sep=crcr]{%
-3.32113074130221	0.00932224741748588\\
-3.31160491468423	0.289846308893928\\
-3.30857384962019	0.350667674477197\\
-3.29848075474477	0.582615268329546\\
-3.29282794772663	0.648778029730409\\
-3.28878220227901	0.675535399344926\\
-3.28429836881038	0.702595112118929\\
-3.27953671308113	0.723456790123468\\
-3.26865434743431	0.798689846308904\\
-2.80998920312644	0.836281179138331\\
-2.56805290382034	0.862635424540194\\
-2.35374643639717	0.928495842781561\\
-1.93971525954419	0.964121945074328\\
-1.16786213025245	0.999244142101285\\
};
\addplot [color=mycolor3,solid,line width=1.2pt,mark=o,mark size=1.5,mark options={solid}]
  table[row sep=crcr]{%
-3.32113074130221	0.00932224741748588\\
};
\addlegendentry{DC, $\lambda = 160 \,\mu$s};

\end{axis}
\end{tikzpicture}%

%% file: figures/perf/th_2_2040.tex
%
%
\definecolor{mycolor1}{rgb}{0.06293,0.47369,0.85544}%
\definecolor{mycolor2}{rgb}{0.03837,0.67427,0.74355}%
\definecolor{mycolor3a}{rgb}{0.6473,0.7456,0.4188}%
\definecolor{mycolor4a}{rgb}{0.785843,0.735567,0.36327}%
\definecolor{mycolor3}{rgb}{0.53003,0.74911,0.46611}%
\definecolor{mycolor4}{rgb}{0.97390,0.73140,0.26665}%
\definecolor{mycolor1a}{rgb}{0.21162,0.18978,0.57768}%
\definecolor{mycolor2a}{rgb}{0.07227,0.48867,0.84670}%
\begin{tikzpicture}
\tikzstyle{every node}=[font=\scriptsize]
\pgfplotsset{every x tick label/.append style={font=\tiny, yshift=0.5ex}}
\begin{axis}[%
width=0.951\fwidth,
height=\fheight,
at={(0\fwidth,0\fheight)},
scale only axis,
xmin=0,
xmax=10,
xlabel={X2/buffer configurations},
xtick={1,2,3,4,5,6,7,8,9,10},
xticklabels={$0.1$ $1$, $1$ $1$, $10$ $1$, $0.1$ $10$, $1$ $10$, $10$ $10$, $0.1$ $100$, $1$ $100$, $10$ $100$, ms MB},
xticklabel style={text width=6, align=center}, xmajorgrids,
ymin=40,
ymax=300,
ylabel={$\hat{S}_{PDCP}$ [Mbit/s]},
ymajorgrids, ylabel shift = -5 pt, yticklabel shift = -3 pt,
axis background/.style={fill=white},
legend columns=2,
legend style={legend cell align=left,align=left,draw=white!15!black,at={(0.05\fwidth,0.98\fheight)},anchor=south west}
]
\addplot [color=mycolor1,solid,line width=1.2pt,mark size=1.5pt,mark=o,mark options={solid}]
  table[row sep=crcr]{%
1	279.063308384326\\
2	278.533190372703\\
3	270.762503580931\\
};
\addplot [color=mycolor1,solid,line width=1.2pt,mark size=0.5pt,mark=*,mark options={solid}, forget plot]
  table[row sep=crcr]{%
4	289.830279768407\\
5	289.578219259803\\
6	281.42850739019\\
};
\addplot [color=mycolor1,solid,line width=1.2pt,mark size=0.5pt,mark=*,mark options={solid}, forget plot]
  table[row sep=crcr]{%
7	290.133206401059\\
8	289.551555545966\\
9	281.823171991166\\
};
\addlegendentry{DC, $\lambda = 20 \,\mu$s};

\addplot [color=mycolor2,solid,line width=1.2pt,mark size=1.5pt,mark=asterisk,mark options={solid}]
  table[row sep=crcr]{%
1	196.659394831824\\
2	196.249892366201\\
3	189.299339781997\\
};
\addplot [color=mycolor2,solid,line width=1.2pt,mark size=1.5pt,mark=asterisk,mark options={solid}, forget plot]
  table[row sep=crcr]{%
4	201.088879809934\\
5	200.646087057132\\
6	194.848783864509\\
};
\addplot [color=mycolor2,solid,line width=1.2pt,mark size=1.5pt,mark=asterisk,mark options={solid}, forget plot]
  table[row sep=crcr]{%
7	201.08889865134\\
8	200.646096477835\\
9	194.848795995114\\
};
\addlegendentry{DC, $\lambda = 40 \,\mu$s};

\addplot [color=mycolor1a,solid,line width=1.2pt,mark size=0.5pt,mark=*,mark options={solid}]
  table[row sep=crcr]{%
1 106.785614728653\\
2 106.619802810933\\
3 106.245801943528\\
};
\addplot [color=mycolor1a,solid,line width=1.2pt,mark size=0.5pt,mark=*,mark options={solid}, forget plot]
  table[row sep=crcr]{%
4 106.860627608393\\
5 106.699054877929\\
6 106.54228697362\\
};
\addplot [color=mycolor1a,solid,line width=1.2pt,mark size=0.5pt,mark=*,mark options={solid}, forget plot]
  table[row sep=crcr]{%
7 106.860627608393\\
8 106.699054877929\\
9 106.54228697362\\
};
\addlegendentry{DC, $\lambda = 80 \,\mu$s};

\addplot [color=mycolor2a,solid,line width=1.2pt,mark size=1.5pt,mark=x,mark options={solid}]
  table[row sep=crcr]{%
1 53.4199239837223\\
2 52.8792266113488\\
3 52.5776508582627\\
};
\addplot [color=mycolor2a,solid,line width=1.2pt,mark size=1.5pt,mark=asterisk,mark options={solid}, forget plot]
  table[row sep=crcr]{%
4 53.4199239837223\\
5 52.8792266113488\\
6 52.5776508582627\\
};
\addplot [color=mycolor2a,solid,line width=1.2pt,mark size=1.5pt,mark=asterisk,mark options={solid}, forget plot]
  table[row sep=crcr]{%
7 53.4199239837223\\
8 52.8792266113488\\
9 52.5776508582627\\
};
\addlegendentry{DC, $\lambda = 160 \,\mu$s};

\addplot [color=mycolor3,dashed,line width=1.2pt,mark size=0.5pt,mark=*,mark options={solid}]
  table[row sep=crcr]{%
1	262.833174536411\\
2	262.178698301453\\
3	262.770805121608\\
};
\addplot [color=mycolor3,dashed,line width=1.2pt,mark size=1.1pt,mark=square,mark options={solid}, forget plot]
  table[row sep=crcr]{%
4	268.156871343831\\
5	267.593864499285\\
6	267.510789360134\\
};
\addplot [color=mycolor3,dashed,line width=1.2pt,mark size=1.1pt,mark=square,mark options={solid}, forget plot]
  table[row sep=crcr]{%
7	268.397552688197\\
8	267.738995802522\\
9	267.752083287549\\
};
\addlegendentry{HH, $\lambda = 20 \,\mu$s};

\addplot [color=mycolor4,dashed,line width=1.2pt,mark size=1.5pt,mark=x,mark options={solid}]
  table[row sep=crcr]{%
1	184.118885837104\\
2	183.777653799541\\
3	183.413155819095\\
};
\addplot [color=mycolor4,dashed,line width=1.2pt,mark size=1.5pt,mark=x,mark options={solid}, forget plot]
  table[row sep=crcr]{%
4	186.957905705981\\
5	186.987820433086\\
6	185.247888628536\\
};
\addplot [color=mycolor4,dashed,line width=1.2pt,mark size=1.5pt,mark=x,mark options={solid}, forget plot]
  table[row sep=crcr]{%
7	186.957905705981\\
8	186.987820433086\\
9	185.247888628536\\
};
\addlegendentry{HH, $\lambda = 40 \,\mu$s};

\addplot [color=mycolor3a,dashed,line width=1.2pt,mark size=1.1pt,mark=square,mark options={solid}]
  table[row sep=crcr]{%
1 103.955844320906\\
2 104.546529611766\\
3 103.086042826011\\
};
\addplot [color=mycolor3a,dashed,line width=1.2pt,mark size=1.1pt,mark=square,mark options={solid}, forget plot]
  table[row sep=crcr]{%
4 104.442618391248\\
5 104.976830041456\\
6 103.362441693328\\
};
\addplot [color=mycolor3a,dashed,line width=1.2pt,mark size=1.1pt,mark=square,mark options={solid}, forget plot]
  table[row sep=crcr]{%
7 104.442618391248\\
8 104.976830041456\\
9 103.362441693328\\
};
\addlegendentry{HH, $\lambda = 80 \,\mu$s};

\addplot [color=mycolor4a,dashed,line width=1.2pt,mark size=2.6pt,mark=diamond,mark options={solid}]
  table[row sep=crcr]{%
1 53.3573144717902\\
2 52.8957457514293\\
3 51.9746730336081\\
};
\addplot [color=mycolor4a,dashed,line width=1.2pt,mark size=1.5pt,mark=x,mark options={solid}, forget plot]
  table[row sep=crcr]{%
4 53.4187977379307\\
5 52.9722875697036\\
6 51.9841103053596\\
};
\addplot [color=mycolor4a,dashed,line width=1.2pt,mark size=1.5pt,mark=x,mark options={solid}, forget plot]
  table[row sep=crcr]{%
7 53.4187977379307\\
8 52.9722875697036\\
9 51.9841103053596\\
};
\addlegendentry{HH, $\lambda = 160 \,\mu$s};

\end{axis}
\end{tikzpicture}%

%% file: figures/perf/th_a_80_b_10_x2.tex
%
%
\definecolor{mycolor1}{rgb}{0.01651,0.42660,0.87863}%
\definecolor{mycolor2}{rgb}{0.14527,0.70976,0.66463}%
\definecolor{mycolor3}{rgb}{0.07794,0.50399,0.83837}%
\definecolor{mycolor4}{rgb}{0.34817,0.74243,0.54727}%
\definecolor{mycolor5}{rgb}{0.03434,0.59658,0.81985}%
\definecolor{mycolor6}{rgb}{0.57086,0.74852,0.44939}%
\definecolor{mycolor7}{rgb}{0.02666,0.66420,0.76072}%
\definecolor{mycolor8}{rgb}{0.75249,0.73840,0.37681}%
\begin{tikzpicture}
\tikzstyle{every node}=[font=\scriptsize]
\pgfplotsset{every x tick label/.append style={font=\tiny, yshift=0.5ex}}

\begin{axis}[%
width=0.951\fwidth,
height=\fheight,
at={(0\fwidth,0\fheight)},
scale only axis,
xmin=0.5,
xmax=3.5,
xlabel={$D_{X2}$ [ms]},
xtick={1,2,3},
xticklabels={$0.1$, $1$, $10$},
xticklabel style={text width=6, align=center}, xmajorgrids,
ymajorgrids, ylabel shift = -5 pt, yticklabel shift = -3 pt,
ymin=102,
ymax=109,
ylabel={$\hat{S}_{PDCP}$ [Mbit/s]},
ymajorgrids, ylabel shift = -5 pt, yticklabel shift = -3 pt,
axis background/.style={fill=white},
legend columns=2,
legend style={legend cell align=left,align=left,draw=white!15!black,at={(0.05\fwidth,0.98\fheight)},anchor=south west}
]
\addplot [color=mycolor1,solid,line width=1.2pt,mark size=2.6pt,mark=diamond,mark options={solid}]
  table[row sep=crcr]{%
1	106.860627608393\\
2	106.699054877929\\
3	106.54228697362\\
};
\addlegendentry{DC, $s=2$m/s};

\addplot [color=mycolor3,solid,line width=1.2pt,mark size=1.0pt,mark=triangle,mark options={solid}]
  table[row sep=crcr]{%
1	107.106924150481\\
2	106.978486082912\\
3	106.505470903795\\
};
\addlegendentry{DC, $s=4$m/s};

\addplot [color=mycolor5,solid,line width=1.2pt,mark size=1.0pt,mark=triangle,mark options={solid,rotate=180}]
  table[row sep=crcr]{%
1	107.927527329243\\
2	107.766572269939\\
3	107.159255714261\\
};
\addlegendentry{DC, $s=8$m/s};

\addplot [color=mycolor7,solid,line width=1.2pt,mark size=1.0pt,mark=triangle,mark options={solid,rotate=270}]
  table[row sep=crcr]{%
1	108.308742724314\\
2	108.409654895397\\
3	106.324934627615\\
};
\addlegendentry{DC, $s=16$m/s};

\addplot [color=mycolor2,dashed,line width=1.2pt,mark size=1.5pt,mark=+,mark options={solid}]
  table[row sep=crcr]{%
1	104.442618391248\\
2	104.976830041456\\
3	103.362441693328\\
};
\addlegendentry{HH, $s=2$m/s};

\addplot [color=mycolor4,dashed,line width=1.2pt,mark size=1.5pt,mark=o,mark options={solid}]
  table[row sep=crcr]{%
1	104.351786426431\\
2	104.622229194511\\
3	103.344209314059\\
};
\addlegendentry{HH, $s=4$m/s};

\addplot [color=mycolor6,dashed,line width=1.2pt,mark size=1.5pt,mark=asterisk,mark options={solid}]
  table[row sep=crcr]{%
1	103.149662364008\\
2	102.984971041309\\
3	102.600358936605\\
};
\addlegendentry{HH, $s=8$m/s};

\addplot [color=mycolor8,dashed,line width=1.2pt,mark size=0.5pt,mark=*,mark options={solid}]
  table[row sep=crcr]{%
1	103.611564357035\\
2	103.587488763368\\
3	102.198991799163\\
};
\addlegendentry{HH, $s=16$m/s};

\end{axis}
\end{tikzpicture}%

%% file: figures/perf/rrc_80_B_10_speed.tex
%
%
\definecolor{mycolor1}{rgb}{0.01651,0.42660,0.87863}%
\definecolor{mycolor2}{rgb}{0.14527,0.70976,0.66463}%
\definecolor{mycolor3}{rgb}{0.07794,0.50399,0.83837}%
\definecolor{mycolor4}{rgb}{0.34817,0.74243,0.54727}%
\definecolor{mycolor5}{rgb}{0.03434,0.59658,0.81985}%
\definecolor{mycolor6}{rgb}{0.57086,0.74852,0.44939}%
\definecolor{mycolor7}{rgb}{0.02666,0.66420,0.76072}%
\definecolor{mycolor8}{rgb}{0.75249,0.73840,0.37681}%
\begin{tikzpicture}
\tikzstyle{every node}=[font=\scriptsize]
\pgfplotsset{every x tick label/.append style={font=\tiny, yshift=0.5ex}}

\begin{axis}[%
width=0.951\fwidth,
height=\fheight,
at={(0\fwidth,0\fheight)},
scale only axis,
xmin=0.5,
xmax=3.5,
xlabel={$D_{X2}$ [ms]},
xtick={1,2,3},
xticklabels={$0.1$, $1$, $10$},
xticklabel style={text width=6, align=center}, xmajorgrids,
ymin=4000,
ymax=12000,
ylabel={RRC throughput [bit/s]},
ymajorgrids, ylabel shift = -5 pt, yticklabel shift = -3 pt,
axis background/.style={fill=white},
legend columns=2,
legend style={legend cell align=left,align=left,draw=white!15!black,at={(0.1\fwidth,0.98\fheight)},anchor=south west}
]
\addplot [color=mycolor1,solid,line width=1.2pt,mark=diamond,mark options={solid}]
  table[row sep=crcr]{%
1	5512.66954068819\\
2	5468.48354158857\\
3	4371.5498896224\\
};
\addlegendentry{DC, $s =2$~m/s};

\addplot [color=mycolor2,dashed,line width=1.2pt,mark=+,mark options={solid}]
  table[row sep=crcr]{%
1	10777.6639054363\\
2	10499.8802450844\\
3	8534.12585464779\\
};
\addlegendentry{HH, $s =2$~m/s};

\addplot [color=mycolor3,solid,line width=1.2pt,mark=triangle,mark options={solid}]
  table[row sep=crcr]{%
1	5954.44870756334\\
2	5907.32793087574\\
3	4642.78093343619\\
};
\addlegendentry{DC, $s =4$~m/s};

\addplot [color=mycolor4,dashed,line width=1.2pt,mark=o,mark options={solid}]
  table[row sep=crcr]{%
1	10952.9507534999\\
2	10627.6201948363\\
3	8495.07202947351\\
};
\addlegendentry{HH, $s =4$~m/s};

\addplot [color=mycolor5,solid,line width=1.2pt,mark=triangle,mark options={solid,rotate=180}]
  table[row sep=crcr]{%
1	6696.08781924404\\
2	6565.80469566959\\
3	5092.69977101461\\
};
\addlegendentry{DC, $s =8$~m/s};

\addplot [color=mycolor6,dashed,line width=1.2pt,mark=asterisk,mark options={solid}]
  table[row sep=crcr]{%
1	11388.2295066441\\
2	11176.7192657598\\
3	8481.97753811186\\
};
\addlegendentry{HH, $s =8$~m/s};

\addplot [color=mycolor7,solid,line width=1.2pt,mark=triangle,mark options={solid,rotate=270}]
  table[row sep=crcr]{%
1	7853.62177680839\\
2	7786.73661774196\\
3	5905.27671229914\\
};
\addlegendentry{DC, $s =16$~m/s};

\addplot [color=mycolor8,dashed,line width=1.2pt,mark=*,mark options={solid}]
  table[row sep=crcr]{%
1	11211.5663381013\\
2	10910.0671826381\\
3	8459.37868907288\\
};
\addlegendentry{HH, $s =16$~m/s};

\end{axis}
\end{tikzpicture}%

%% file: figures/perf/x2pdcpRatio_x_1_b_10.tex
%
%
\definecolor{mycolor1}{rgb}{0.01651,0.42660,0.87863}%
\definecolor{mycolor2}{rgb}{0.14527,0.70976,0.66463}%
\definecolor{mycolor3}{rgb}{0.07794,0.50399,0.83837}%
\definecolor{mycolor4}{rgb}{0.34817,0.74243,0.54727}%
\definecolor{mycolor5}{rgb}{0.03434,0.59658,0.81985}%
\definecolor{mycolor6}{rgb}{0.57086,0.74852,0.44939}%
\definecolor{mycolor7}{rgb}{0.02666,0.66420,0.76072}%
\definecolor{mycolor8}{rgb}{0.75249,0.73840,0.37681}%
\begin{tikzpicture}
\tikzstyle{every node}=[font=\scriptsize]
\pgfplotsset{every x tick label/.append style={font=\tiny, yshift=0.5ex}}

\begin{axis}[%
width=0.951\fwidth,
height=\fheight,
at={(0\fwidth,0\fheight)},
scale only axis,
xmin=0.5,
xmax=4.5,
xlabel={UE speed $s$ [m/s]},
xmajorgrids,
xtick={1,2,3,4},
xticklabels={$2$, $4$, $8$, $16$},
ymin=0,
ymax=1.6,
ylabel={$\hat{S}_{X2}/\hat{S}_{PDCP}$},
ymajorgrids, ylabel shift = -5 pt, yticklabel shift = -3 pt,
legend columns=2,
axis background/.style={fill=white},
legend style={legend cell align=left,align=left,draw=white!15!black,at={(0.1\fwidth,0.4\fheight)},anchor=west}
]
\addplot [color=mycolor1,solid,line width=1.2pt,mark size=1.5pt,mark=star,mark options={solid}]
  table[row sep=crcr]{%
1	1.41271627433061\\
2	1.39493625268708\\
3	1.4454931157743\\
4	1.5053027164776\\
};
\addlegendentry{DC, $\lambda=20\,\,\mu$s};

\addplot [color=mycolor3,solid,line width=1.2pt,mark size=1.5pt,mark=star,mark options={solid}]
  table[row sep=crcr]{%
1	1.04775756259896\\
2	1.01854527742555\\
3	1.07292453927942\\
4	1.19094110962294\\
};
\addlegendentry{DC, $\lambda=40\,\,\mu$s};

\addplot [color=mycolor5,solid,line width=1.2pt,mark size=1.5pt,mark=+,mark options={solid}]
  table[row sep=crcr]{%
1	0.959594379511032\\
2	0.941672104818571\\
3	0.96242326479685\\
4	1.06013950095952\\
};
\addlegendentry{DC, $\lambda=80\,\,\mu$s};

\addplot [color=mycolor7,solid,line width=1.2pt,mark size=1.5pt,mark=o,mark options={solid}]
  table[row sep=crcr]{%
1	0.93253932101603\\
2	0.910924295052717\\
3	0.940467904532066\\
4	0.822297836799396\\
};
\addlegendentry{DC, $\lambda=160\,\,\mu$s};

\addplot [color=mycolor2,dashed,line width=1.2pt,mark size=1.5pt,mark=+,mark options={solid}]
  table[row sep=crcr]{%
1	0.184955373967971\\
2	0.198507982794856\\
3	0.207997176934134\\
4	0.332443233028718\\
};
\addlegendentry{HH, $\lambda=20\,\,\mu$s};

\addplot [color=mycolor4,dashed,line width=1.2pt,mark size=1.5pt,mark=o,mark options={solid}]
  table[row sep=crcr]{%
1	0.180714191553949\\
2	0.189529574177409\\
3	0.196768012248116\\
4	0.318380460006693\\
};
\addlegendentry{HH, $\lambda=40\,\,\mu$s};

\addplot [color=mycolor6,dashed,line width=1.2pt,mark size=1.5pt,mark=asterisk,mark options={solid}]
  table[row sep=crcr]{%
1	0.180232221458091\\
2	0.203434446425622\\
3	0.195743034310453\\
4	0.34330377253605\\
};
\addlegendentry{HH, $\lambda=80\,\,\mu$s};

\addplot [color=mycolor8,dashed,line width=1.2pt,mark size=0.5pt,mark=*,mark options={solid}]
  table[row sep=crcr]{%
1	0.171943776979895\\
2	0.208942087617335\\
3	0.179296676678791\\
4	0.327768674959025\\
};
\addlegendentry{HH, $\lambda=160\,\,\mu$s};

\end{axis}
\end{tikzpicture}%

%% file: chapters/conclusion.tex
\chapter{Conclusions And Future Work}\label{chap:conclusion}

This thesis introduced dual connectivity with fast switching at the PDCP layer as an architecture for LTE-5G tight integration, and presented a performance evaluation of the system, comparing it with the baseline of hard handover between RATs. 

After having illustrated the main technologies which are expected to be part of the next 5G standard, the main challenges of mmWave communications were described. The main proposals for LTE-5G tight integration were then reviewed, along with the state of the art in the LTE mobile protocol stack, dual connectivity and handover procedures.

Besides, the ns--3 simulator was presented as the tool used for the performance evaluation. In particular, the mmWave module developed by NYU was illustrated, with specific attention to its main strengths in modeling the mmWave channel. The LTE module was also examined. 

The network procedures and the architecture for integration were presented next, together with a discussion on their implementation in ns--3. Details were given on the necessary control signalling and on the modifications at the RRC layer, on the definition of a dual-connected network device in the simulator, and on the implementation of handover between LTE and 5G. This was the first main contribution of this Thesis. The second is the definition of a simulation scenario to compare fast switching and hard handover, and the simulation campaign that allowed to describe the performance of the system with several metrics. 

Results showed that, given a suitable latency on the X2 interface, the fast switching solution is able to provide a lower latency for RLC PDU transmissions. Moreover, it guarantees lower packet losses and RRC traffic, so that the overhead due to control traffic is reduced. On the other hand, it sometimes performs retransmissions of already successfully transmitted packets, and needs a minimum rate on the X2 interfaces comparable to the sum of the rates of the UEs that the mmWave eNB wants to support. 

The handover, on the other hand, has the benefit of requiring a smaller level of integration and coordination between the two RATs, a lower computational load on the LTE eNB and a lower utilization of the X2 link, but the performance in terms of latency is worse than with the fast switching setup, because of the service interruption and buffering required during the handover operation. Moreover, one of the main limitations of the handover solution is the RRC traffic, which is much higher with respect to the fast switching system.

Therefore, it is possible to conclude that the fast switching option is preferable, but its deployment must be carefully designed.

\section{Future Work}
As future work, it will be interesting to implement the coordinator in a node different from the LTE eNB, and study architectural solutions to manage the relation of the coordinator with the core network on one hand and the radio access on the other. 

Moreover, more refined algorithms for handover can be implemented and tested, in order to minimize the number of handovers and switches, by trying to predict the behavior of the channel. Enhanced procedures for secondary cells handover will be proposed and studied, in order to minimize service interruption time of secondary cells. 

Finally, the simulation framework implemented for the evaluations of this Thesis is very flexible and can be adapted also to other studies, such as, for example, the evaluation of dual connectivity to increase the throughput, control and user plane split or diversity and the performance of multipath TCP algorithms.